  \documentclass{aa}
   \usepackage{graphicx}
  \begin{document}


  \title{On a precessing jet-nozzle scenario with a common helical 
   trajectory-pattern for blazar 3C345}
     
  \author{S.J.~Qian\inst{1}}
   \institute{
    National Astronomical Observatories, Chinese Academy of Sciences,
    Beijing 100012, China}
 \date{Compiled by using A\&A latex}
 \abstract{Based on the possible existence of the jet precession, 
     the kinematics and flux evolution of superluminal components 
     in blazar 3C345 were interpreted in the framework of the precessing 
     jet-nozzle scenario with a precessing common helical trajectory-pattern.}
     {This study is to show that the jet in 3C345 precesses with a period of
     7.3\,yr and the superluminal knots move consistently along a precessing
     common helical trajectory-pattern in their inner jet regions, while in the
     outer jet regions they follow their own individual tracks. The 
     trajectory-transits can extend to core distances of $\sim$1.2\,mas, or
     traveled distances of $\sim$300\,pc (for example for knots C4 and C9).}
     {Through model-fitting of the observed kinematic behavior of the  
    superluminal components, their bulk Lorentz factor and Doppler 
    factor are  derived as continuous functions of time which were used 
    to investigate their flux evolution.} 
    {It is found that the light-curves of the superluminal components
     observed at 15, 22 and 43\,GHz can be  well explained by 
    their Doppler boosting effect or model-fitted in terms of their Doppler
    boosting profiles ($\propto[\delta(t)]^{3+\alpha}$, $\alpha$--spectral 
    index) associated with their superluminal motion. Additionally, 
     flux fluctuations on shorter time-scales also exist
     due to variations in knots' intrinsic flux density and spectral index.} 
    { The close association of the flux evolution 
    with the  Doppler-boosting effect is important, not only firmly validating
    the precessing jet-nozzle scenario being fully appropriate to explain the
    kinematics and emission properties of superluminal components in QSO 3C345,
     but also strongly supporting the traditional
   common pointview: superluminal components are physical entities 
   (shocks or plasmons) participating relativistic motion towards us with 
   acceleration/deceleration along helical trajectories. 
   Finally, we have proposed both the  precessing nozzle scenarios 
   with a single jet and double jets (this paper and Qian [2022b]) 
    to understand the 
   VLBI-phenomena measured in 3C345. VLBI-observations with higher resolutions
   deep into the core regions (core distances $<$0.1\,mas) are
    required to test them.}
  \maketitle
  \section{Introduction}
   3C345 (z=0.595) is a prototypical quasar emanating emission over the entire
   electromagnetic spectrum from radio, infrared/optical/UV and  X-rays 
   to high-energy $\gamma$ rays. It is also  one of the best-studied blazars.
    3C345 is a remarkable compact flat-spectrum radio source which was 
    one of the firstly discovered quasars to have a relativistic jet, emanating
    superluminal components steadily. Its flaring  activities in multifrequency
    bands  (from radio to $\gamma$-rays) are closely connected with the jet
    activity and ejection of superluminal components (Biretta et al.
    \cite{Bi86}, Hardee \cite{Ha87},  Babadzhantants et al. \cite{Ba95},
    Zensus \cite{Ze97}, Klare \cite{Kl03},
     Klare et al. \cite{Kl05}, Jorstad et al. \cite{Jo05}, \cite{Jo13},
     \cite{Jo17}, Lobanov \& Roland \cite{Lo05}, Qian \cite{Qi09},
     Schinzel et al. \cite{Sc10}, Schinzel \cite{Sc11}). \\
      Since 1991 (Qian et al. \cite{Qi91a}, \cite{Qi91b}), we have started to 
    interpret the VLBI-phenomenon in QSO 3C345, explaining the kinematics of 
    its superluminal components  in terms of a precessing jet-nozzle scenario. 
    \begin{itemize}
    \item (1) In the recent work (Qian, \cite{Qi22a})  the kinematics of 
    twenty-seven superluminal components observed during a 38-year period was 
    explained in detail. We proposed the hypothesis that these 
    knots might be possibly divided into two groups (group-A [13knots] and 
    group-B [14 knots]),
    which were ejected from  their precessing nozzles of a
    double-jet structure (jet-A and jet-B), respectively. Interestingly,
    it was found that both nozzles precess with the same period 7.3\,yr in
    the same direction (anticlockwise seen in the line of sight).
    The whole kinematic behaviors of these knots observed by 
     VLBI-observations (including trajectories,
     coordinates, apparent speed) can be  consistently well model-fitted and 
    explained. Their Lorentz and Doppler factor were derived as continuous
    functions of time. Thus their flux evolution can be investigated.
    \item (2) In the works of Qian (\cite{Qi22b},\cite{Qi23}), the flux
     evolution of the five knots of group-A (C4, C5, C9, C10 and C22) and 
    the five knots of  group-B (C19, C20, C21, B5 and B7) were investigated.
    It was found that their radio light-curves can be well fitted by their
    Doppler-boosting profiles. That is, the variability timescales of the knots
    are consistent with those of their Doppler factor and  they light curves
    are well fitted by Doppler-boosting effect
     ($\propto[{\delta(t)}]^{3+\alpha}$; $\alpha$--spectral index, 
    ${S_{\nu}}\propto{\nu^{-\alpha}}$).  
     \item (3) We also found that the knots of group-A have a precessing
     common helical trajectory along which they moved according to their
     precession phases (or ejection times, see Figure 2 below) in the
     inner-jet regions. Thus their curved trajectories and the swing of their
     ejection position angle can be well explained. Although their inner 
     trajectories followed the common helical pattern, in the outer-jet regions
     they move along their own individual trajectories. The transitions between
     the common trajectory and individual trajectory have been determined for
      these superluminal knots (see Table 2 below). Mostly, the precessing
      common trajectory can extend to traveled distance of $\sim$50--300\,pc.
     \item (4) An important phenomenon was trajectory pattern periodically 
    recurred. For example: the observed trajectories of knots  C5, C9 and 
    C22 showed similar curved shapes with similar ejection precession phases
    C5[5.83\,rad, 1980.80], C9[5.54\,rad+4$\pi$, 1995.06]
     and C22[5.28\,rad+8$\pi$,2009.36]). This strongly demonstrates 
    that both the precessing common trajectory and the nozzle precession are 
     really existing.
   \item (5) Based on the model-fitting of  the kinematics of the knots in 
    terms of the precessing double-jet nozzle scenario,
    the Doppler-factor of the knots as function of time was determined 
    and  used to explain the flux evolution of the knots by Doppler boosting
     effect. Thus by using our precessing 
    jet-nozzle scenario, the entire properties of the knots in 3C345 observed
     (kinematics and flux evolution) can all-sidedly and consistently be
     interpreted;
    \item (7) However, the double precessing jet scenario would involve
     a supermassive binary black hole and with its double relativistic jets 
     which have not been investigated theoretically in detail (Artymovicz
    \cite{Ar98}, Lobanov \& Roland \cite{Lo05},  Begelman et al. \cite{Be80},
     Blandford \& Znajek \cite{Bl77}, Blandford \& Payne \cite{Bl82},
     Meier \& Nakakura \cite{Me06}, Shi \& Krolik \cite{Sh15},
     Vlahakis \& K\"onigl \cite{Vl03}, \cite{Vl04}). This scenario 
      needs to be tested in the future by higher-resolution VLBI-observations
     deep into the core. Thus
    the double jet scenario was introduced only as an working hypothesis to 
    interprete the VLBI-phenomena in 3C345. In this paper, we would propose
    an alternative possibility [i.e. a single-jet precessing jet-nozzle
    scenario] to interprete the VLBI phenomena in 3C345.
    \end{itemize}
    \begin{figure*}
    \centering
    \includegraphics[width=8cm,angle=90]{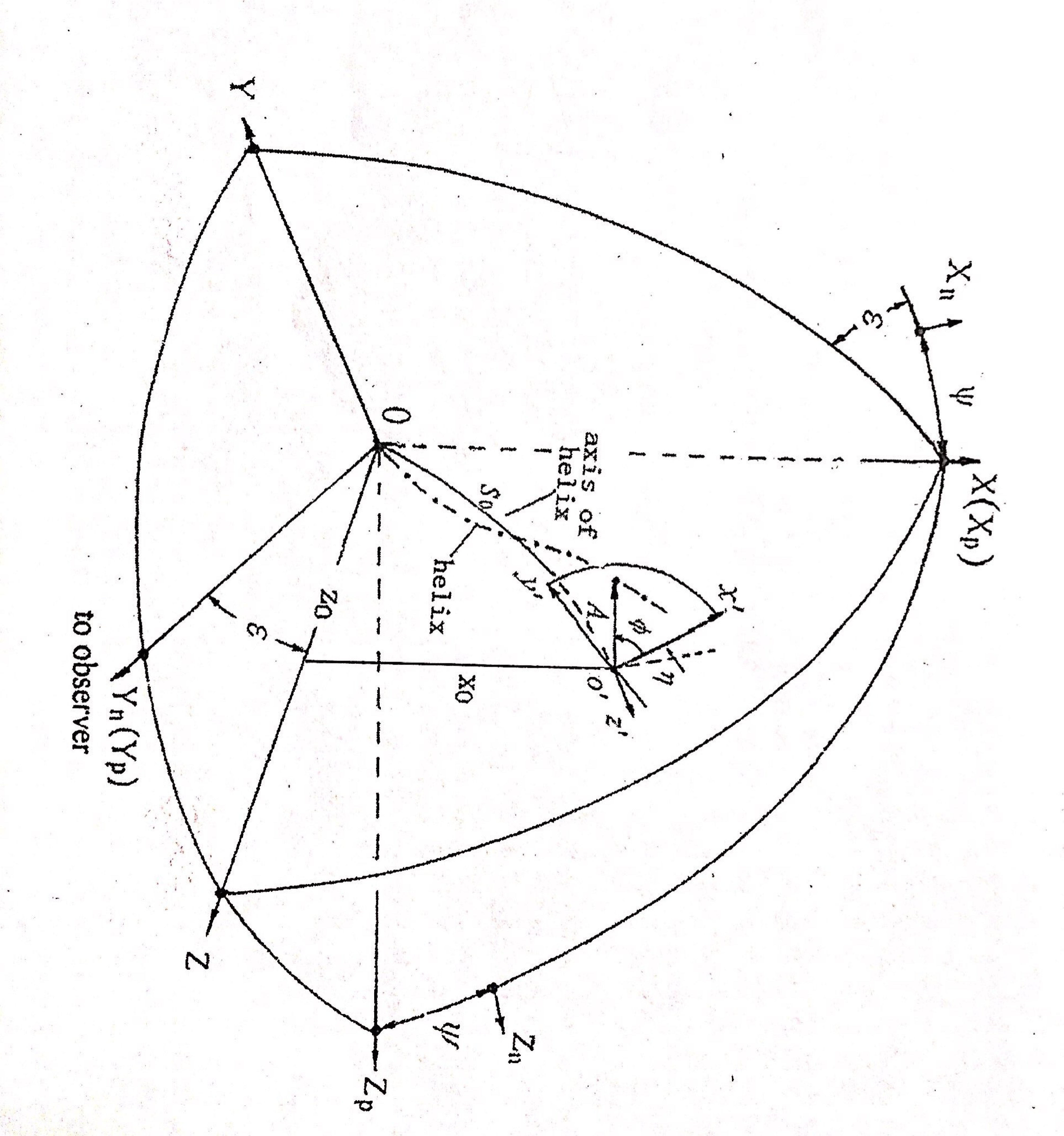}
    \caption{Geometry of the precessing jet-nozzle scenario for 3C345. 
    The jet-axis is defined in the $(X,Z)$-plane by parameters
    ($\epsilon$, $\psi$) and function $\it{x_0(z_0)}$. The common helical 
    trajectory pattern is defined by functions A(Z) and $\phi$(Z) given 
    in Section 2.3 (Figure 2). Because $\epsilon$ is very small, the change 
    in $\psi$ indicates a rotation around the line of sight.}
    \end{figure*}
    \section{The precessing jet nozzle scenario with a precessing common 
    helical trajectory pattern}
      The precessing jet-nozzle scenario has been previously described
     (cf. Qian \cite{Qi22a}, \cite{Qi22b}, \cite{Qi23}; Qian et al. 
    \cite{Qi09}, \cite{Qi91a}) in detail for 
    investigating the VLBI-kinematics of superluminal components
    on parsec scales in the QSO 3C345. The formulism of its geometry is
    recaptulated as  follows.\\
    A special geometry consisting of four coordinate systems is 
    shown in Figure 1. We assume that the superluminal components move along
    helical trajectories around the curved jet axis (i.e., the axis of
    the helix).\\
    We use coordinate system ($X_n,Y_n,Z_n$) to define the plane of the sky 
    ($X_n,Z_n$) and the direction of observer ($Y_n$), with $X_n$-axis pointing
    toward the negative right ascension and $Z_n$-axis toward the north pole.\\
    The coordinate system ($X,Y,Z$) is used to locate the 
    curved jet-axis in the 
    plane ($X,Z$), where $\epsilon$ represents the angle between $Z$-axis and
    $Y_n$-axis and $\psi$ the angle between $X$-axis and $X_n$-axis. 
    Thus parameters $\epsilon$ and $\psi$ are used to define the plane where 
    the jet-axis locates relative to the coordinate system ($X_n,Y_n,Z_n$).\\
    We use coordinate system (${\it{x}}'$,${\it{y}}'$,${\it{z}}'$) along the
     jet-axis to define the helical trajectory pattern for a knot, introducing
    parameters $A(s_0)$ (amplitude) and $\phi(s_0)$ (phase), where $s_0$
    represents the arc-length along the axis of helix (or curved jet-axis).
    ${\it{z}}'$-axis is along the tangent of the axis of helix.
    ${\it{y}}'$-axis is parallel to the $Y$-axis and $\eta$ is the angle
    between ${\it{x}}'$-axis and $X$-axis (see Figure 1).\\
     \subsection{Expressions for defining the jet axis} 
     In general we assume that the jet-axis can be defined by a function
     $x_0(z_0)$ in the $(X,Z)$-plane as follows.
    \begin{equation}
    {x_0}=p({z_0}){{z_0}^{\zeta}}
    \end{equation}
     where
    \begin{equation}
    p({z_0})={p_1}+{p_2}[1+\exp(\frac{{z_t}-{z_0}}{{z_m}})]^{-1}
    \end{equation}
     $\zeta$, $p_1$, $p_2$, $z_t$ and $z_m$ are constants. The exponential
     term is devised for describing the jet-axis gradually curving toward
     the north, as the trajectory of knot C4 shows on large-scales.
    \begin{equation}
    {s_0}=\int_{0}^{z_0}{\sqrt{1+(\frac{d{x_0}}{d{z_0}})^2}}{d{z_0}}
    \end{equation}
    Therefore, the helical trajectory of a knot can be described in the (X,Y,Z)
    system as follows.
    \subsection{Formulas for calculation of kinematic parameters}
    \begin{equation}
    X({s_0})=A({s_0}){\cos{\phi({s_0})}}{\cos{\eta({s_0})}}+{x_0}
    \end{equation}
    \begin{equation}
    Y({s_0})=A({s_0}){\sin{\phi({s_0})}}
    \end{equation}
    \begin{equation}
    Z({s_0})=-A({s_0}){\cos{\phi({s_0})}}{\sin{\eta({s_0})}}+{z_0}
    \end{equation}
    where $\tan{\eta({s_0})}$=$\frac{d{x_0}}{d{z_0}}$. The projection of
   the helical trajectory on the sky-plane (or the apparent trajectory)
     is represented by
   \begin{equation}
    {X_n}={X_p}{\cos{\psi}}-{Z_p}{\sin{\psi}}
    \end{equation}
    \begin{equation}
    {Z_n}={X_p}{\sin{\psi}}+{Z_p}{\cos{\psi}}
    \end{equation}
    where
    \begin{equation}
       {X_p}=X({s_0})
    \end{equation}
    \begin{equation}
     {Z_p}={Z({s_0})}{\sin{\epsilon}}-{Y({s_0})}{\cos{\epsilon}}
    \end{equation}
    (All coordinates and amplitude (A) are measured in units of mas).
    Introducing the functions
    \begin{equation}
    {\Delta}=\arctan[(\frac{dX}{dZ})^2+(\frac{dY}{dZ})^2]^{-\frac{1}{2}}
    \end{equation}
   \begin{equation}
    {{\Delta}_p}=\arctan(\frac{dY}{dZ})
    \end{equation}
    \begin{equation}
    {{\Delta}_s}=\arccos[(\frac{dX}{d{s_0}})^2+(\frac{dY}{d{s_0}})^2+
                        (\frac{dZ}{d{s_0}})^2]^{-\frac{1}{2}}
    \end{equation}
    we can then calculate the viewing angle $\theta$, apparent transverse
    velocity ${\beta}_a$, Doppler factor $\delta$ and the elapsed time T,
    at which the knot reaches distance $z_0$ as follows: 
    \begin{equation}
     {\theta}=\arccos[{\cos{\epsilon}}(\cos{\Delta}+
               \sin{\epsilon}\tan{{\Delta}_p})]
    \end{equation}
    \begin{equation}
     {\Gamma}=(1-{\beta}^2)^{-\frac{1}{2}}
    \end{equation}
    \begin{equation}
    {\delta}=[{\Gamma}(1-{\beta}{\cos{\theta}})]^{-1}
    \end{equation}
    \begin{equation}
     {{\beta}_{app}}={{\beta}{\sin{\theta}}/(1-{\beta}{\cos{\theta}})}
    \end{equation}
    \begin{equation}
    T=\int^{{s_0}}_{0}{\frac{(1+z)}{{\Gamma\delta}{v}{\cos{{\Delta}_s}}}}
                       {d{s_0}}
    \end{equation}
   \subsection{Precessing common helical trajectory pattern}
    The amplitude and phase of the precessing comm trajectory pattern (helical 
   trajectory pattern) for superluminal  knots 
   are defined as follows.
    \begin{equation}
    {A({Z})} = {{A_0}[\sin({\pi}{Z}/{Z_1})]^{1/2}} 
    \end{equation}
    \begin{equation}
    {\phi}({Z})={{\phi}_0}-{({Z}/{Z_2})^{1/2}}
     \end{equation}
      $A_0$ represents the amplitude coefficient of the common helical
     trajectory pattern and  ${\phi}_0$ is the precession phase of an 
     individual knot, which is related to its ejection time $t_0$:\\
     \begin{equation}
      {\phi_0}=4.28+\frac{2\pi}{T_0}({t_0}-1979.0)
      \end{equation}
     where $T_0$=7.30\,yr--precession period of the jet-nozzle.
   \subsection{Adopted values for the precessing jet nozzle scenario}   
   As shown above, the precessing common helical trajectory which 
   the superluminal
   components follow is defined by parameters ($\epsilon$, $\psi$) , and 
   formulas (1)-(2) and (19)-(21) and the corresponding parameters in them.
   The following values are adopted:\\
    \\
      $\epsilon$=0.0349\,rad=$2^{\circ}$; $\psi$=0.125\,rad=$7.16^{\circ}$; 
      $\zeta$=2.0, $p_1$=0; $p_2$=1.34$\times{10^{-4}}$; $z_t$=66\,mas; 
     $z_m$=6\,mas; $A_0$=0.605\,mas, $Z_1$=396\,mas and $Z_2$=3.58\,mas.\\
    \\ 
     We apply the concordant cosmological model (Spergel et al. \cite{Sp03}, 
    Hogg \cite{Hog99}) with $\Omega_{\lambda}$=0.73, $\Omega_m$=0.27 and
    $H_0$=71 km$s^{-1}$$Mpc^{-1}$. Thus the luminosity distance  of 3C345
    is $D_L$=3.49\,Gpc, angular-diameter distance 
       $D_a$=1.37\,Gpc, 1mas=6.65\,pc
    and 1mas/yr=34.6\,c. 1\,c is equivalent to an angular speed 0.046\,mas/yr. 
     \subsection{Basic equations for Doppler boosting effct and flux evolution}
      In order to investigate the relation between  the flux evolution of
      superluminal components and their Doppler boosting effect during their
      accelerated/decelerated motion along helical trajectories,
      the observed flux density of superluminal components can be described
      as follows.
     \begin{equation}
   {S_{obs}}(\nu,t)={S_{int}}(\nu,t){\times}{\delta(t)}^{3+\alpha(\nu,t)}
     \end{equation}
     $S_{obs}(\nu,t)$$\propto{\nu^{-\alpha(\nu,t)}}$--the observed flux density
    and $\alpha(\nu,t)$--spectral index at the observing frequency $\nu$,
    $S_{int}(\nu,t)$-- intrinsic flux density, $\delta(t)$--Doppler factor.
    Note that both variations in intrinsic flux and spectral index 
    with time and frequency can give rise to variations in the observed flux
    densities.\\
     We also use normalized flux density ${S_{obs,N}}(\nu,t)$ which is 
    defined as
     \begin{equation}
       {S_{obs,N}}(\nu,t)={S_{obs}}(\nu,t)/{S_{obs,max}}
     \end{equation}
      $S_{obs,max}$--the observed maximum flux density.\\
    In the precessing nozzle scenario the Doppler factor $\delta(t)$ of 
    superluminal knots depends on their  motion along helical
    trajectories which are produced by the precessing common trajectory at
    corresponding precession phases (or ejection epochs).  As shown in the
     previous works, the kinematics of the superluminal components can well be 
    model-fitted, and  their bulk Lorentz factor $\Gamma(t)$ and
    Doppler factor $\delta(t)$ as continuous functions of time can be derived,
    which can be used to investigate the relation between the Doppler boosting
    effect and their flux evolution. 
     \begin{figure*}
     \centering
     \includegraphics[width=6.5cm,angle=-90]{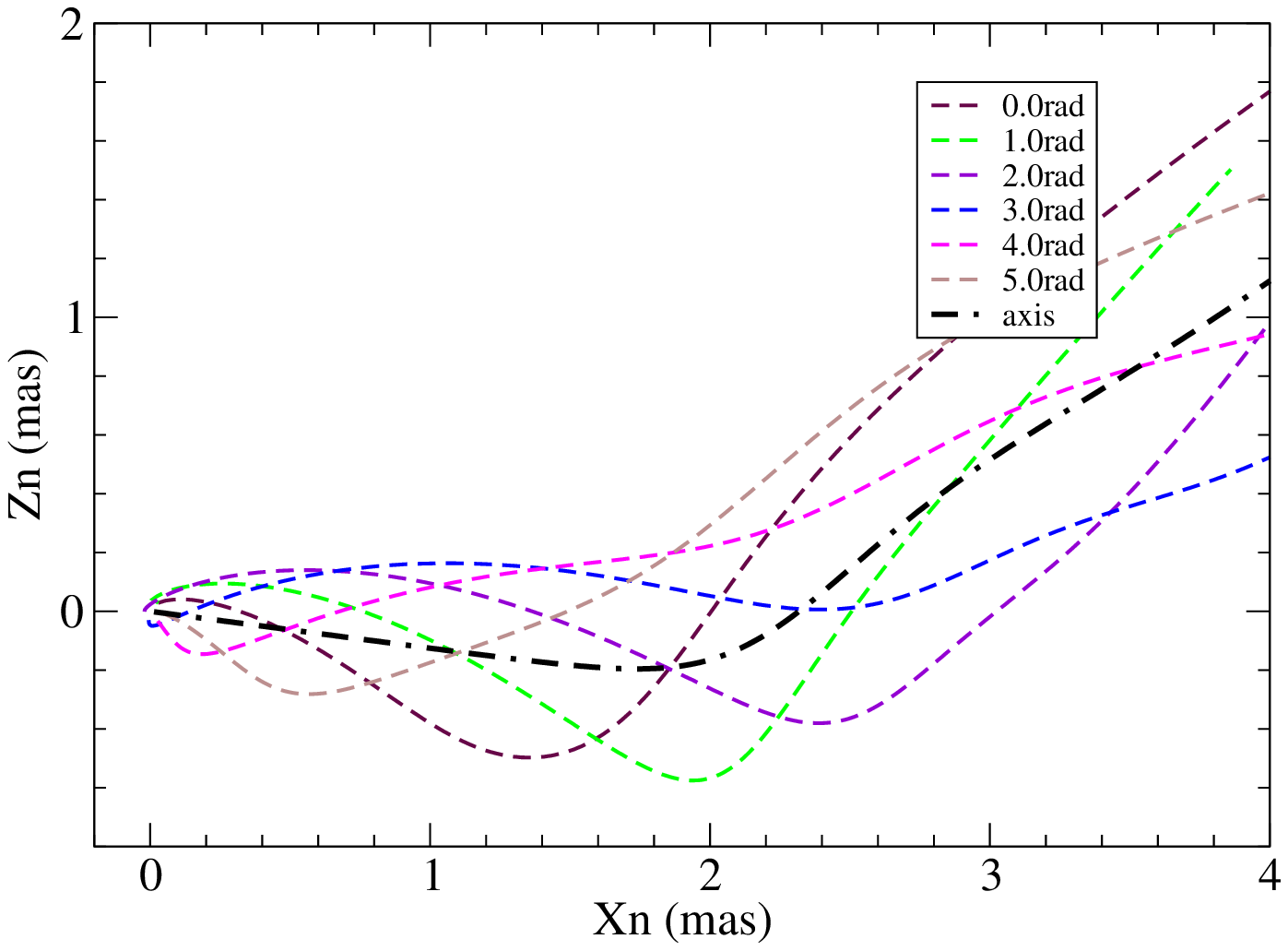}
     \caption{Apparent distribution of the precessing common helical
     trajectory for different precession phases of the
     superluminal components in 3C345: $\phi_0$=0.0, 1.0,
    2.0, 3.0, 4.0 and 5.0\,rad, showing the apparent curvature of trajectories 
    around the apparent jet-axis (black dashed line).}
     \end{figure*}
    \begin{figure*}
    \centering
    \includegraphics[width=5cm,angle=-90]{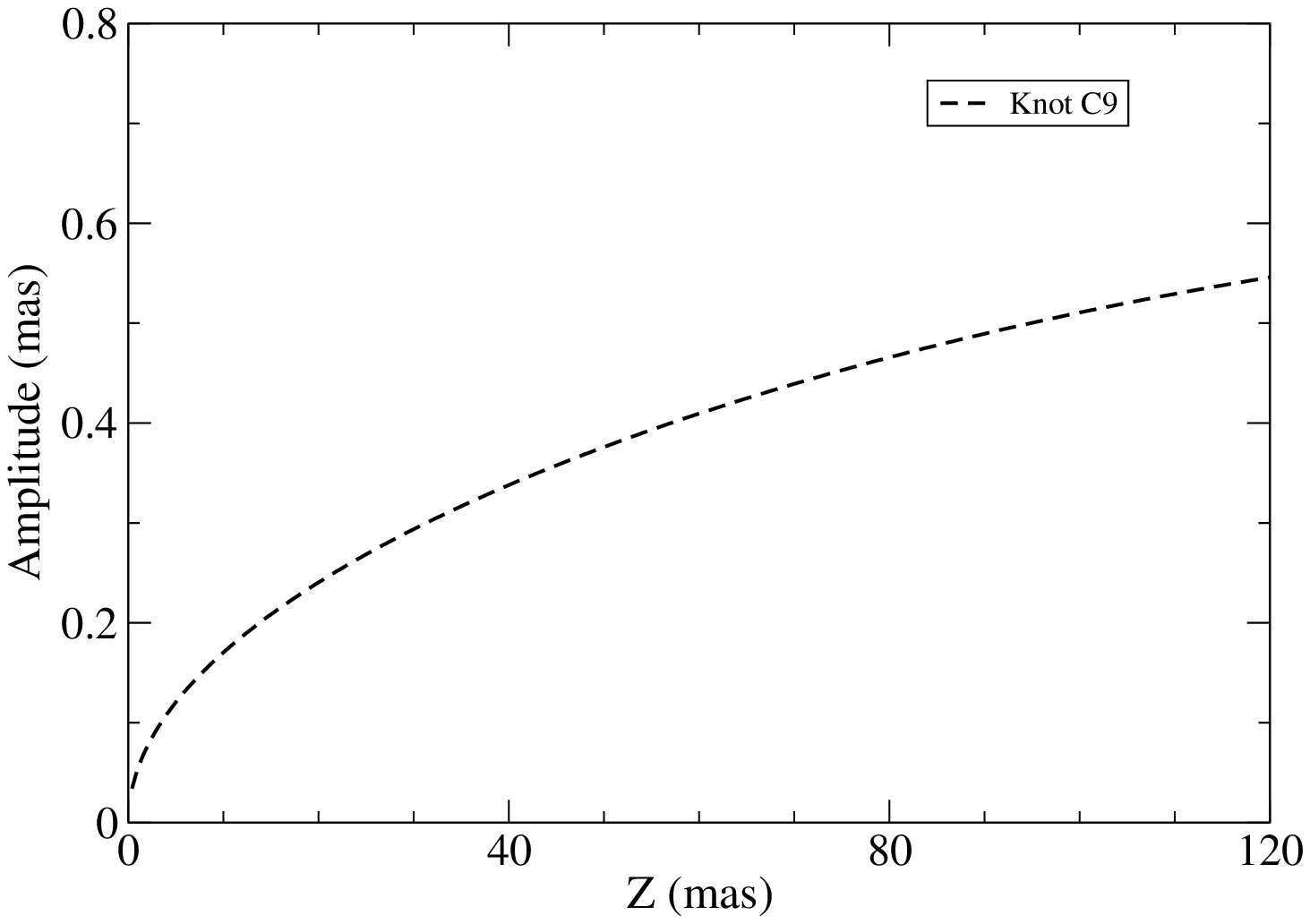}
    \includegraphics[width=5cm,angle=-90]{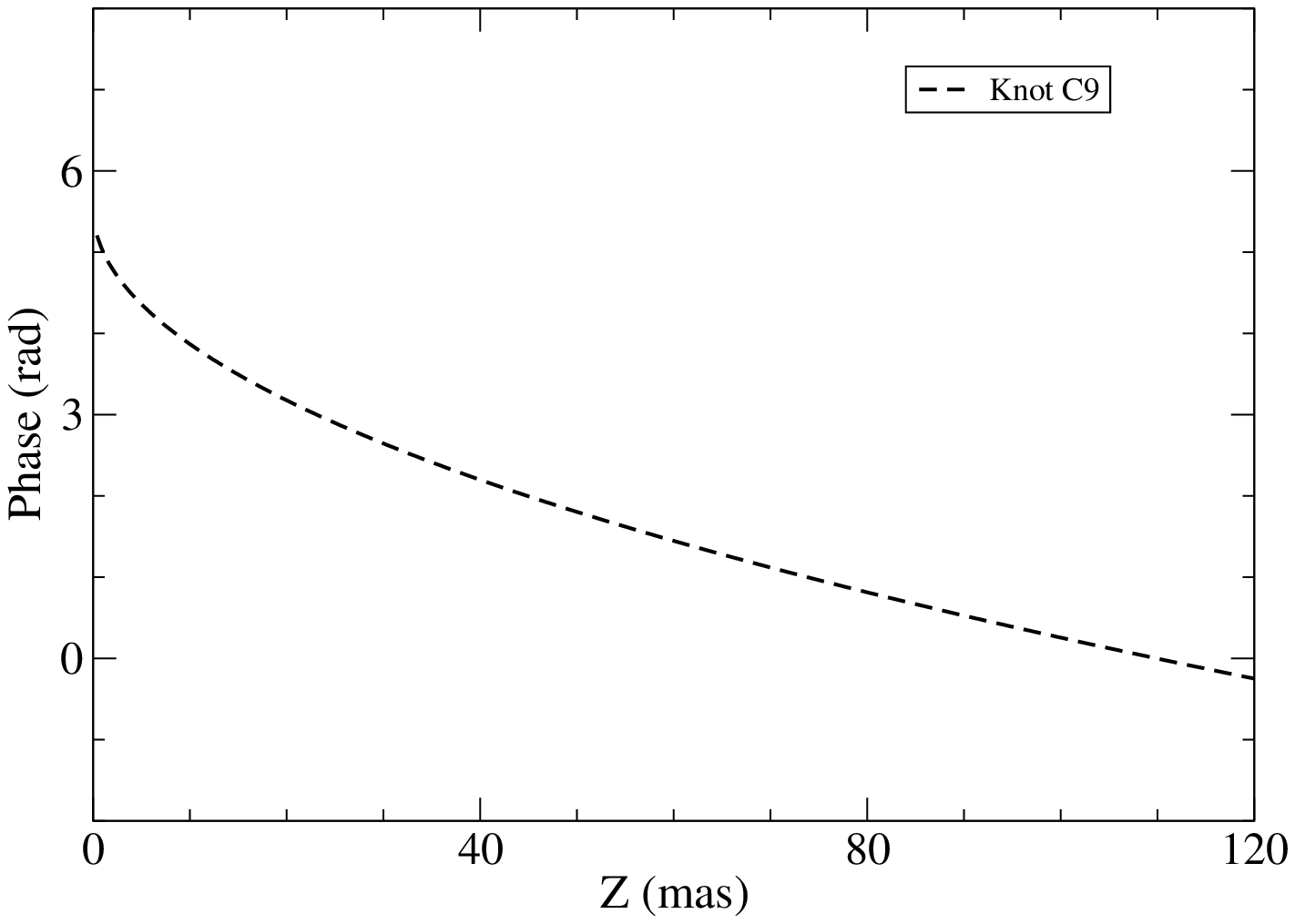}
    \caption{Knot C9: The amplitude A(Z) and phase $\phi$(Z) defining
     the precessing common helical trajectory pattern (equations (19) and (20)
     in the text.}
    \end {figure*}
    \begin{figure*}
    \centering
    \includegraphics[width=9cm,angle=-90]{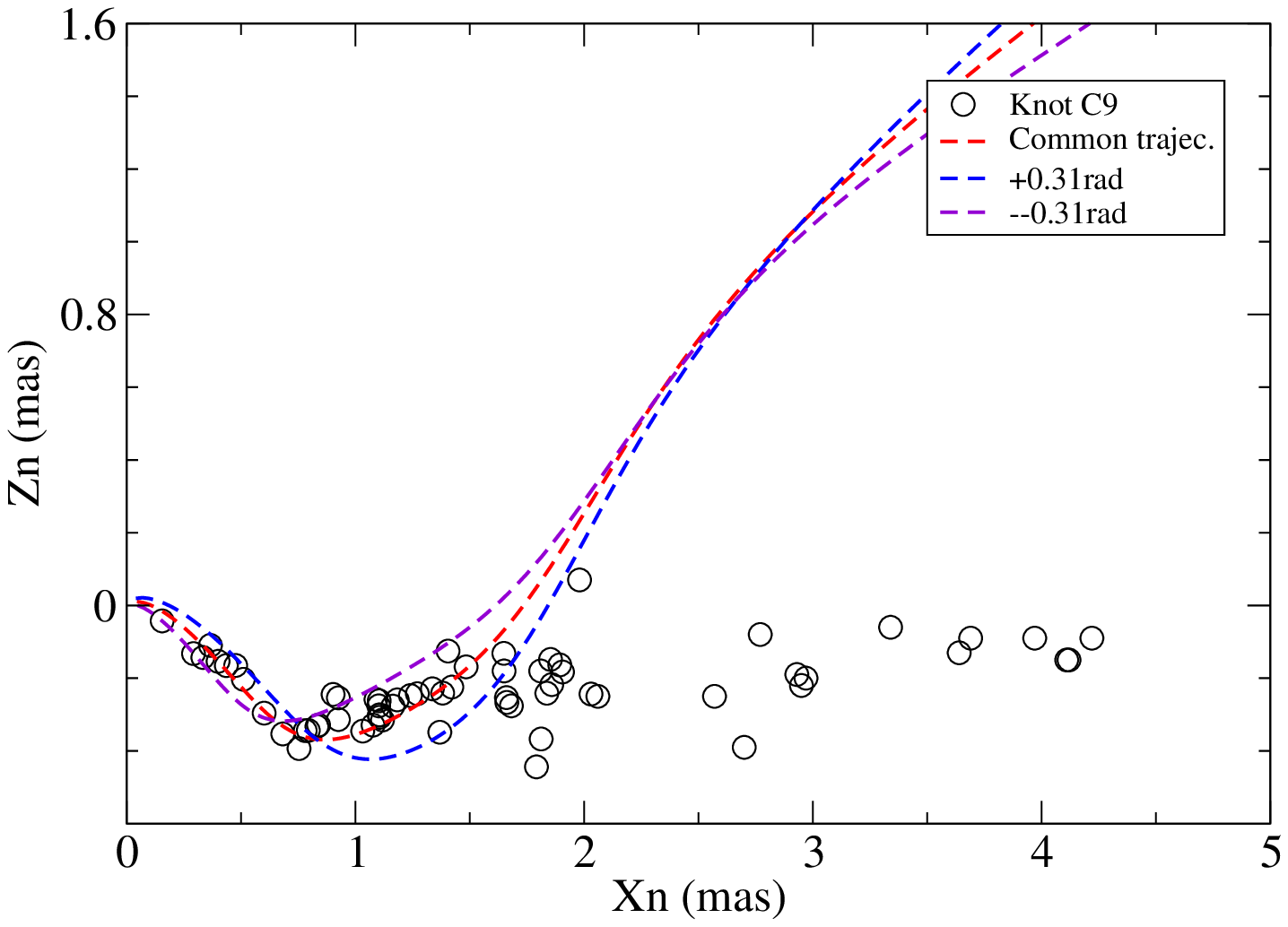}
    \caption{Knot C9: model-fitting of the observed trajectory in terms of 
    the precessing common helical trajectory pattern (red line with
    $\phi_0$=4.28\,rad; within $X_n{\leq}$1.2\,mas). The blue and violent 
    curves with $\phi_0$=4.28$\pm$0.31\,rad show the uncertainty range
    of $\pm$5\% of the precession period $T_0$ (=7.3\,yr). Beyond 
    $X_n$=1.2\,mas  its trajectory started to deviate northward from the 
     precessing common pattern, following its own individual track.}
    \end{figure*}
    \begin{figure*}
    \centering
    \includegraphics[width=5cm,angle=-90]{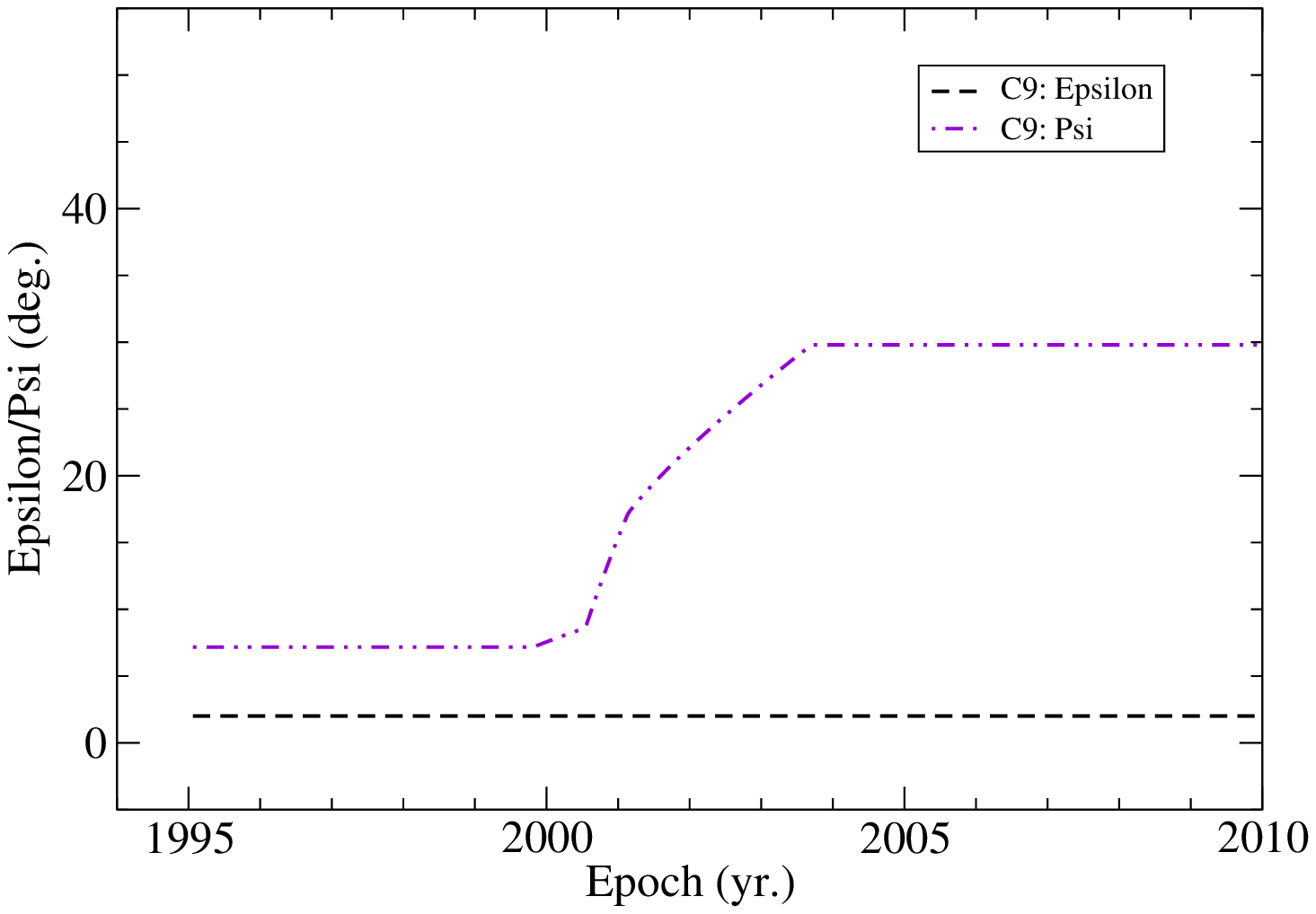}
    \includegraphics[width=5cm,angle=-90]{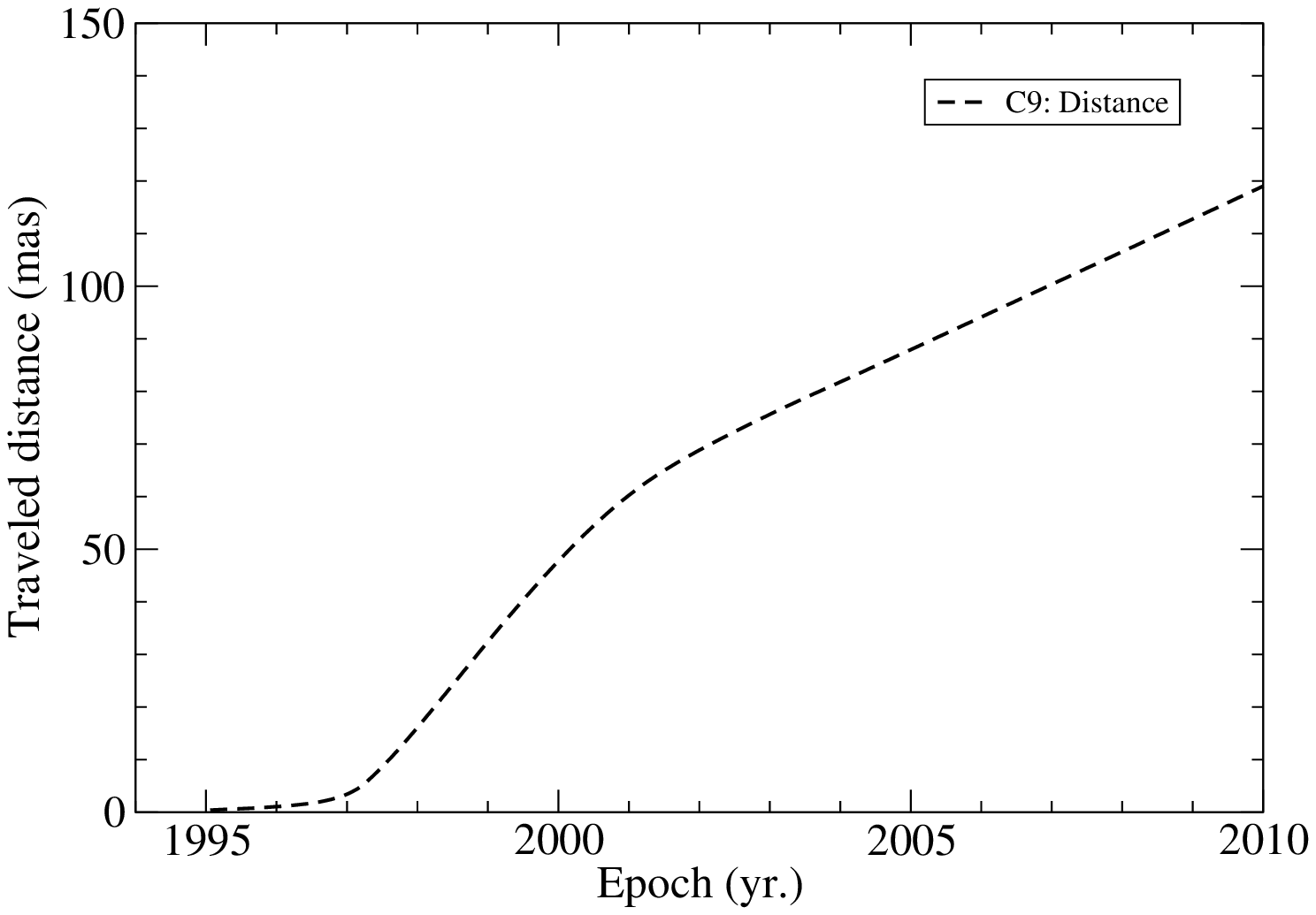}
    \caption{Knot C9. The modeled parameters $\epsilon(t)$ and $\psi(t)$
    (left panel) and the travelled distance Z(t) along the Z-axis
     (right panel), showing $\epsilon$=const.=$2^{\circ}$ and 
    $\psi$=$7.16^{\circ}$ before 1999.80 (corresponding $r_n{\leq}$1.25\,mas,
    $X_n{\leq}$1.21\,mas and traveled distance Z$\leq$44.8\,pc),
    when it moved along the precessing common pattern
    (in its inner trajectory-region). After 1999.80
     $\psi$ increased and it started to move along its own individual 
    track (in its outer trajectory-region) extending to the travelled
    distance $\sim$104\,mas$\sim$692\,pc; corresponding core distance
    $r_n{\sim}$4.1\,mas (see Fig.6). }
    \end{figure*}
    \begin{figure*}
    \centering
    \includegraphics[width=5cm,angle=-90]{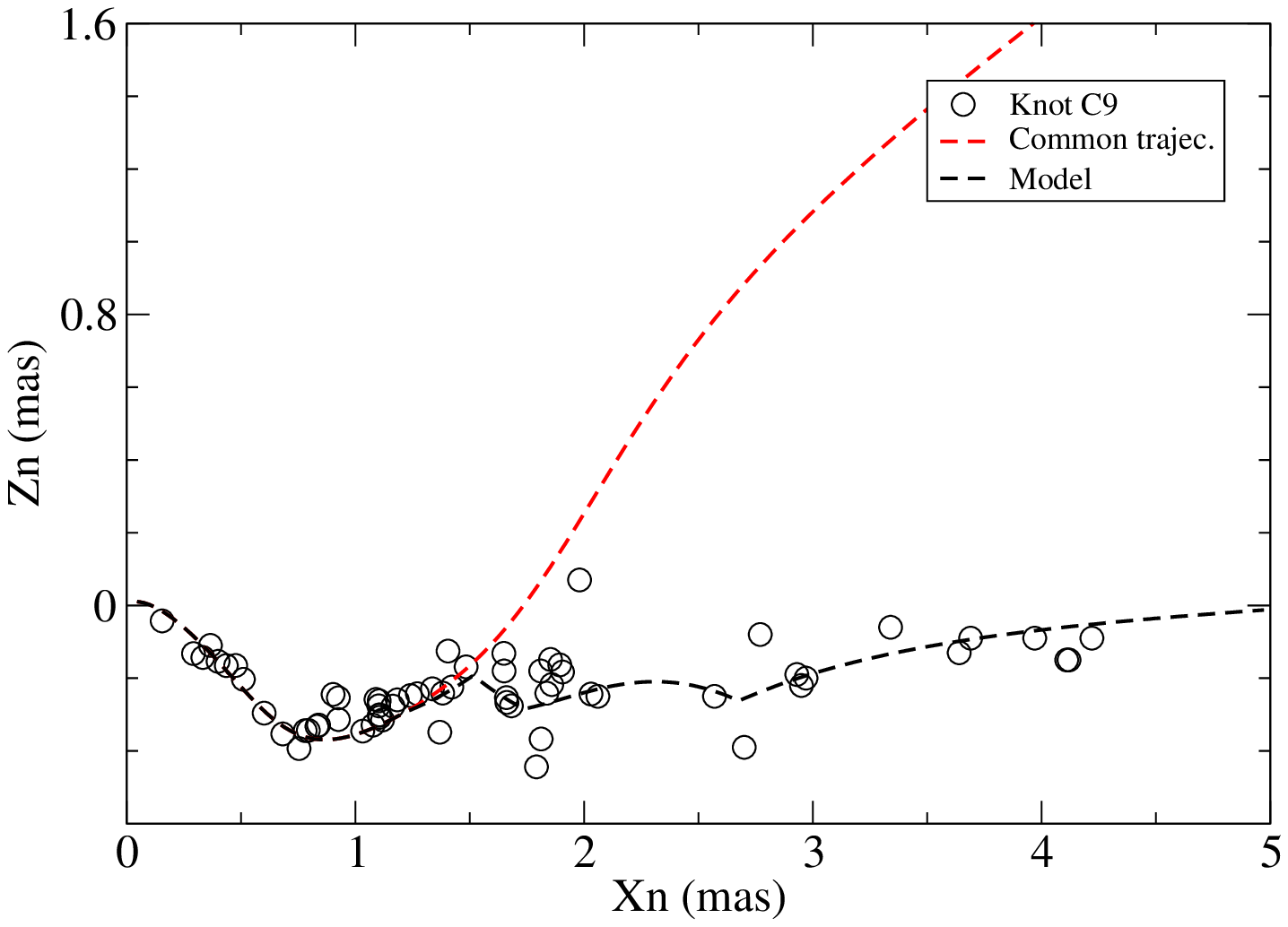}
    \includegraphics[width=5cm,angle=-90]{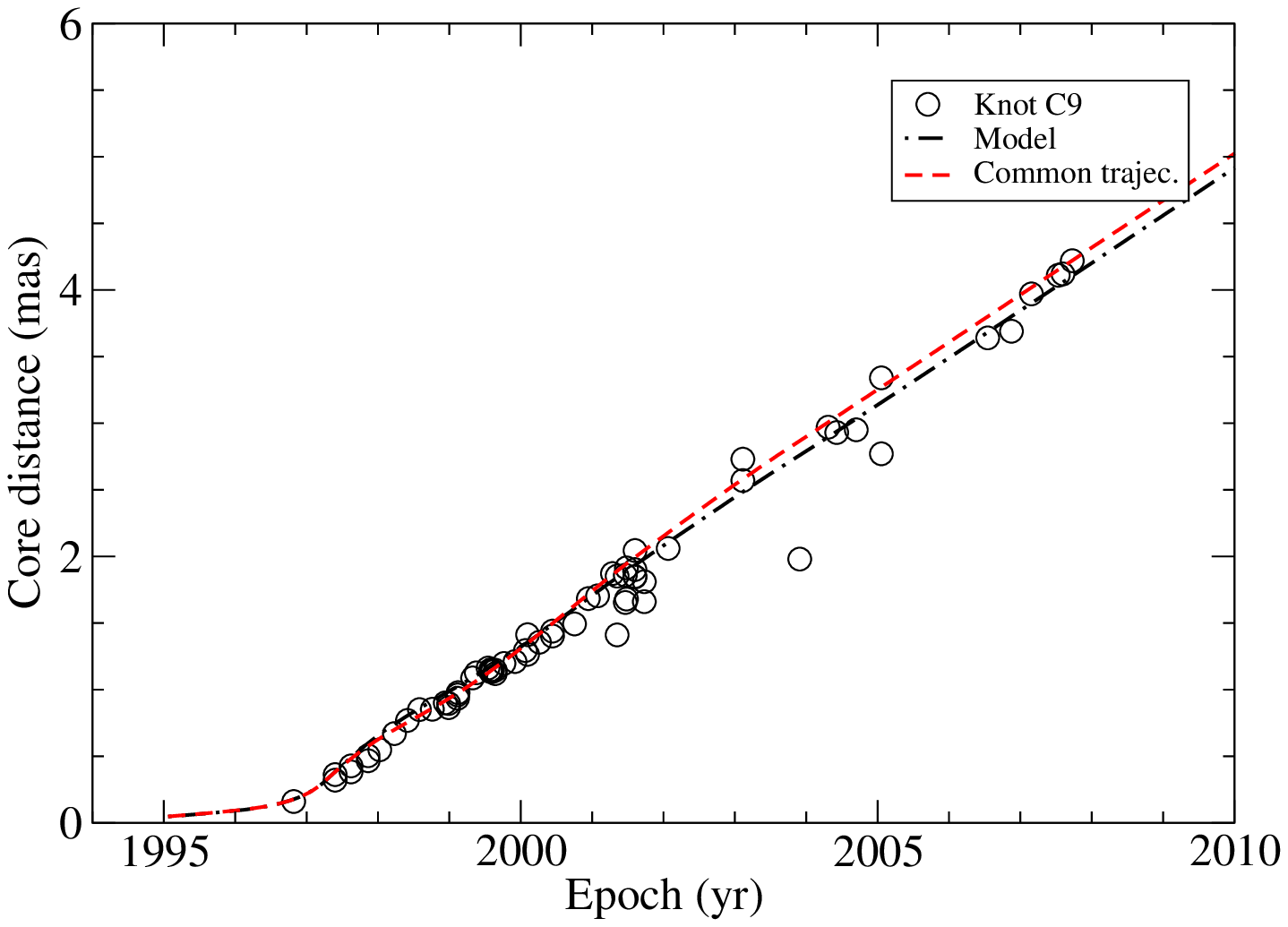}
    \includegraphics[width=5cm,angle=-90]{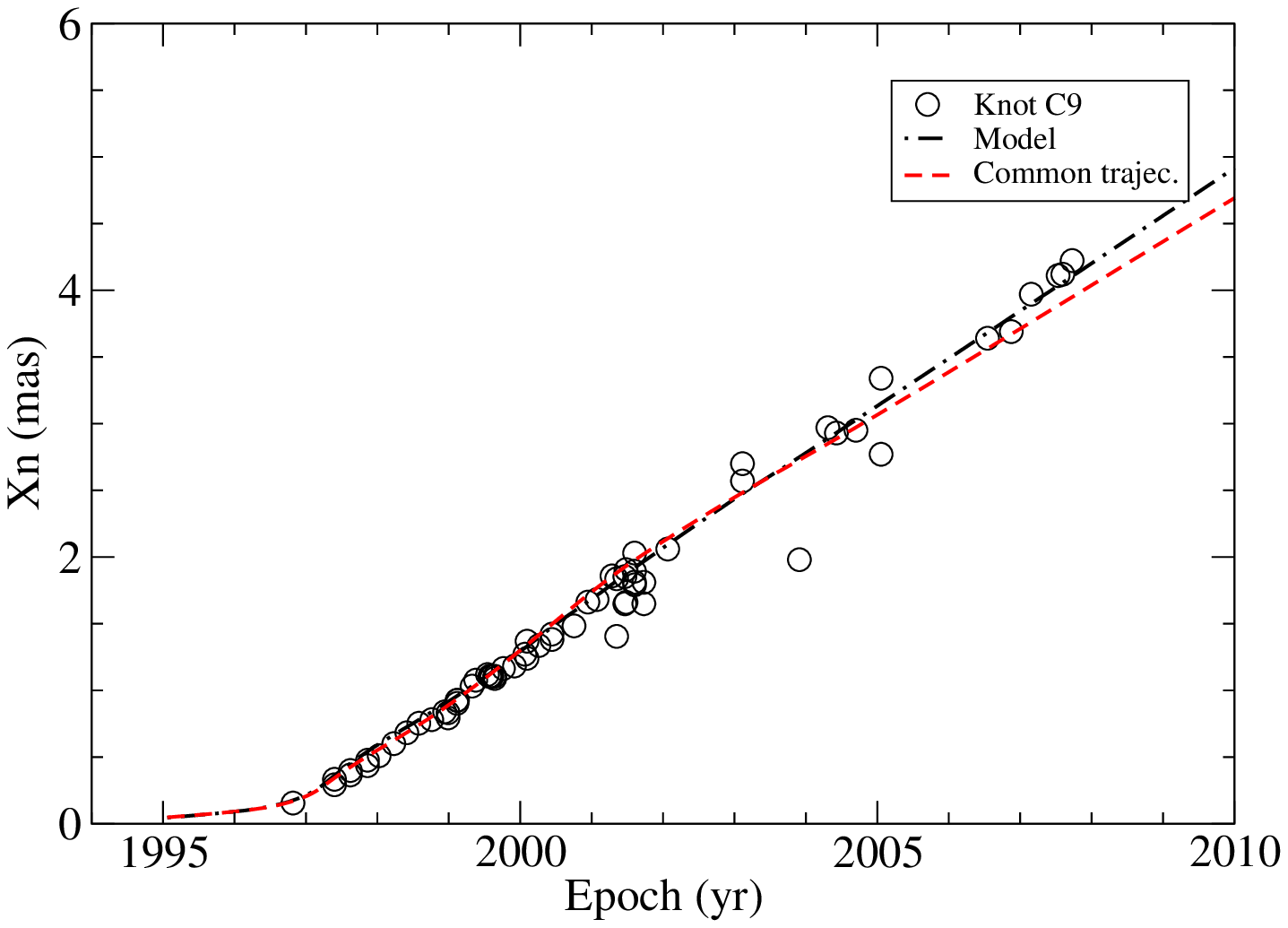}
    \includegraphics[width=5cm,angle=-90]{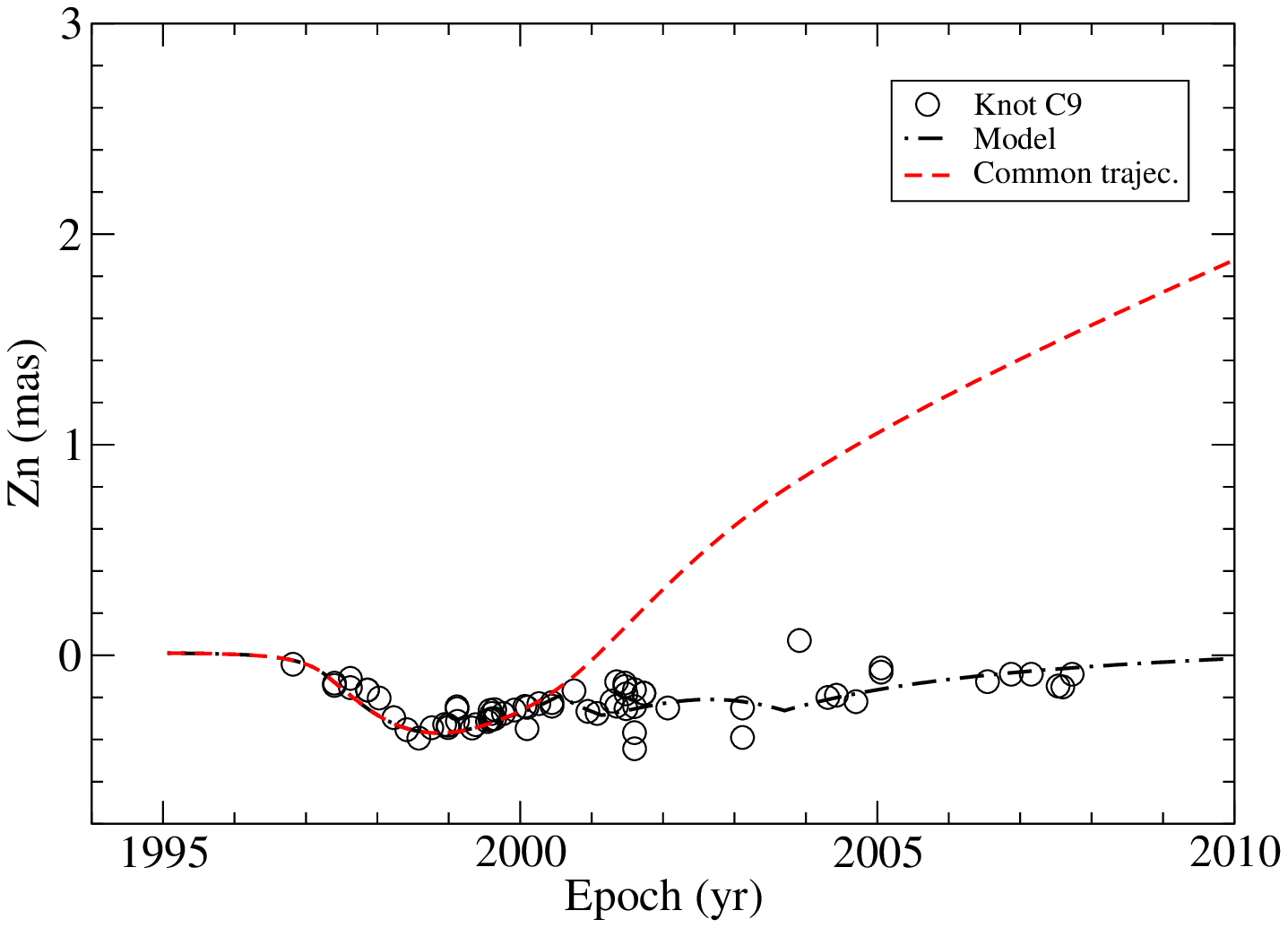}
    \caption{Knot C9: Model-fitting of its whole kinematics extending to
    core distance $r_n{\sim}$4.2\,mas: whole trajectory $Z_n(X_n)$, core 
     distance $r_n$(t), coordinates $X_n$(t) and  $Z_n$(t). The transition 
    between the precessing common helical trajectory pattern and its own 
    individual track occurred at core distance $r_n{\simeq}$1.25\,mas (or 
     $X_n{\simeq}$1.21\,mas; 1999.80). It can be seen that the observed 
    trajectory deviates prominently from the precessing common trajectory
     mainly in  the direction of $Z_n$ (declination).} 
    \end{figure*}
    \begin{figure*}
    \centering
    \includegraphics[width=5cm,angle=-90]{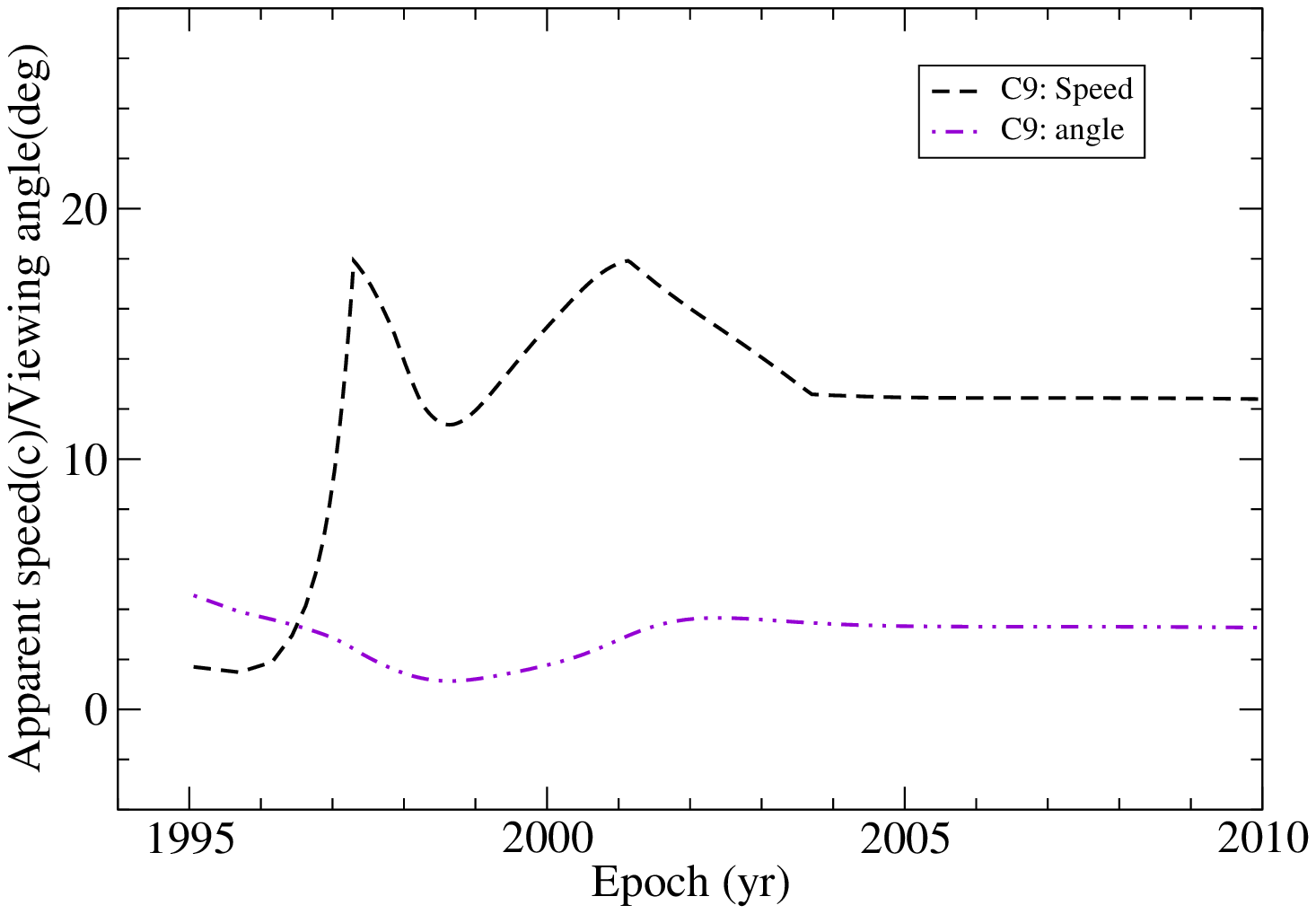}
    \includegraphics[width=5cm,angle=-90]{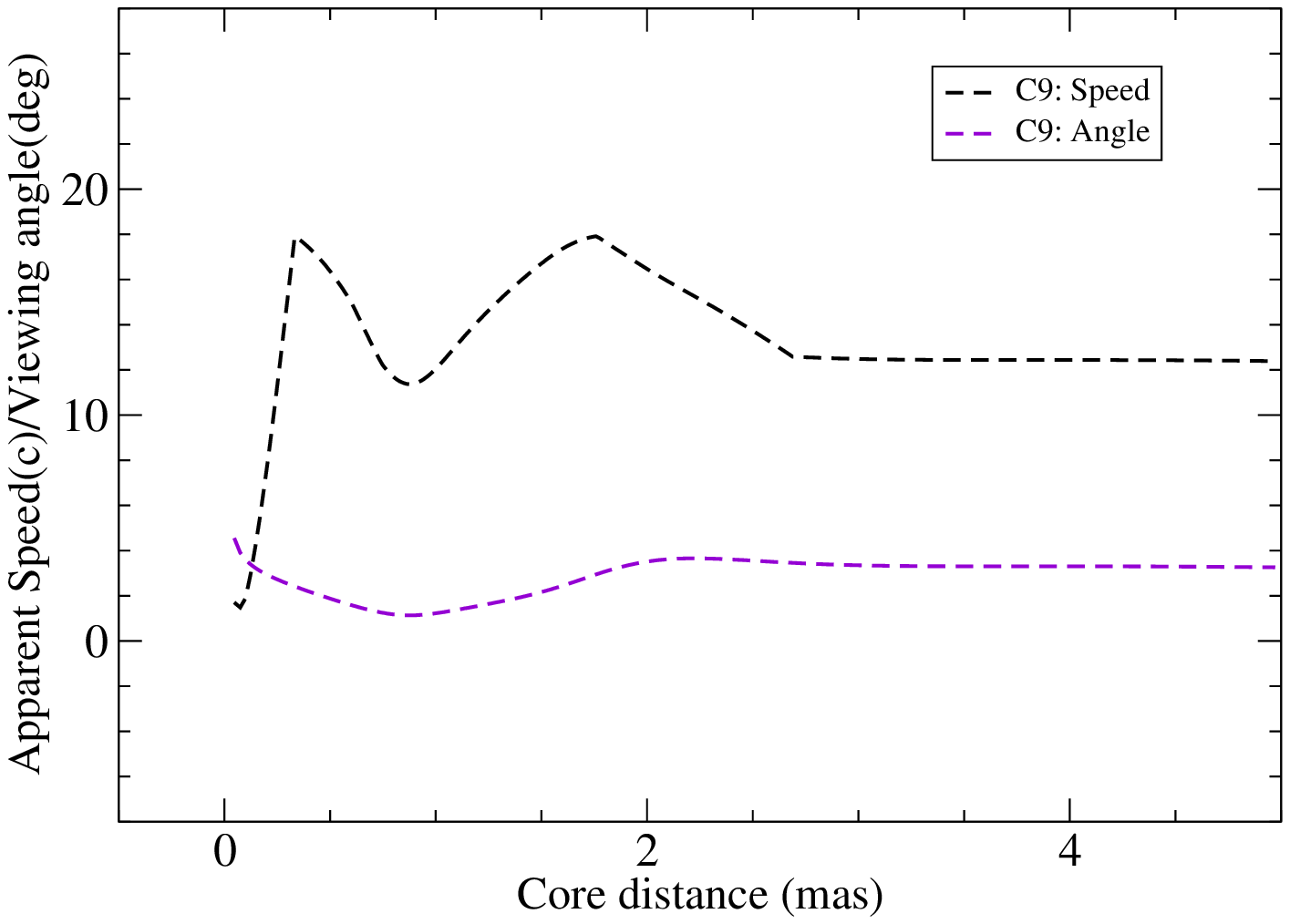}
   \includegraphics[width=5cm,angle=-90]{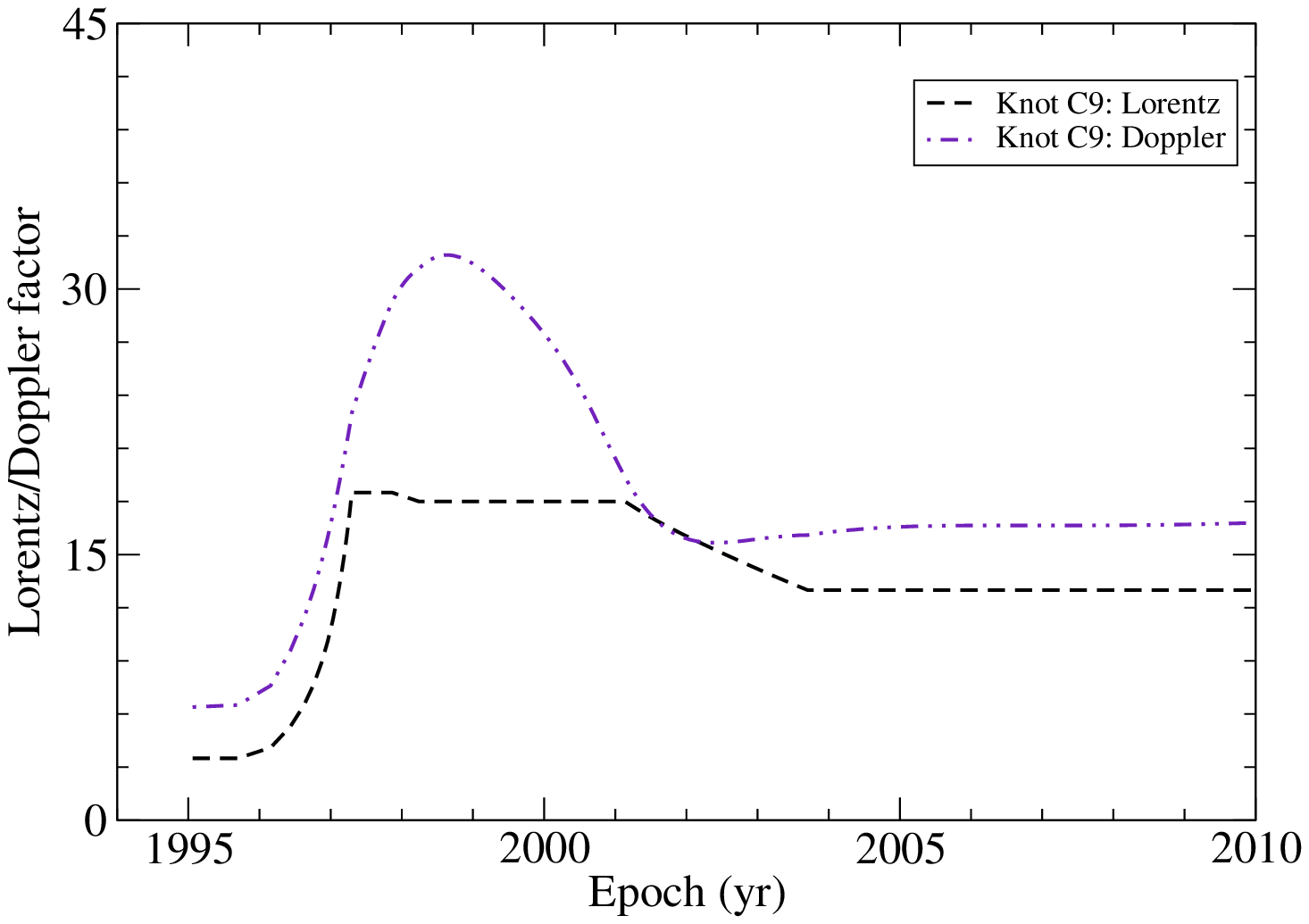}
   \includegraphics[width=5cm,angle=-90]{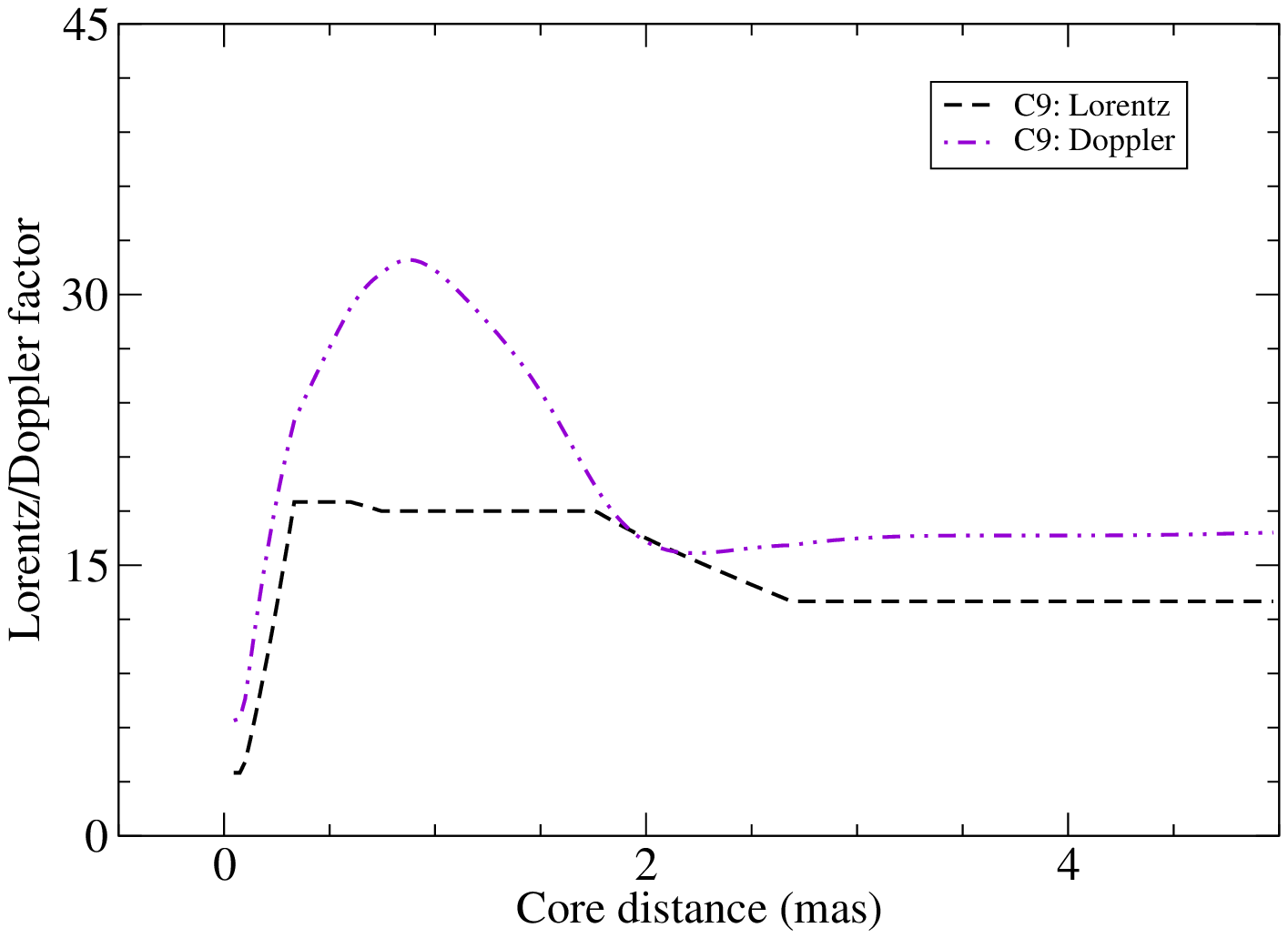}
    \caption{Knot C9: The model-derived appparent speed 
    $\beta_{app}$/viewing angle $\theta$ as functions of time and core distance
    (upper panels), and the  model-derived bulk Lorentz factor 
   $\Gamma$/Doppler factor $\delta$ (bottom panels) which are
    used to model-fitting of its observed flux evolution. During 1997--2001 
    ($r_n{\sim}$0.33--1.75\,mas) knot C9 underwent an apparent 
    acceleration-deceleration-reacceleration process induced by the change
    in its viewing angle, creating a smooth broad
    profile of its Doppler factor which is well coincident with the light
    curves  measured at 15-22-43\,GHz. (see Table 1 and Figure 8).  }
    \end{figure*}
   \begin{table*}
   \caption{Kont C9. The model-derived kinematic parameters for five epochs
    during the period of its acceleration/deceleration and Doppler boosting
     (Figures 7 and 8):  t (epoch), $r_n$ (core distance in mas), 
    $X_n$ (coordinate in mas), $\beta_{app}$
    (apparent speed in c), $\Gamma$ (bulk Lorentz factor), $\delta$ (Doppler 
    factor), $\theta$ (viewing angle in deg.), $Z_m$ (traveled distance in mas),
    $Z_p$ (traveled distance in pc). The maximum Doppler factor 
     ($\delta_{max}$=31.9 at 1998.63) is coincident with the minimum viewing
     angle $\theta_{min}$=$1.14^{\circ}$ and minimum apparent speed
    $\beta_{app,min}$=11.4c. }
   \begin{flushleft}
   \centering
   \begin{tabular}{lllllllll}
   \hline
   t & $r_n$ & $X_n$ & $\beta_{app}$ & $\Gamma$ & $\delta$ & $\theta$ &
                $Z_m$ & $Z_p$  \\
   \hline
   1996.96 & 0.21 & 0.20 & 8.1 & 10.1 & 16.0 & 2.91 & 3.20 & 21.3   \\
   1997.30 & 0.33 & 0.32 & 17.9 & 18.5 & 23.0 & 2.42 & 6.00 & 39.9   \\
   1998.63 & 0.87 & 0.80 & 11.4 & 18.0 & 31.9 & 1.14 & 26.4 & 175.6   \\
   2001.13 & 1.76 & 1.74 & 17.9 & 18.0 & 19.4 & 2.94 & 61.6 & 409.6  \\
   2003.83 & 2.73 & 2.72 & 12.6 & 13.0 & 16.2 & 3.44 & 80.8 & 537.3  \\
   \hline
   \end{tabular}
   \end{flushleft}
   \end{table*}
   \begin{figure*}
   \centering
   \includegraphics[width=6cm,angle=-90]{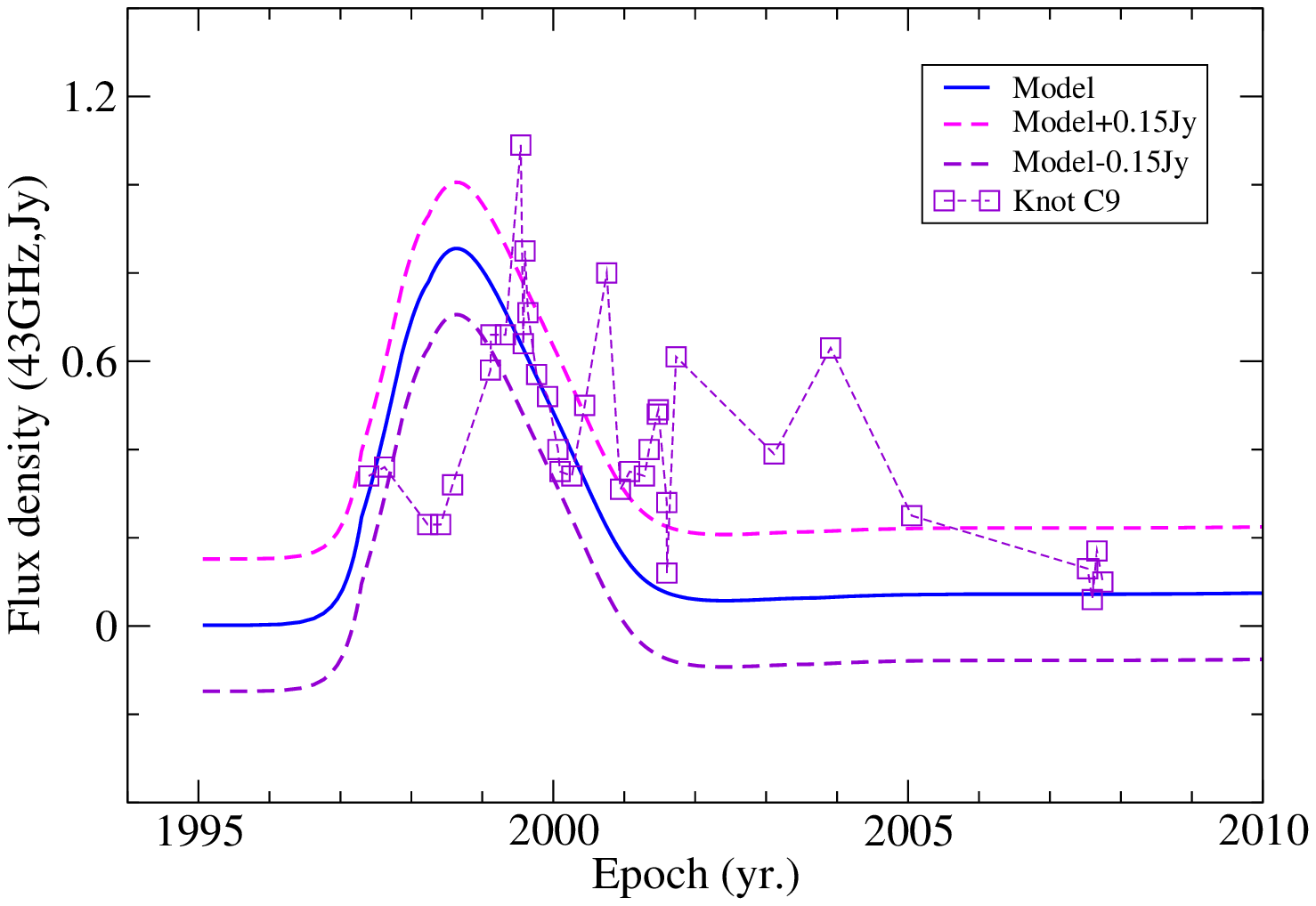}
   \includegraphics[width=6cm,angle=-90]{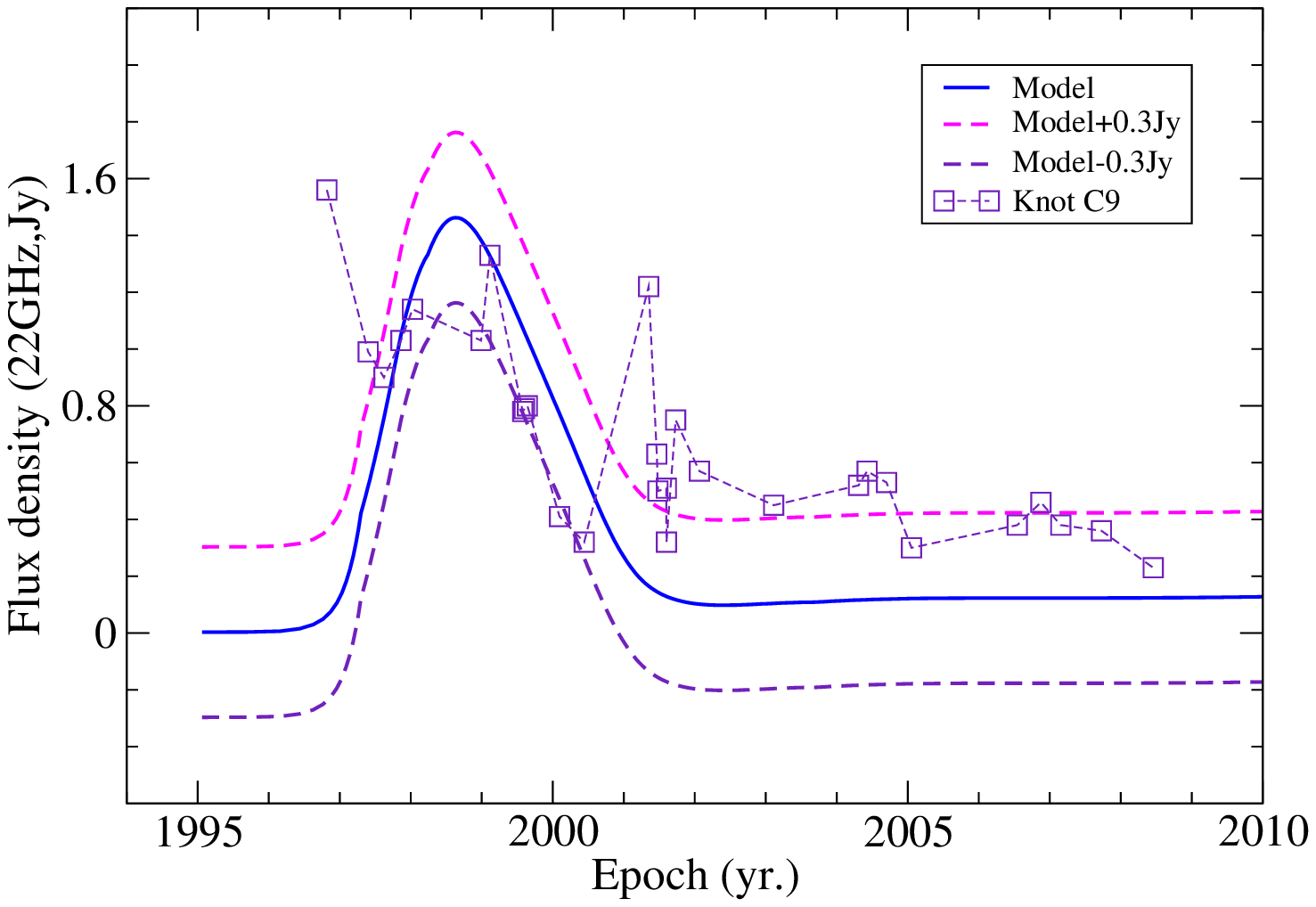}
   \includegraphics[width=7.5cm,angle=-90]{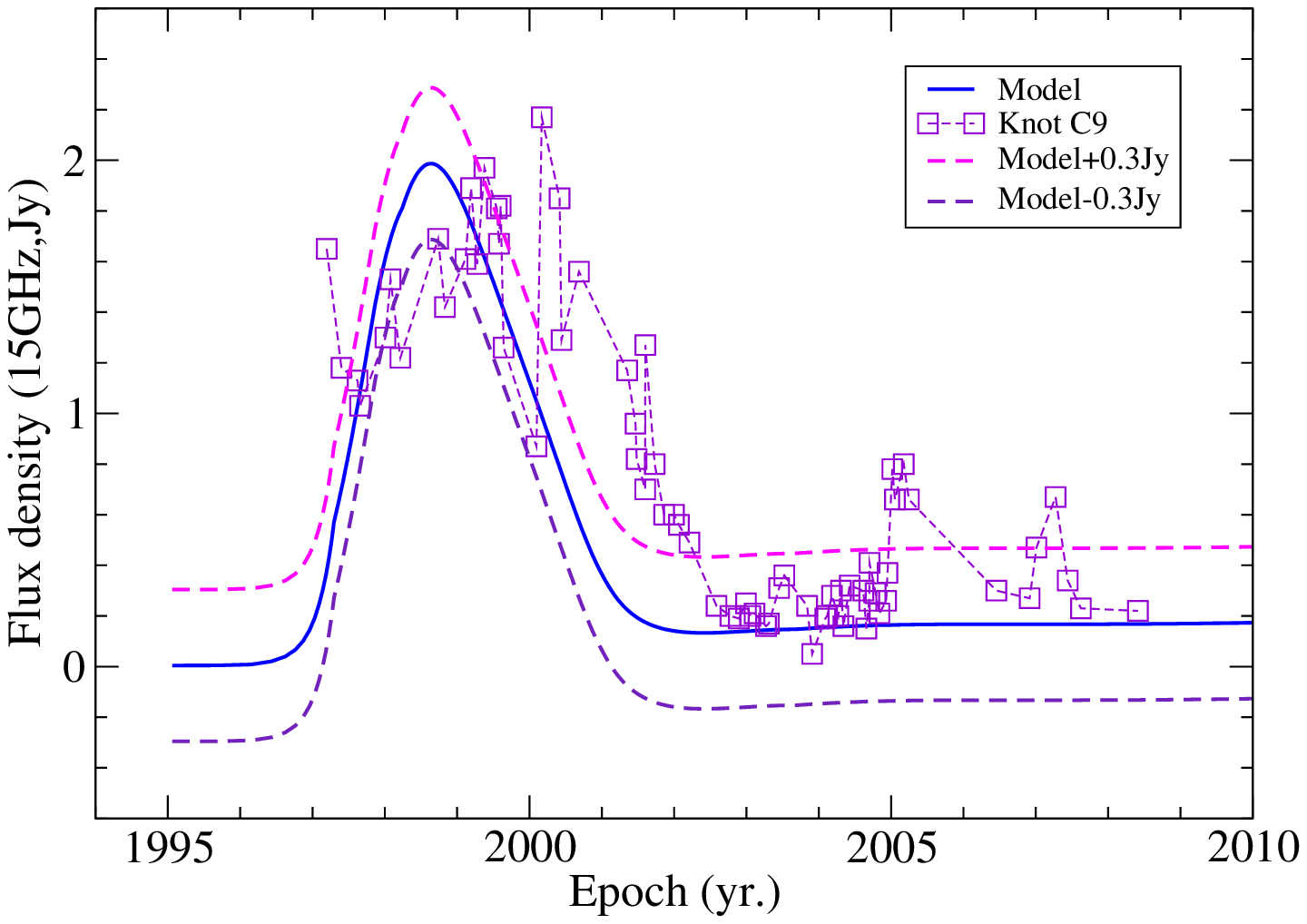}
   \caption{Knot C9: model fitting of the flux evolution measured at 43, 22 and
    15\,GHz. The model-derived Doppler boosting profiles (black lines), which
    are associated with the change in its viewing angle, can well fit the 
   light curves, especially at 15\,GHz (lower panel). Its intrinsic variations
    in flux densitty  and spectral index give rise to the fluctuations 
    with shorter timescales at different times. The blue and violent lines
    show the range of these fluctuatons. The intrinsic base-level flux 
    densities at 15, 22 and 43\,GHz are 3.83, 2.82 and 1.65$\mu$Jy and spectral
    index $\alpha$=0.80. Maximum Doppler factor $\delta_{max}$=31.9 at 
    1998.63.}
   \end{figure*}
   \section{Interpretation of kinematics and flux evolution for knot C9}
    As show in the previous paper (Qian \cite{Qi22b}), the flux evolution
     associated with its Doppler boosting effect is very important
    and encourages the application of our precessing nozzle scenario to
    study the phenomena in  3C345 and other blazars. Knot C9 was a typical
    and exceptionally instructive example of applying the scenario for a
     satisfying interpretation of its VLBI-kinematics and flux evolution. 
    Here  we recaptulate the main  results obtained in Qian  (\cite{Qi22a}, 
    \cite{Qi22b}) and supplement some new results on its flux evolution
    and spectral features. All the results for model-fitting of its kinematics 
    and flux evolution are shown in Figures 4--8.
     \subsection{Model-fits to the kinematics of knot C9}
     The results of model-fitting of its kinematics can be described   as 
     follows.\\
     (1) In the precessing  jet-nozzle scenario its precession phase and 
     ejection epoch are adopted as $\phi_0$=5.54+4$\pi$ and $t_0$=1995.06.
      The amplitude and phase of its helical trajectory are shown in Figure 3
      (see equations (19) and (20)).\\
     (2) For knot C9 the parameters $\epsilon$ and $\psi$ as function of time
      are modeled as shown in Figure 5 (left), indicating that knot C9
       moved along the precessing common trajectory  before 1999.94. 
      After 1999.94 parameter $\psi$ started to increase 
     (here $\epsilon$=const.=$2^{\circ}$), and the modeled
       jet-axis started to deviate from the direction defined in the scenario
      and knot C9 moved along its own individual track. 
       Thus its  precessing  common trajectory  extends to core
      distance $r_n$=1.25\,mas and the corresponding  traveled distance 
     $\sim$298\,pc (or $\sim$45\,mas; see Figure 5 (right) and Table 2).\\
     (3) The curved trajectory observed within $X_n$$\sim$1.5\,mas for C9 
     is the projection of its precessing helical common trajectory, which
      fits the observational data very well as shown in Figure 4. Beyond 
      $X_n$$\sim$1.5\,mas the observed trajectory clearly deviated from the 
      precessing helical common trajectory and knot C9 moved along its own 
      individual trajectory. In the Figure two trajectories defined by
      $\phi_0$=5.54$\pm$0.31\,rad are also shown, demonstrating that within 
      $X_n\sim$1.5\,mas the observed trajectory section was very accurately 
     fitted by the processing  common helical trajectory (in the range of
     $\pm$5\% of the precession period, or $\pm$0.36\,yr.). \\
   (4) In Figure 6 the model-fitting results of its entire kinematics 
     (during 1996--2008; or in the core-distance range $r_n$=[0--4.2\,mas]) 
     are  presented: the entire trajectory $Z_n(X_n)$, core separation 
    $r_n(t)$, coordinates $X_n(t)$ and $Z_n(t)$. It can be seen that all these
    properties are very well fitted by the model, but the precessing common
    trajectory only applies to the inner trajectory section within 
    $r_n$$\stackrel {<}{_\sim}$1.25\,mas. The deviation of its outer
    track occurred mainly in the direction of $Z_n$ (declination).\\
    (5) The model-derived apparent velocity $\beta_{app}(t)$,
    viewing angle $\theta(t)$, bulk Lorentz factor $\Gamma(t)$ and Doppler
    factor $\delta(t)$ are shown as functions of both time and core distance 
    in Figure 7.  It accelerates  to $\beta_{app}{\sim}$18c near the core
    ($r_n{\sim}$0.33\,mas; 1997.3) and  decelerates to $\sim$11.4c at
    $r_n{\sim}$0.87\,mas (1998.6). Then it re-accelerates to 18.0\,c at 
    $r_n{\sim}$1.76\,mas (2001.13) and re-decelerates to 12.6\,c at
     $r_n{\sim}$2.73\,mas  (2003.83). [Note: within $r_n{\sim}$1.25\,mas (or
     before 1999.80, the observed trajectory of C9 is well fitted by the 
     precessing common helical trajectory of our scenario]. 
     It worths emphasizing that our
     model fitting results on the change in apparent speed 
    associated with the twisted trajectory  are very well consistent with 
    the measurments by Jorstad et al. (\cite{Jo05}).   \\
    (6) As shown in Figure 7 the model-derived Doppler factor profile is 
     closely  related to  the change in apparent speed $\beta_{app}$. The 
     maximum Doppler factor (31.9 at 1998.63) is coincident with the minimum
      apparent speed 
     (11.4\,c) and minimum viewing angle ($\theta$=$1.14^\circ$). That is,
     the Doppler factor profile is closely related to its motion along the
     common helical trajectory (mainly during 1997--2000; see Table 1).\\
   \subsection{Doppler boosting effect and flux evolution of knot C9}
    The model-derived Doppler factor $\delta(t)$ as a continuous 
    function of time is shown in Figure 7 has a broad smooth bump
    structure  during 1997--2002, providing  a distinctly smooth
    Doppler boosting profile to study  the Doppler boosting effect
    in the flux density variations observed in knot C9. This is 
    rare and extremely valuable opportunity to test our precessing
    nozzle scenario: whether our scenario is able to fully 
    interprete the entire phenomena measured by VLBI observations,
    including both kinematics and radiation properties.\\
      The light curves measured at 15GHz, 22GHz and 43GHz are well fitted 
     by the derived Doppler boosting effect as shown in Figure 8.
     The model-derived broad (1997--2001) Doppler boosting profiles
      ${S_{int}}[\delta(t)]^{3+\alpha}$ are well coincident with the observed
      light curves.
     \footnote{The intrinsic base-level flux densities at the three 
      frequencies are: $S_{int}$=3.83$\times{10^{-6}}$\,Jy, 
      2.82$\times{10^{-6}}$\,Jy, and 
     1.65$\times{10^{-6}}$\,Jy at 15\,GHz, 22\,GHz and 43\,GHz, respectively; 
     the spectral index is adopted as $\alpha$=0.8.}  
     But due to its intrinsic flux density
     variations  with shorter timescales the data points fluctuate up and down,
     deviating from the Doppler boosting profiles. Moreover, there are more 
     individual flux peaks with short time-scales at different times, for
      example, in the 15GHz light curve at 1997.20, 2000.17, 2001.60, 2005.17 
     and  2007.27.  These short timescale variations reveal the variations in      their spectral index in the spectral range (15\,GHz--22\,GHz--43\,GHz).
     For example, at 1998.23 and 1998.41, the observed 43\,GHz flux densities
      are especially low ($\alpha$(15-43GHz)$\simeq$1.6--1.8), implying the
     intrinsic variations show spectral steepening at this frequency. In 
     contrast, the observed spectral index $\alpha$(15-22GHz)${\simeq}$--1.99,
     implying an inverted spectrum in the (15-22GHz) waveband.
     \footnote{Actually, the "knots C9" measured  at 15, 22 and 43\,GHz are
       not associated with the same source-region, thus the available data
     are not sufficient to investigate the spectral features of knot C9.}
      Thus in order to explain the flux evolution of knot C9 
     both Doppler boosting effect  and intirnsic variations should be taken
     into account.\\
     Here we would like to emphasize the two points: (1) The model which we 
     used for fitting the kinematic behavior of knot C9 is able to derive its
     bulk Lorentz factor, viewing angle and Doppler factor as continuous
     functions of time and  predict its Doppler boosting profile 
     {$[\delta(t)/\delta_{max}]^{3+\alpha}$. Thus the full coincidence of its
    measured  light curves with the Doppler boosting profiles (both in 
    timescale and pattern) may be regarded as a sound argument for the 
     validity of the precessing nozzle scenario with a precessing common
    helical trajectory pattern; (2) The coincidence of the timescales of
    the light curves and that of the Doppler boosting profiles also has 
    distinctive implications, showing that the measured variability timescales
    are compressed by the light-travel-time effects. These effects are 
    effective for both the Doppler boosting profiles (with varying Doppler 
    factors) and  the superposed shorter timescale intrinsic variations
     (with constant Doppler factors) in the case of superluminal components.
     Similarly, in the traveling relativistic shocks propagating in turbulent
     jets, the light-travel-time effects are also significant for 
    understanding the rapid (e.g. intraday) variability observed at 
      centimeter wavelengths 
     in blazars,  solving the problem of extremely high brightness 
      tempratures and the requirement of extremely high bulk Lorentz factors 
     ($\Gamma$$\stackrel {>}{_\sim}$100). Discussions on the application of 
     the light-travel-time effect are referred to: 
      Quirrenbach et al. \cite{Qu89}, \cite{Qu91}, \cite{Qu92},
     Qian et al. \cite{Qi91c}, \cite{Qi96a}, Kraus et al. \cite{Kr99}, 
     \cite{Kr03},  Krichbaum et al. \cite{Kri91}, \cite{Kri02},
       Begelman et al. \cite{Be94},
     Standke et al. \cite{St96}, Marscher \cite{Ma93}, Marscher et al.
     \cite{Ma92},  Gabuzda et al. \cite{Ga99}, \cite{Ga98}, Kochanev \& Gabuzda
      \cite{Ko98}, Wagner et al. \cite{Wa93}, Witzel et al. \cite{Wi88}.
     \begin{figure*}
    \centering
    \includegraphics[width=5cm,angle=-90]{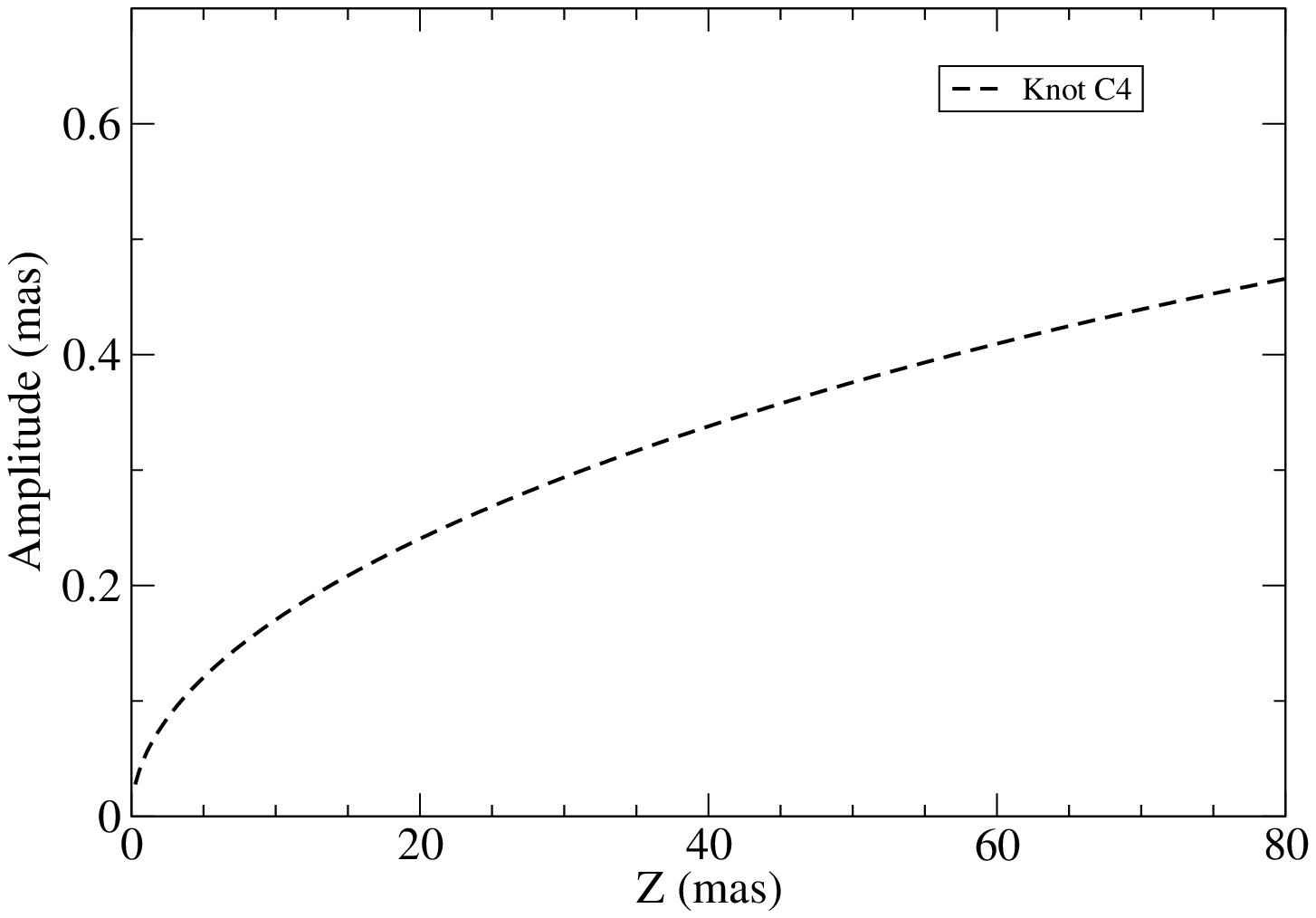}
    \includegraphics[width=5cm,angle=-90]{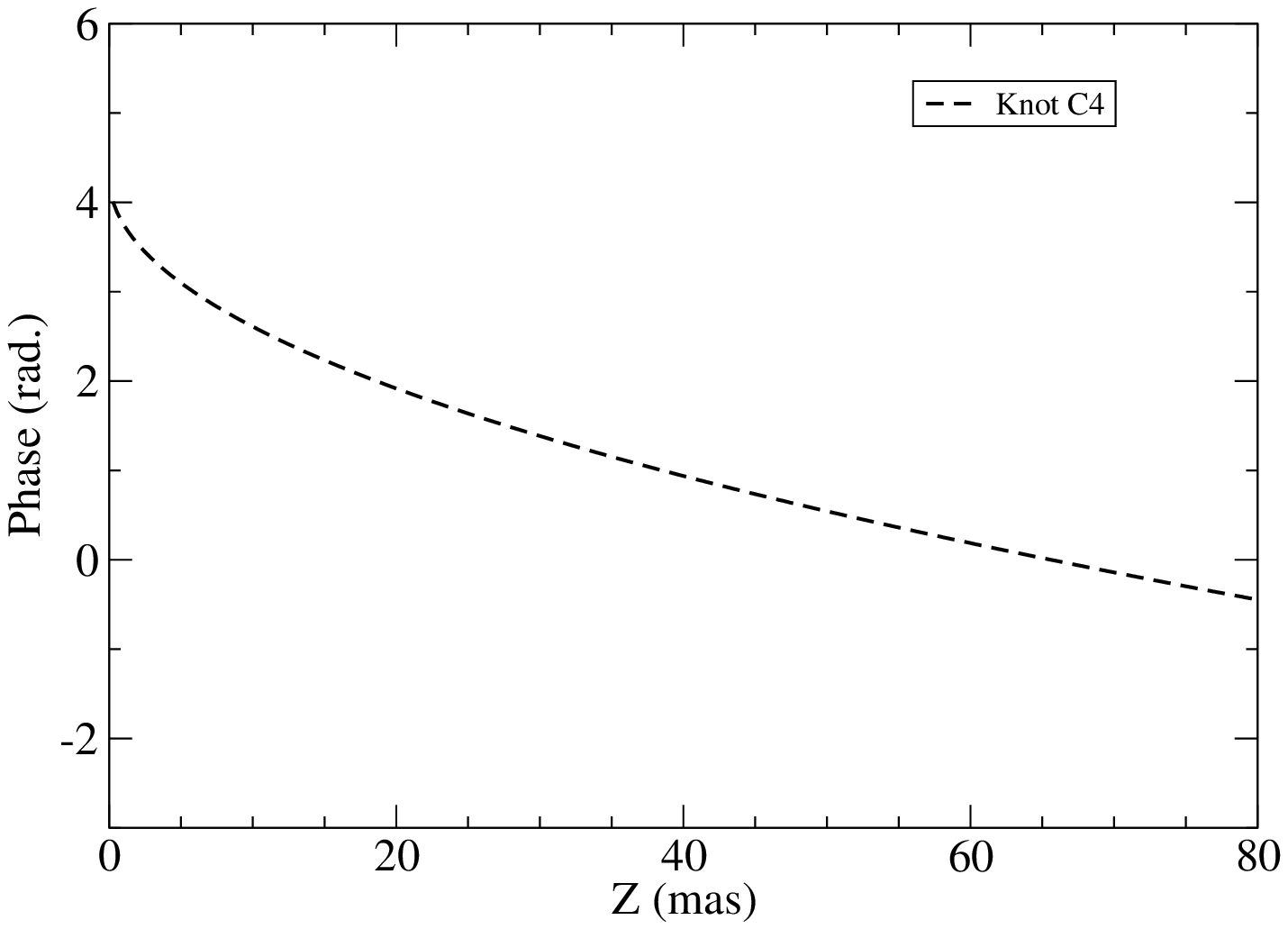}
    \caption{Knot C4: The amplitude A(Z) and phase $\phi$ defining the 
    precessing common helical trajectory pattern with $\phi_0$=4.28\,rad
    and ejection time $t_0$=1979.0.}
    \end{figure*}
    \begin{figure*}
    \centering
    \includegraphics[width=7cm,angle=-90]{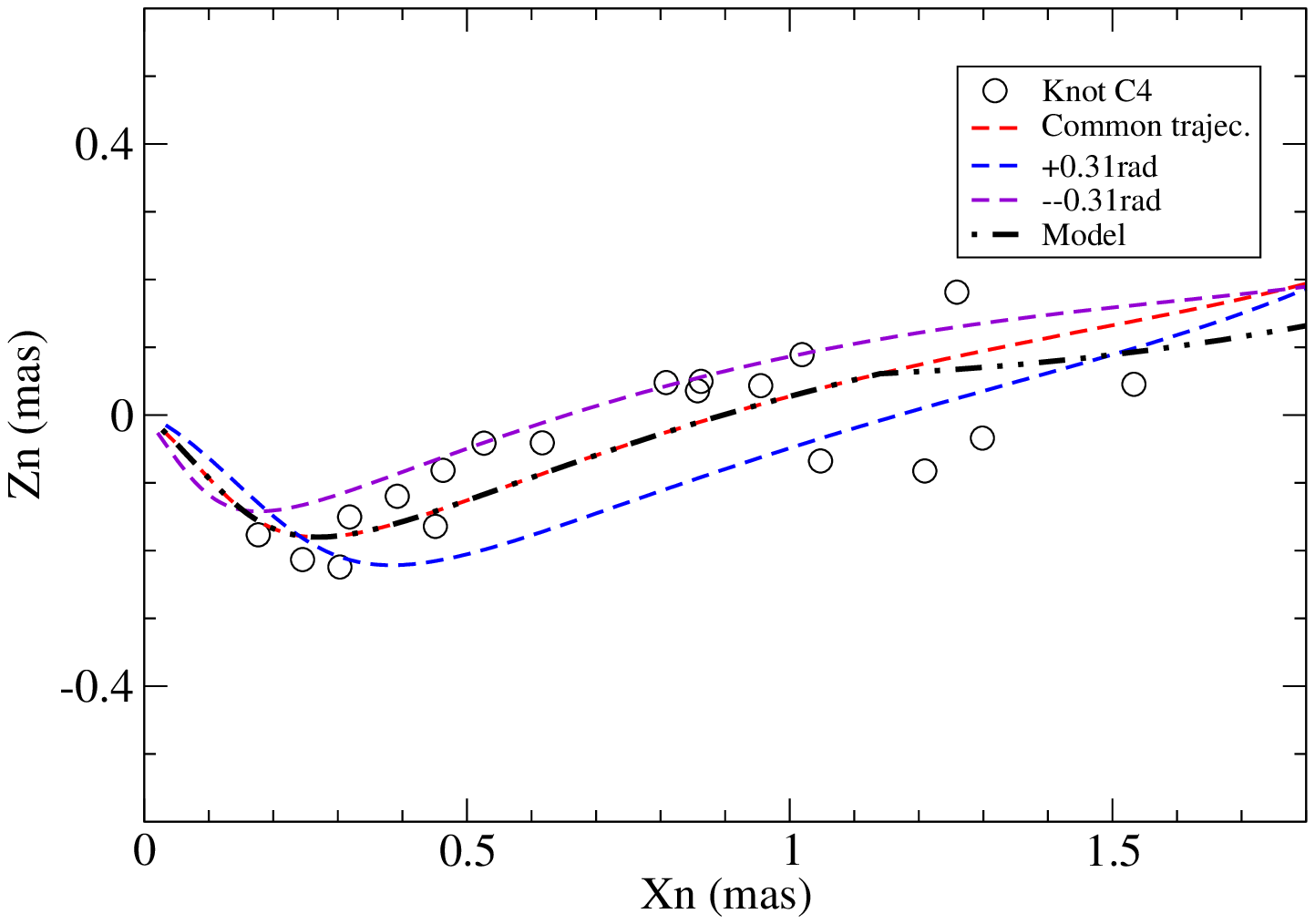}
    \caption{Knot C4: model-fitting of its  inner track by the precessing 
    common helical trajectory (the red line) within $X_n{\leq}$1.14\,mas
     (corresponding traveled distance Z=40.0\,mas or 266.0\.pc). The blue and
    violent lines indicate the uncertainty range $\pm$5\% ($\pm$0.31\,rad) of
     the precession period.  The black dashed line indicates the model-fitting
     of the whole trajectory.}
    \end{figure*}
    \begin{figure*}
    \centering
    \includegraphics[width=5cm,angle=-90]{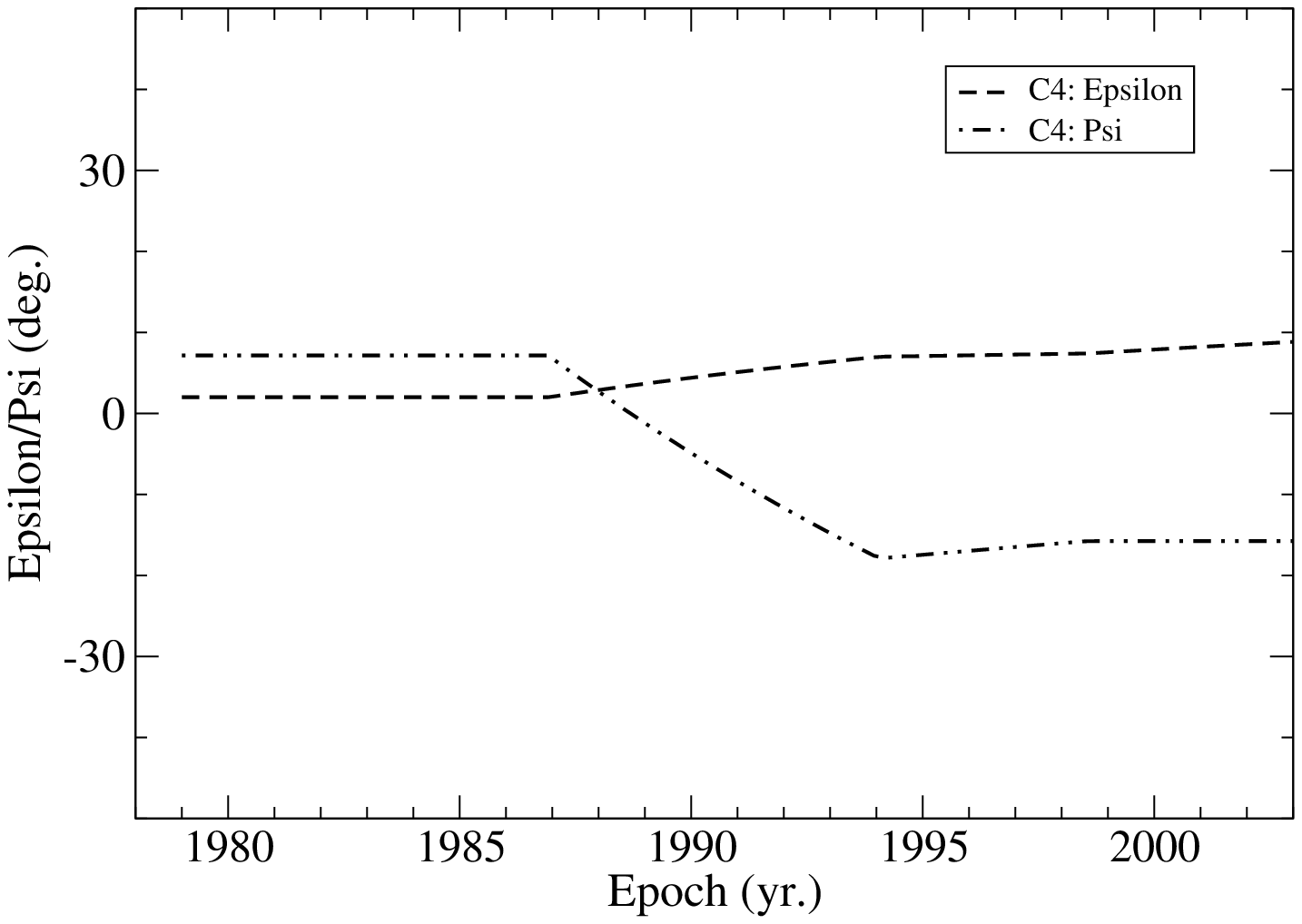}
    \includegraphics[width=5cm,angle=-90]{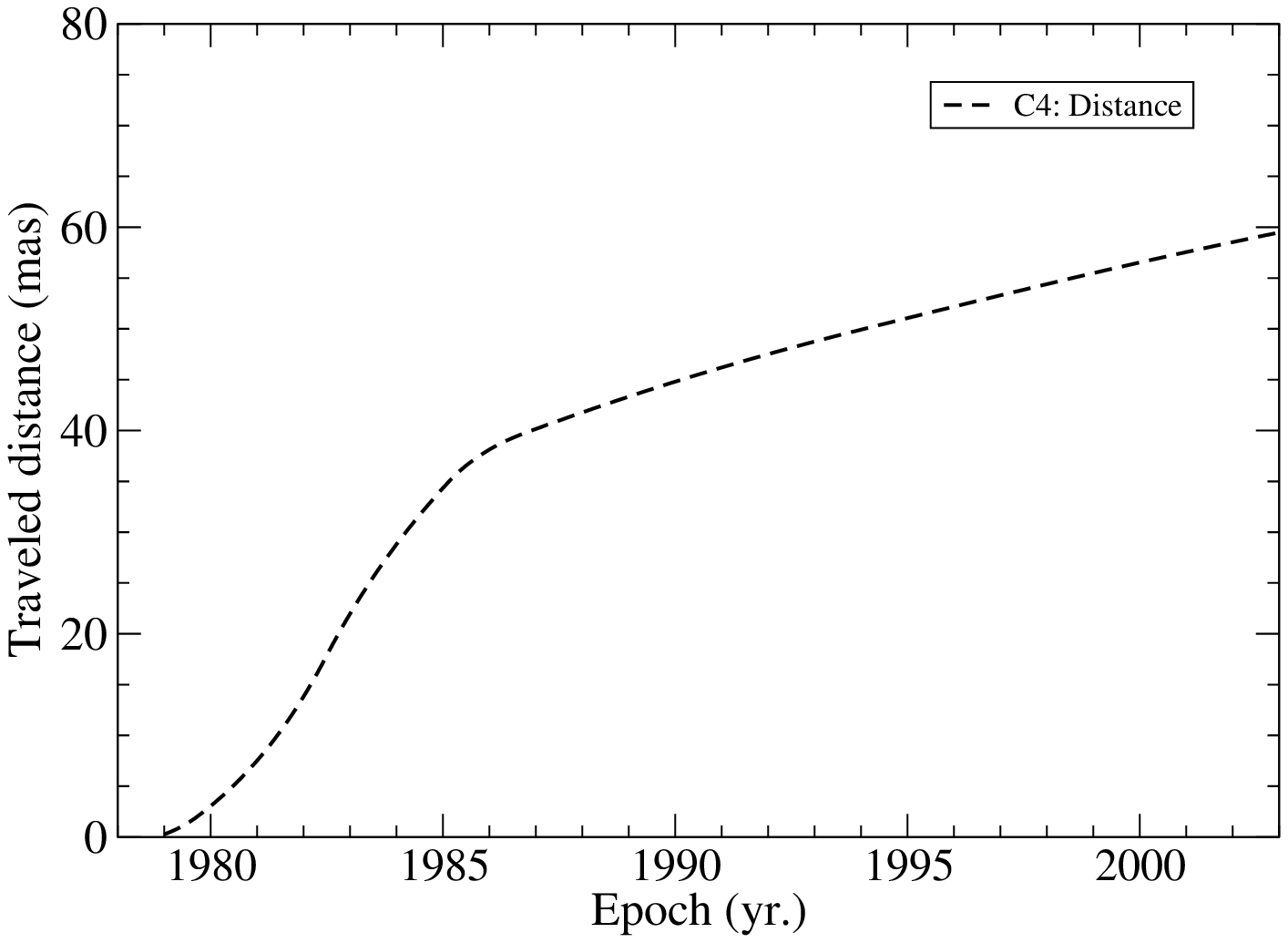}
    \caption{Knot C4. Left panel: functions $\epsilon$(t) and $\psi$(t) 
    defining the jet-axis. Before 1986.91 (corresponding core distance 
    $r_n{\leq}$1.14\,mas) $\epsilon$=$2^{\circ}$ and $\psi$=$7.16^{\circ}$
    knot C4 moved along the precessing common helical traejctory pattern.
    After 1986.91 (or $r_n{>}$1.14\,mas) both $\epsilon$ and $\psi$
    changed and knot C4 moved along its own individual track. Its traveled
    distance Z(t) is shown in the right panel.}
    \end{figure*}
    \begin{figure*}
    \centering
    \includegraphics[width=5cm,angle=-90]{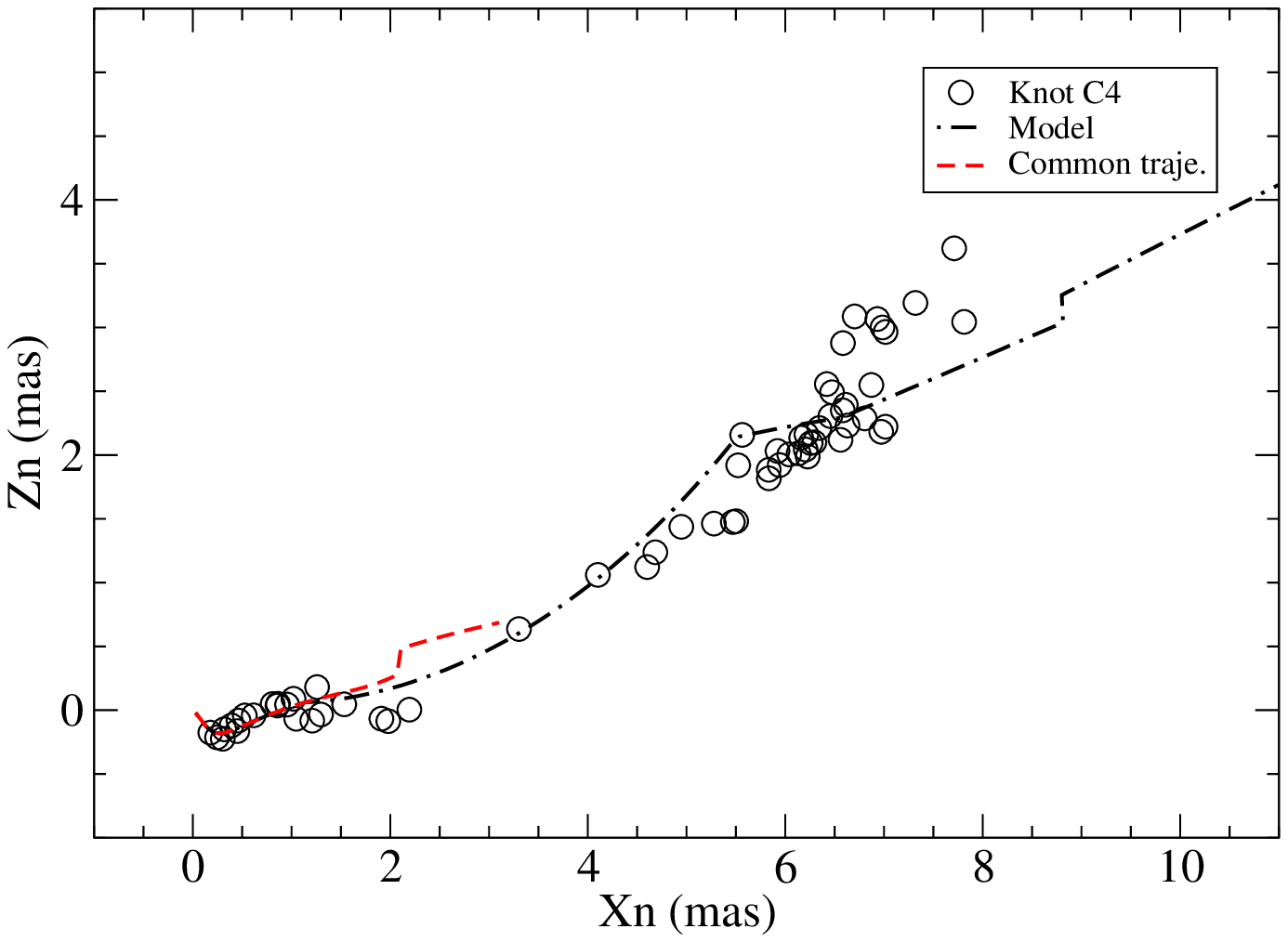}
    \includegraphics[width=5cm,angle=-90]{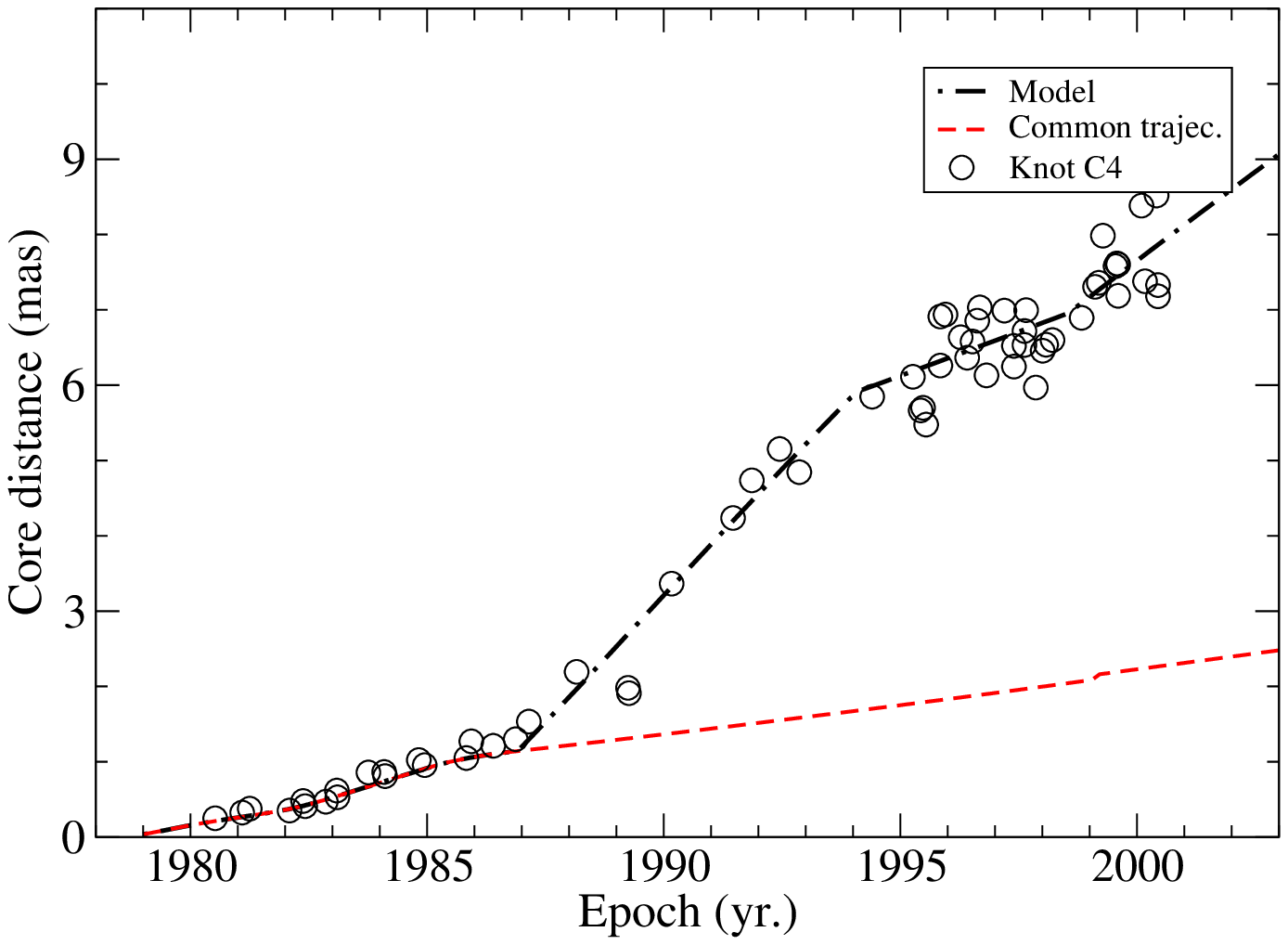}
    \includegraphics[width=5cm,angle=-90]{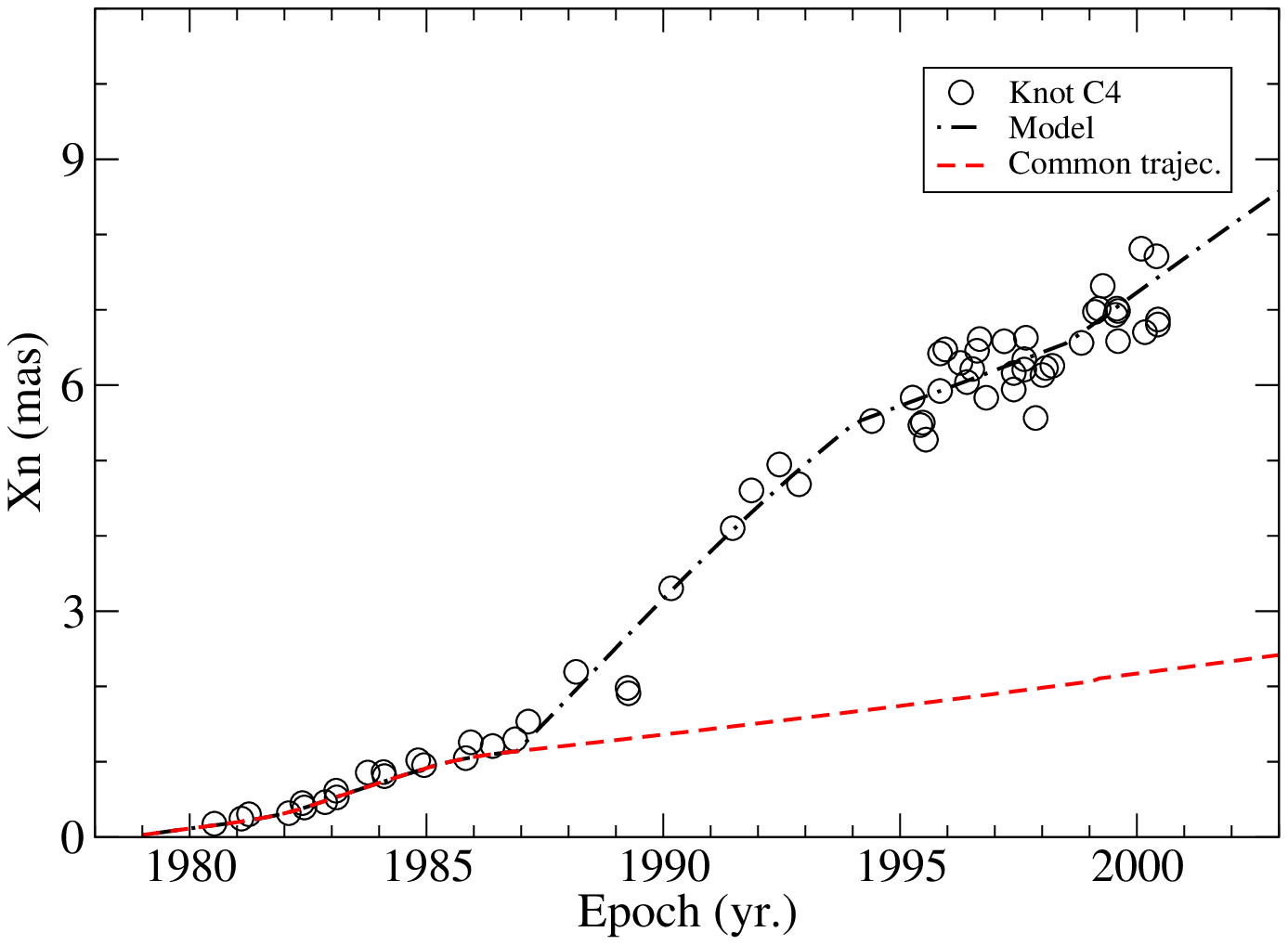}
    \includegraphics[width=5cm,angle=-90]{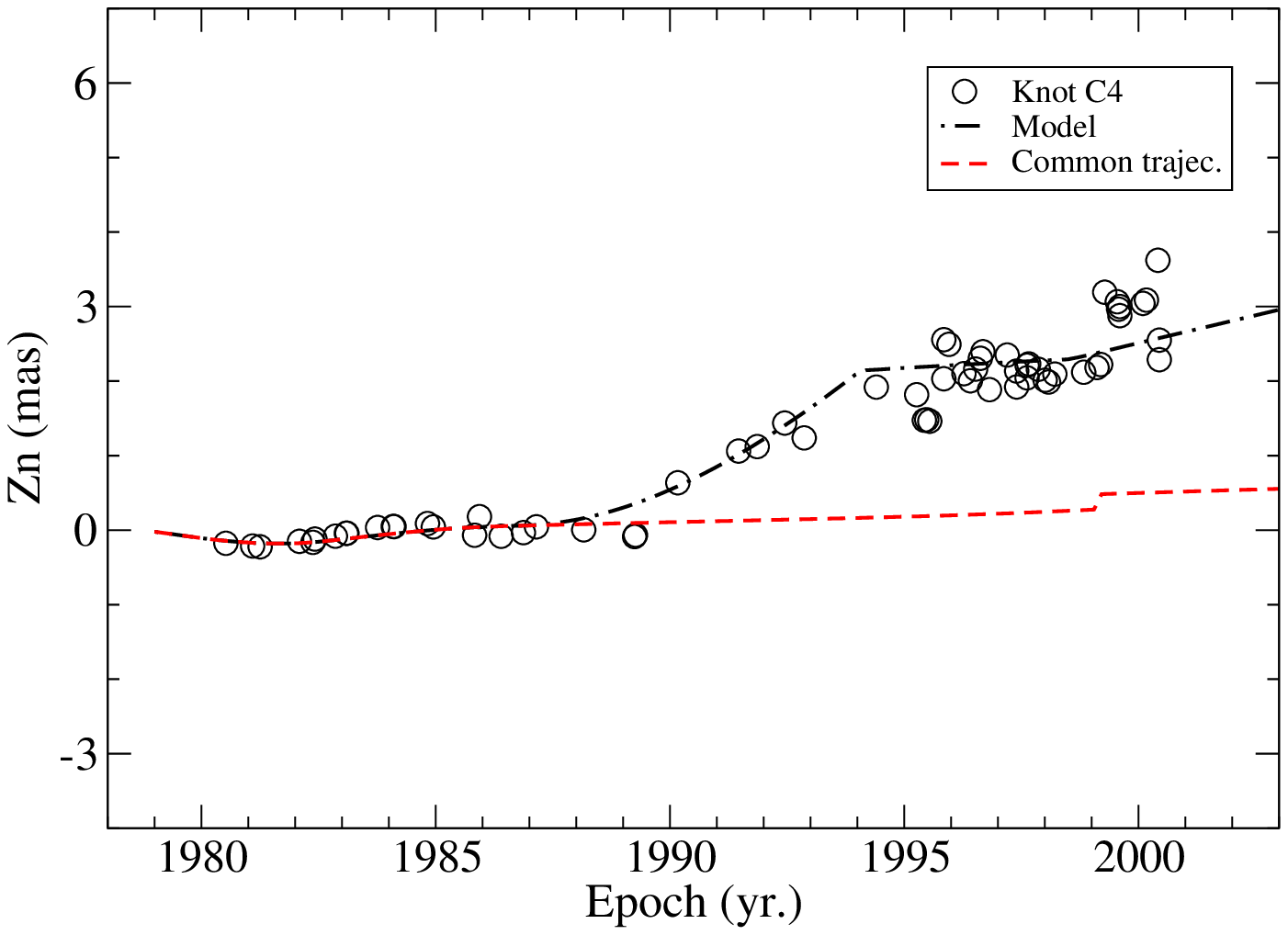}
    \caption{Knot C4: model-fitting of the whole trajectory $Z_n(X_n)$,
    core distance $r_n(t)$,  coordinates $X_n(t)$ and $Z_n(t)$.  The transition
    between the precessing common track and its own individual track occurred
     at $X_n$=1.14\,mas (corresponding traveled distance 
    Z=40.0\,mas=266.0\,pc.). Its whole trajectory extends to core distance 
    $r_n{\sim}$7.53\,mas ($X_n{\sim}$7.11\,mas; corresponding traveled distance
    Z=56.3\,mas=374.2\,pc). The red lines indicate the precessing common track,
    while the black lines indicate the model-fitting of the whole track.  }
    \end{figure*}
    \begin{figure*}
    \centering
    \includegraphics[width=5cm,angle=-90]{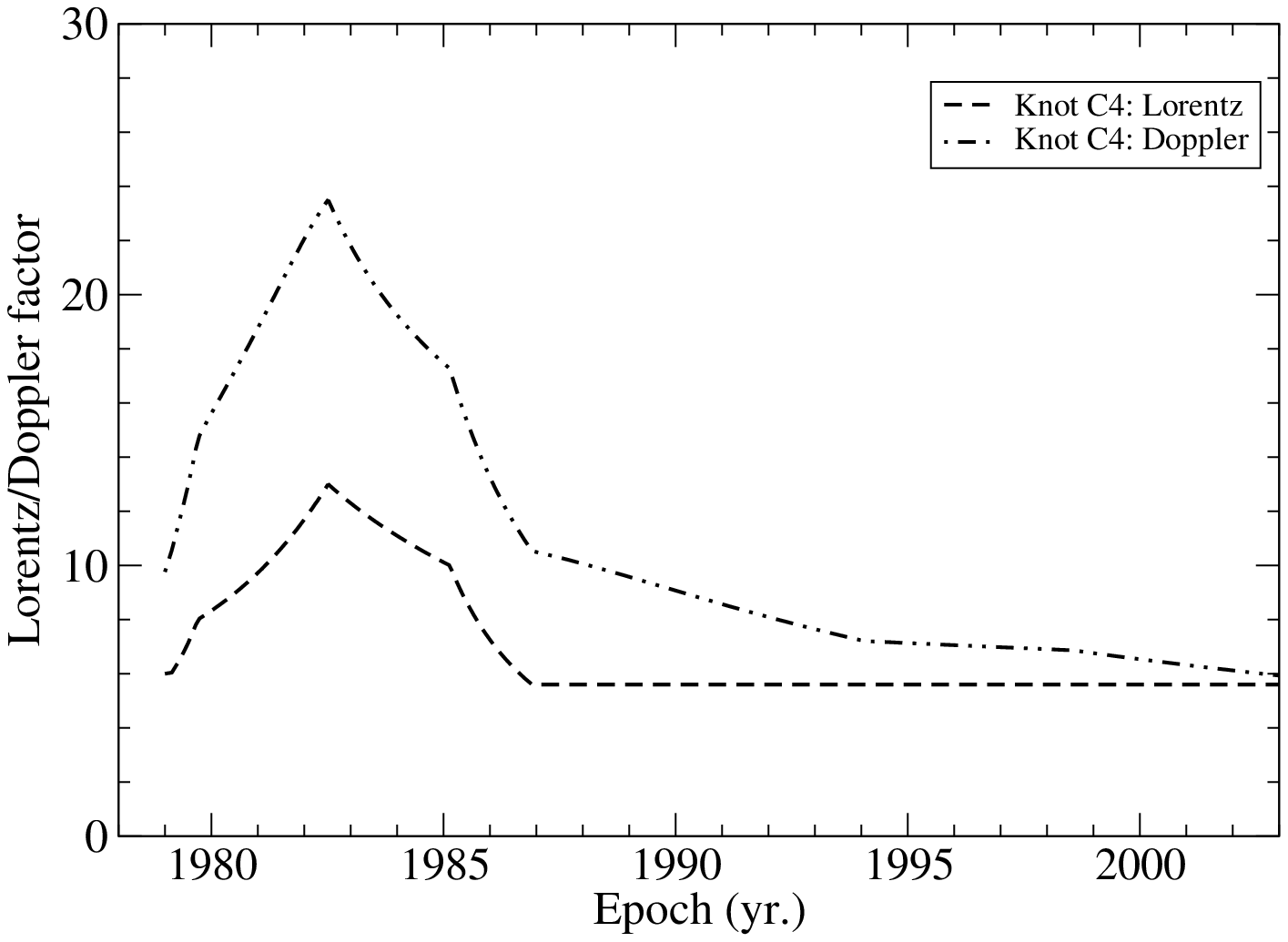}
    \includegraphics[width=5cm,angle=-90]{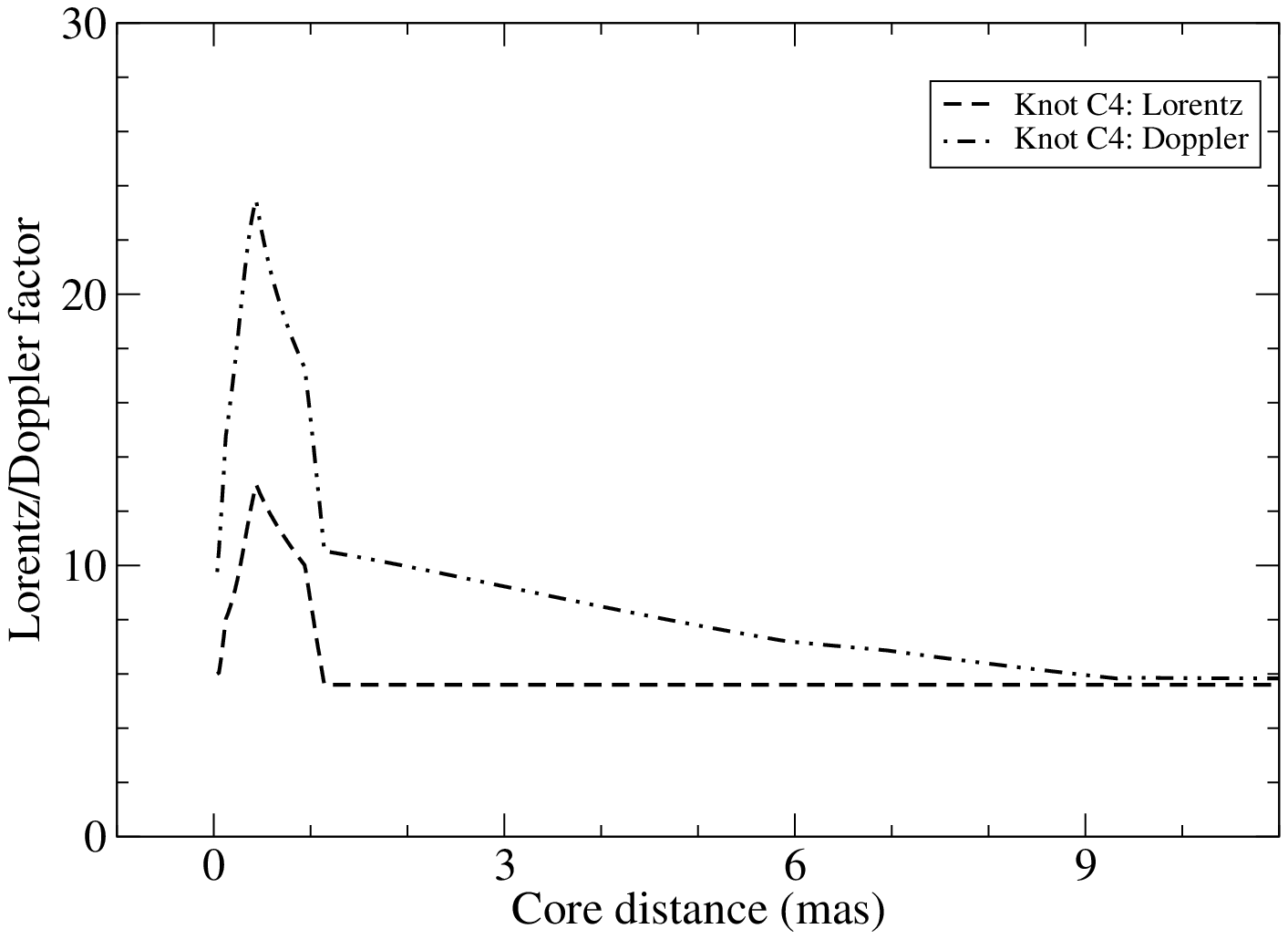}
    \includegraphics[width=5cm,angle=-90]{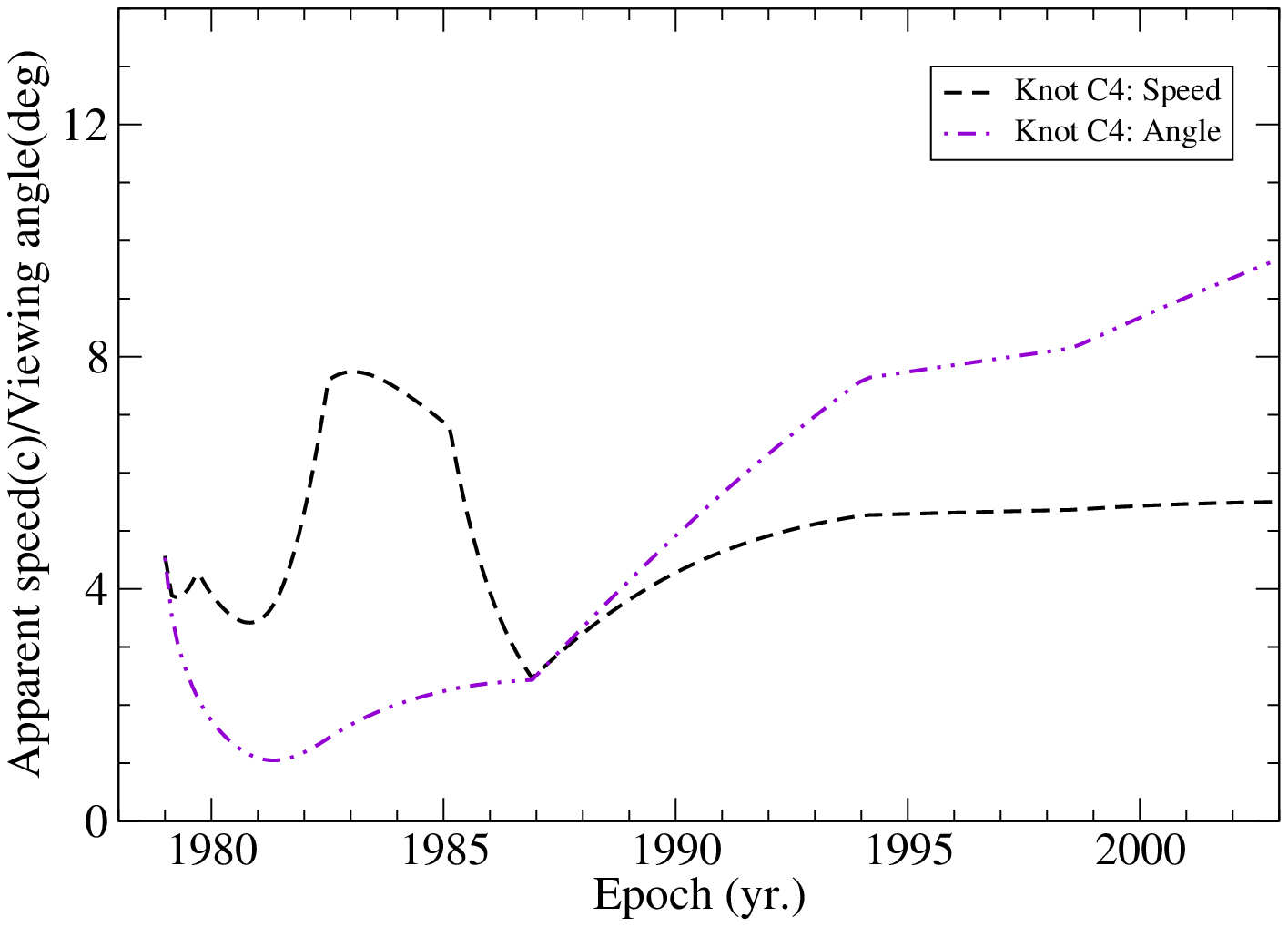}
    \includegraphics[width=5cm,angle=-90]{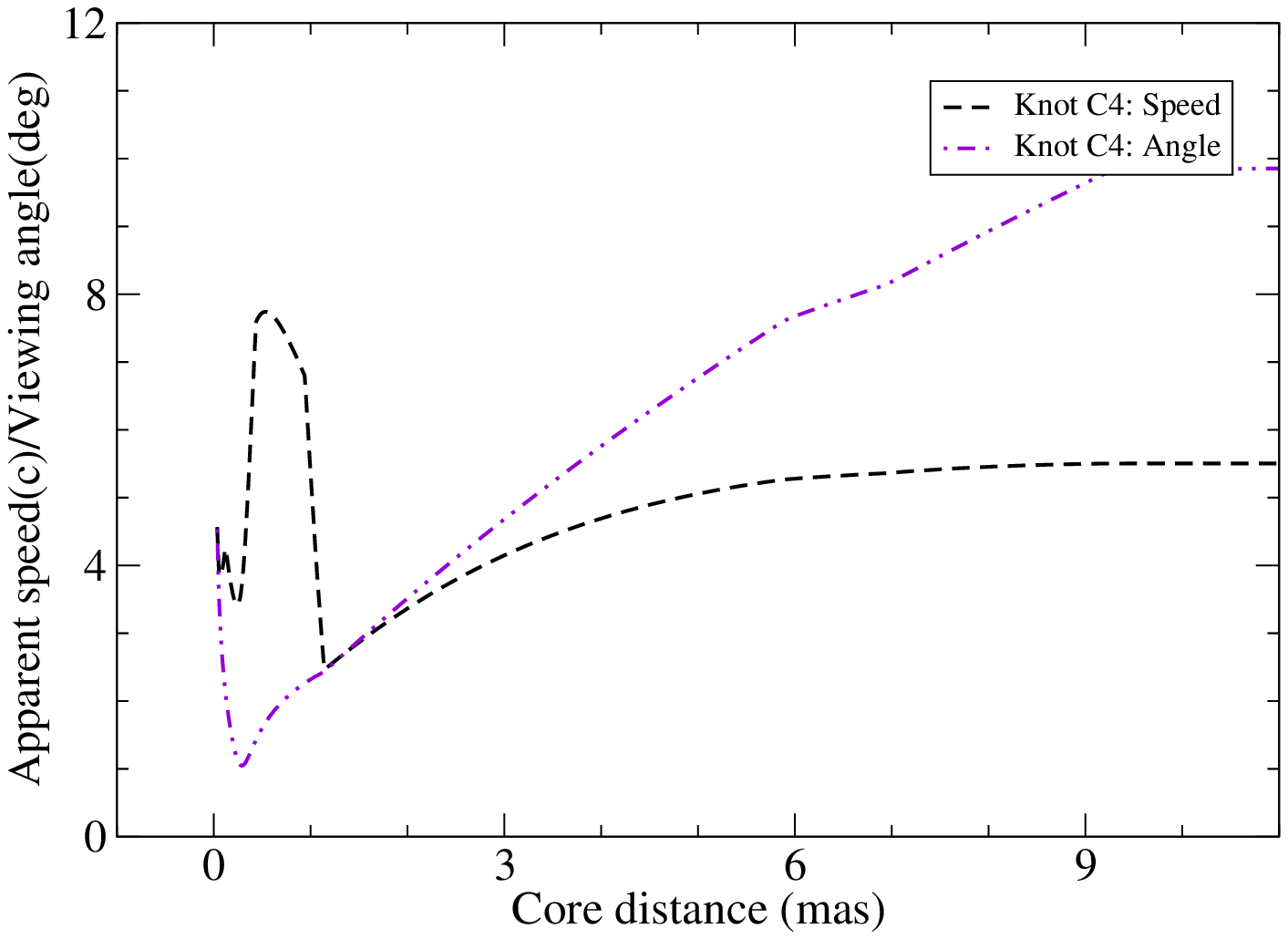}
    \caption{Knot C4: the model-derived bulk Lorentz factor 
    $\Gamma(t)$/Doppler factor $\delta(t)$ and 
     apparent speed $\beta_{app}(t)$/viewing angle $\theta(t)$ as 
     continuous functions of
    time (t) (left panels) and core distance $r_n$ (right panels).}
    \end{figure*}
    \begin{figure*}
    \centering
    \includegraphics[width=5cm,angle=-90]{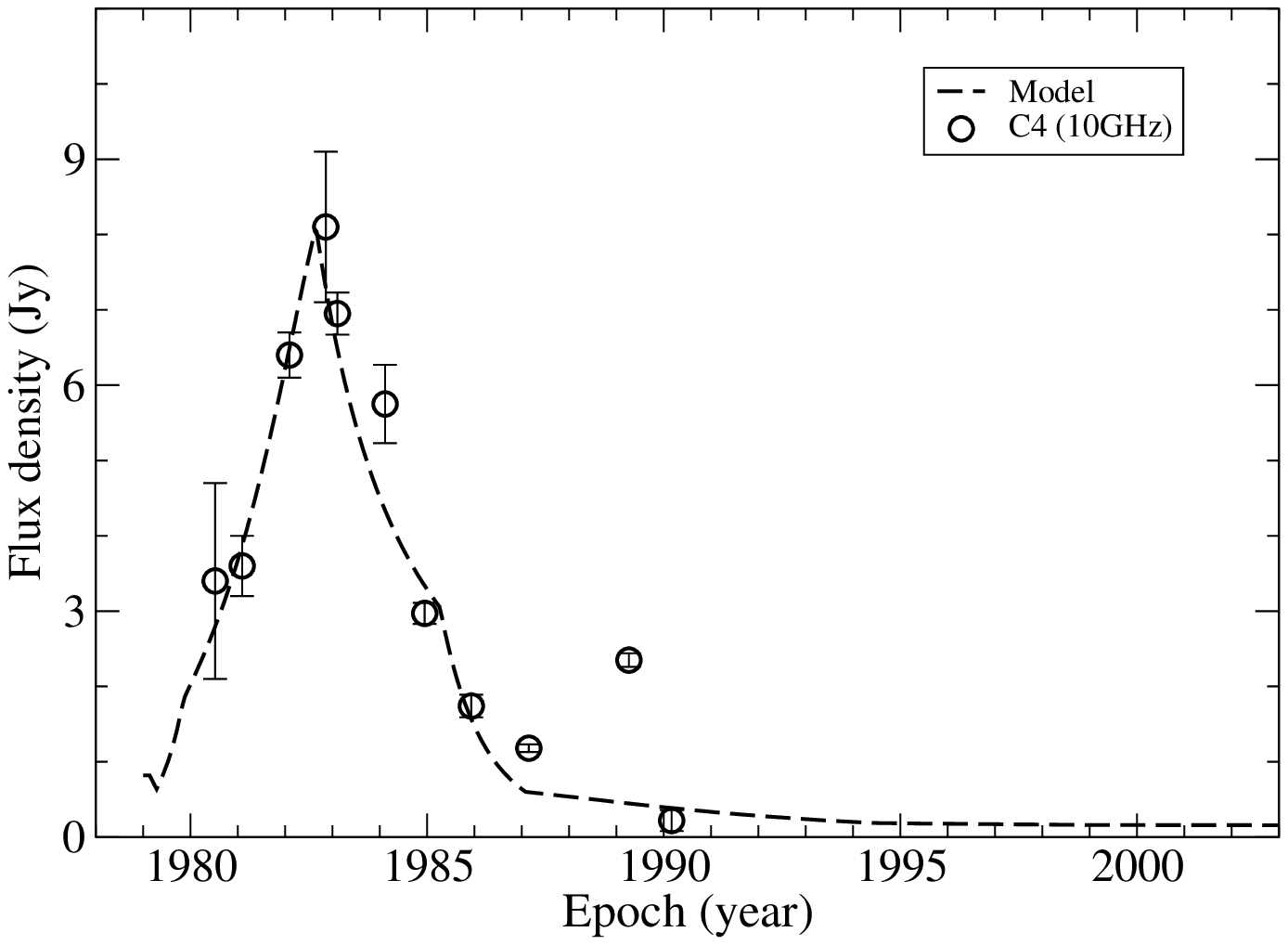}
    \includegraphics[width=5cm,angle=-90]{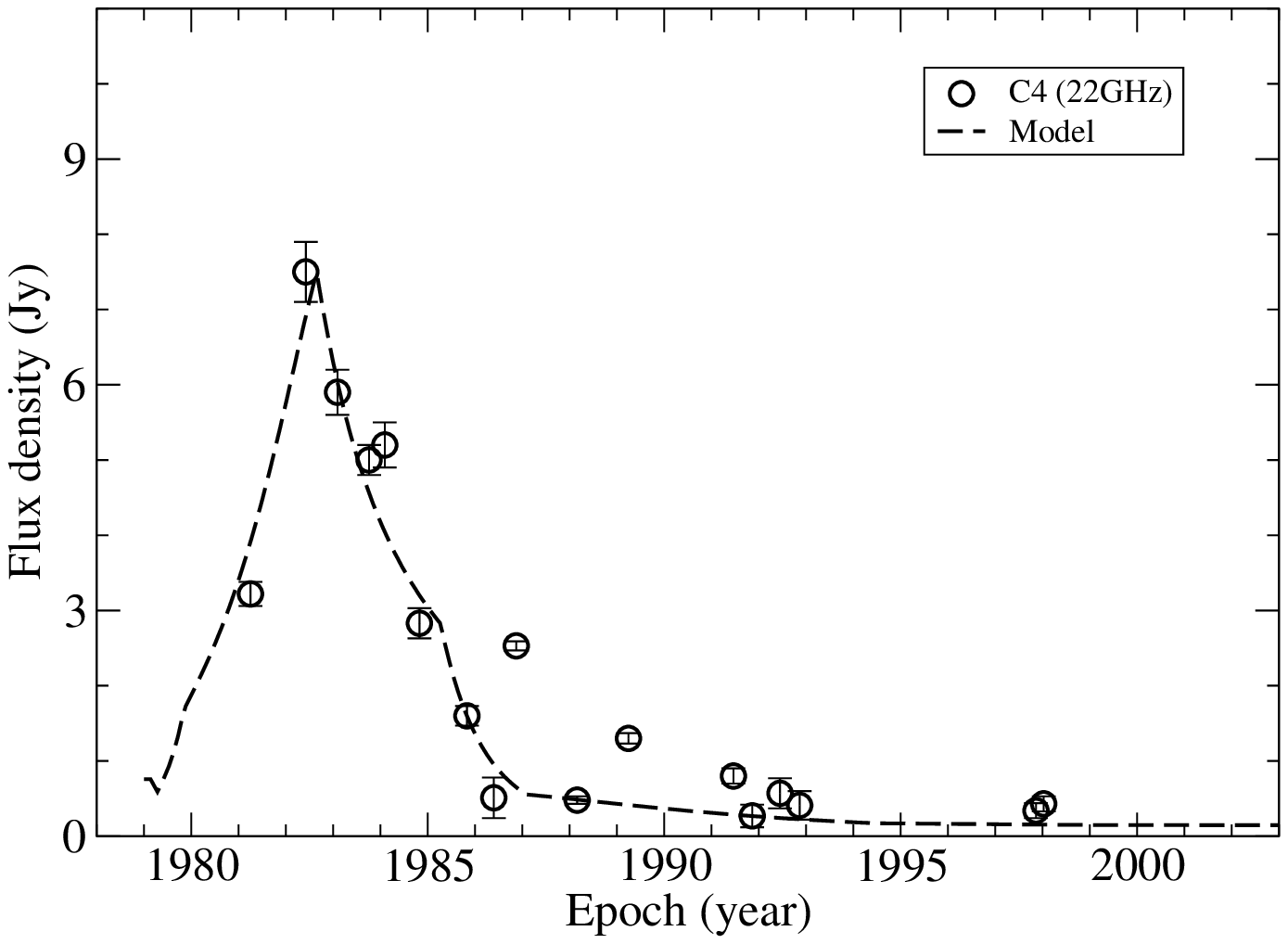}
    \caption{Knot C4: The light curves measured at 10\,GHz and 22\,GHz are well
    fitted by the Doppler boosting profiles.}
    \end{figure*}
    \begin{table*}
    \centering
    \caption{Parameters of the precessing common trajectory fitted to the
    thirteen knots (C9 and C4; C5--C7, C10--C14, C22 and C23): $\phi_0$
    (precession phase,rad), $t_0$ (ejection epoch), $r_{n,com}$ (core separation
    or the extension of the common trajectory from the core),
     $Z_{com,mas}$ (the corresponding 
    travelled distance in mas) and $Z_{com,pc}$ (the corresponding traveled
    distance in pc). References: this paper (C4 and C9); Qian \cite{Qi22b} 
    (C5, C10 and C22); Qian \cite{Qi22a} (C6, C7, C11--C14, C23).}
  \begin{flushleft}
   \centering
  \begin{tabular}{llllll}
  \hline
     Knot & $\phi_0$ & $t_0$ & $r_{n,com}$ & $Z_{com,mas}$ & $Z_{com,pc}$\\
  \hline
 C9 & 5.54+4$\pi$ & 1995.06   & 1.25  &    44.8  & 298\\
 C4 &  4.28      & 1979.00   & 1.14   & 40.0   & 266 \\
 C5 &  5.83    & 1980.80  & 1.25   &  44.8   &     298    \\
 C6 & 5.74+2$\pi$  & 1987.99   & 0.40    &   7.5   &  49.7  \\
 C7  &  6.14+2$\pi$   & 1988.46   & 0.70   & 14.9   & 99.3  \\
 C10 &   6.14+4$\pi$ &   1995.76   & 0.35   & 6.13   & 40.8  \\
 C11 &  5.88+4$\pi$  &  1995.46   & 0.75   &   15.7   &   104  \\
 C12 &  6.30+4$\pi$  & 1995.95   &   0.50   & 9.7   & 64.3  \\
 C13 &  6.50+4$\pi$  &   1996.18  &   0.70  & 14.3  & 95.3  \\
 C14 &   3.16+6$\pi$ & 1999.61   & 0.50   &   18.8   & 125  \\
 C22 &   5.28+8$\pi$  &   2009.36   &   0.20  & 3.12   & 20.7   \\
 C23 &   5.20+8$\pi$   &   2009.27  &  0.20   &   3.4   & 22.6  \\
 \hline
 \end{tabular}
 \end{flushleft}
 \end{table*}
    \section{Interpretation of kinematics and flux evolution for knot
     C4}
    The precessing common trajectory of knot C4 was observed to extend
     to core distance $r_n$=1.14\,mas (corresponding to traveled distance 
     Z=40\,mas=266\,pc). Thus it is a distinct component in 3C345 having the
     second longest extension of precessing common trajectory (only second 
     to knot C9). Its whole traejctory observed extends to 
    $r_n$=7.53\,mas ($X_n$=7.11\,mas) and the corresponding  traveled distance
     Z=56.3\,mas=374.2\,pc. Thus the investigation of its kinematics and flux
     evolution  provides another valuable opportunity to examine our 
     precessing nozzle scenario.  
    \subsection{Knot C4: model fitting of the kinematics}
     The amplitude A(Z) and phase $\phi$(Z) defining its precessing common
    helical trajectory pattern with $\phi_0$=4.28\,rad are shown in Figure 9.
    In Figure 10 the model-fitting of its inner trajectory 
     ($X_n{\leq}$1.14\,mas, corresponding to epoch 1986.91) by the precessing
    common trajectory pattern is shown. The blue and violent lines indicate the
    uncertainty range of $\pm$5\% of the precession period ($\pm$0.31\,rad or
    $\pm$0.36\,yr). It can be seen that the curvature of its trajectory is well
    model fitted.\\
     The functions $\epsilon(t)$ and $\psi(t)$ defining the jet-axis for knot
    C4 and its model-derived traveled distance Z(t) are shown in Figure 11.
    Before 1986.91 (or core distance $r_n{\leq}$1.14\,mas, corresponding
     traveled distance Z=40.0\,mas=266.0\,pc)
    $\epsilon$=$2^{\circ}$  and $\psi$=$7.16^{\circ}$, knot C4 moved along
    the precessing common trajectory pattern. After 1986.91 both $\epsilon$ and
    $\psi$ changed, knot C4 started to move along its own individual track.\\
    The whole kinematics with its entire trajectory length of 374.2\,pc 
     (including trajectory $Z_n(X_n)$, core distance
    $r_n(t)$ and coordinates $X_n(t)$  and $Z_n(t)$) are all well fitted by
     the proposed model. (Figure 12). 
    \subsection{Knot C4: Doppler boosting effect and flux evolution}
    The model-derived bulk Lorentz factor $\Gamma(t)$, Doppler factor 
    $\delta(t)$, apparent speed $\beta_{app}(t)$ and viewing angle $\theta(t)$
    as continuous functions of time (t) are shown in Figure 13. During 
    1980--1987 (or core distance $r_n{<}$1.14\,mas) model-derived Doppler 
   factor and apparent speed have a bump structure induced by the 
   increase/decrease in Lorentz factor $\Gamma$ and the change in the viewing
   angle decreasing and then increasing along a concave curve.
     After 1987 (or $r_n{>}$1.14\,mas) 
   $\Gamma$=constant and $\delta$ decreases, its apparent acceleration is 
    mainly caused by the increase in its viewing angle. \\
   The flux evolution measured at 10\,GHz and 22\,GHz can be  well explained 
   in terms of its Doppler boosting effect. As shown in Figure 14, the 
   model-derived Doppler boosting profiles well fit the observational data at
   both frequencies. There are few intrinsic variations which interfere  the 
   smooth Doppler profiles.
   \subsection{A brief summary on the model-fitting results for C4 and C9}
    The model-fitting results of the kinematics and flux evolution for knots 
    C4 and C9 have significant implications for our precessing jet-nozzle
    scenario. \\
     (1) Firstly, these results confirm the applicability of the scenario to
    interpret the VLBI-phenomena observed for the superluminal components 
    in 3C345:  their kinematics and flux evolution as a whole. Knots C4
     and C9 were well model-fitted to move along the precessing common helical
     trajectory patterns (with precession phases $\phi_0$=4.28\,rad and 
    $\phi_0$=5.54+4$\pi$, respectively) which extend to core distances of 
    $\sim$1.2\,mas or traveled distances of $\sim$300\,pc (Table 2). Thus they 
    are the components in 3C345 having the longest extensions of 
    the precessing common helical trajectory pattern, which provides
     strong arguments  for the 
    existence of jet-nozzle precession and  precessing common trajectory 
    pattern \footnote{These are  two basic assumptions 
    in our precessing
    nozzle scenario.}.  In fact, the curvature in their observed trajectories 
    can be well explained in terms of their helical motion within an 
    uncertainty of $\sim{\pm}$5\%  ($\pm$0.36\,yr) of the  precession period 
    of 7.3\,yr (Figures 4 and 10).    \\
    (2) Except knots C4 and C9, the kinematics and flux evolution of  knots C5,
     C10 and C22 have been well model-fitted as a whole in Qian (\cite{Qi22b})
    ,indicating their motion in the inner-jet regions following their 
     precessing common helical traejctory patterns. There are more components
     (C6--C7,  C11--C14 and C23) for which their kinematics also have been well
     expalined (Qian \cite{Qi22a}), showing their motion in the inner jet
     regions along the
     precessing common helical trajectories with their respective precession
      phases  and different extensions $Z_{com,pc}$ in the range of 
    $\sim$20--250\,pc (Table 2). Obviously, the flux evolution of these knots
     can be interpreted in terms of their Doppler boosting effect as well.\\
    (3) For all these superluminal knots mentioned above, both the 
     model-fitting of their kinematics (curved trajectories, coordinates and 
     apparent sppeds) observed in the inner jet 
     regions and the interpretation of their flux evolution in terms of the 
     model-derived Doppler boosting profiles firmly demonstrate the validity
     of the precessing helical pattern defined by equations (19) and (20)
    [Figures 2, 4 and 10]. Moreover, the recurrence of the 
     inner-trajectory patterns observed during $\sim$30 years (1980--2009)
     for knots C5, C6, C9, C11 and C22 [Qian \cite{Qi22a}] further confirms
     the existence of precessing common helical trajectory pattern in 3C345.\\
    (4) In this paper we will apply the same scenario as described in 
      Section 2 to interpret both the kinematics and flux evolution of knots 
     C8, C20, C21, B5 and B7, demonstrating that the precessing nozzle 
     scenario is still effective, but their motion follows the precessing 
     common helical trajectory  patterns within core distances of 
     $\sim$0.05--0.15\,mas. Therefore, a precessing nozzle scenario with a 
     single jet would be applicable to understanding the whole phenomena in 
     3C345 (both kinematics and emission properties of the superluminal 
     knots).\\
    \section{Interpretation of the kinematics and flux evolution for 
     superluminal knots C8, C20, C21, B5 and B7 }
     The kinematics and flux evolution of these five superluminal knots
     have been investigated in Qian (for knot C8, \cite{Qi22b}; for knots C20,
      C21, B5 and B7, \cite{Qi23}). Here we will apply the same precessing 
     nozzle scenario as proposed above (with the same  precessing common
     helical trajectory pattern defined by equations (19) and (20)) to study
      their kinematic behavior 
     and flux evolution, thus unifying the interpretation of all these 
     superluminal knots' observed properties in 3C345 in a single-jet
      scenario.\\
     Since the transitions between their precessing common trajectory pattern 
    and their individual track occurred
    within core separation $r_n$$\stackrel {<}{_\sim}$0.05--0.15\,mas, thus
    their precessing common trajectories were not observed (for knots
    C20, C21, B5 and B7). Only for knot C8 its transition was observed at
    43GHz (Klare et al. 2003), as already investigated before in Qian
     (\cite{Qi22a}).
    \section{Knot C8: model-fitting of the kinematics and flux evolution}
    Its kinematics was already well model-fitted in Qian (\cite{Qi22a}). Here 
    we recaptulate the results and supplement the explanation of its flux
     evolution in terms of the Doppler boosting effect.
     \begin{figure*}
     \centering
     \includegraphics[width=5cm,angle=-90]{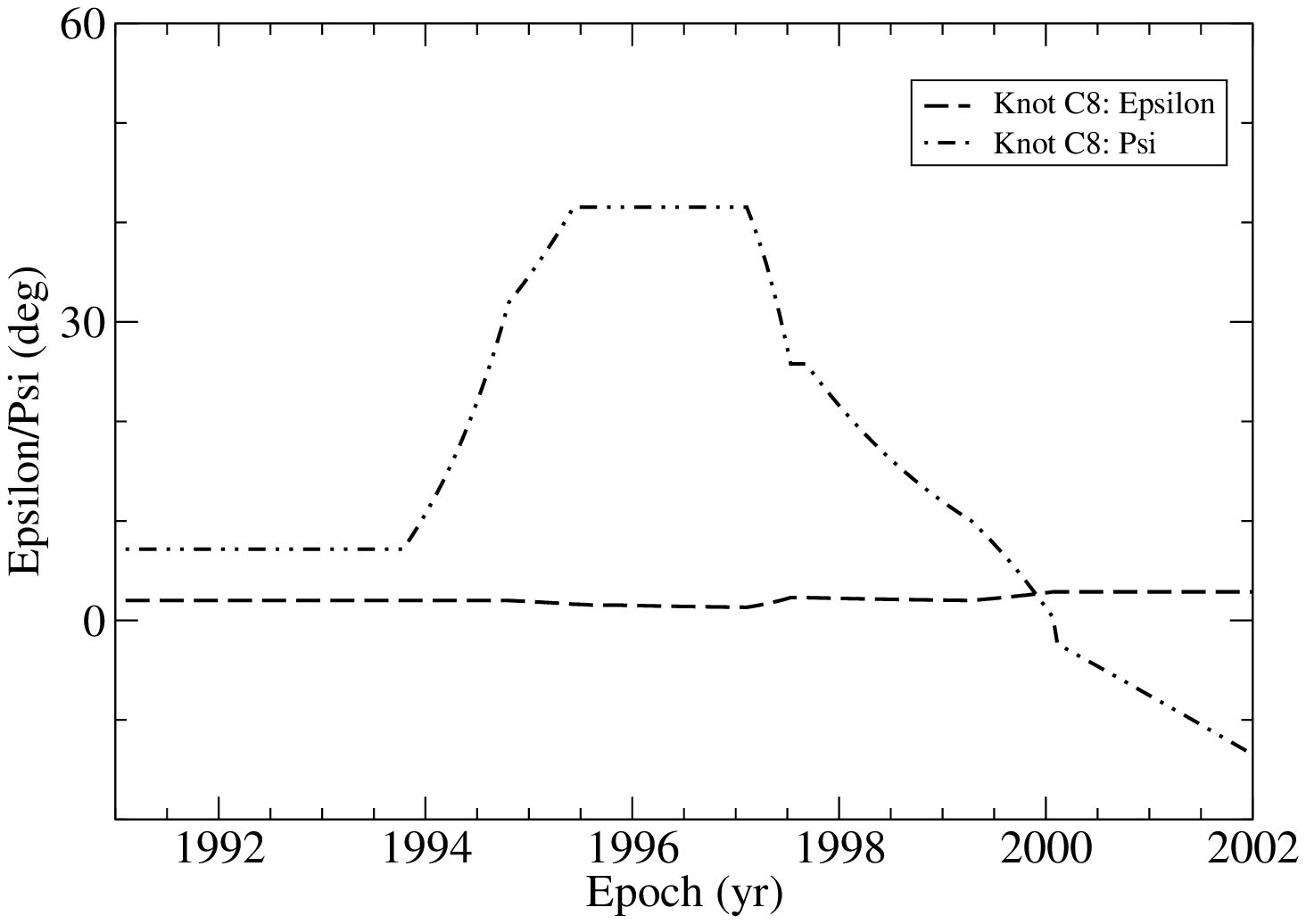}
     \includegraphics[width=5cm,angle=-90]{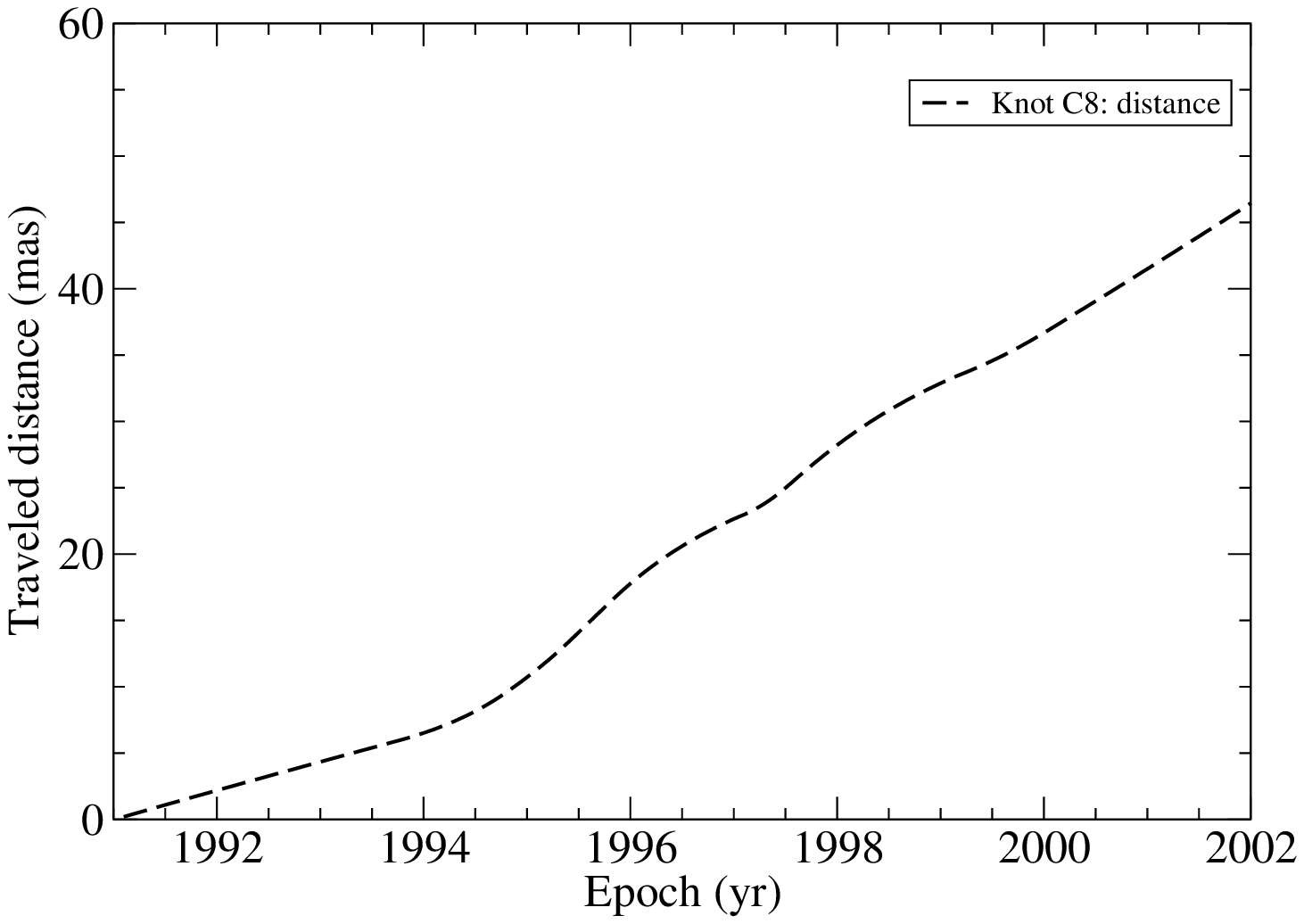}
     \caption{Knot C8 ($\phi_0$=2.13+4$\pi$, $t_0$=1991.10): Functions 
     $\epsilon(t)$ and $\psi(t)$ (left panel) 
     defining the jet-axis and its traveled distance Z(t). Before 1993.78
     (corresponding core distance $r_n$=0.14\,mas, coordinate $X_n$=0.12\,mas
     and traveled distance Z=6.0\,mas=39.9\,pc),
     $\epsilon$=$2^{\circ}$ and $\psi$=$7.16^{\circ}$, knot C8 moved along the
     precessing common helical trajectory pattern. After that $\psi$ 
     increased and  knot C8 started to move along its own individual track.}
     \end{figure*}
    \begin{figure*}
    \centering
    \includegraphics[width=5cm,angle=-90]{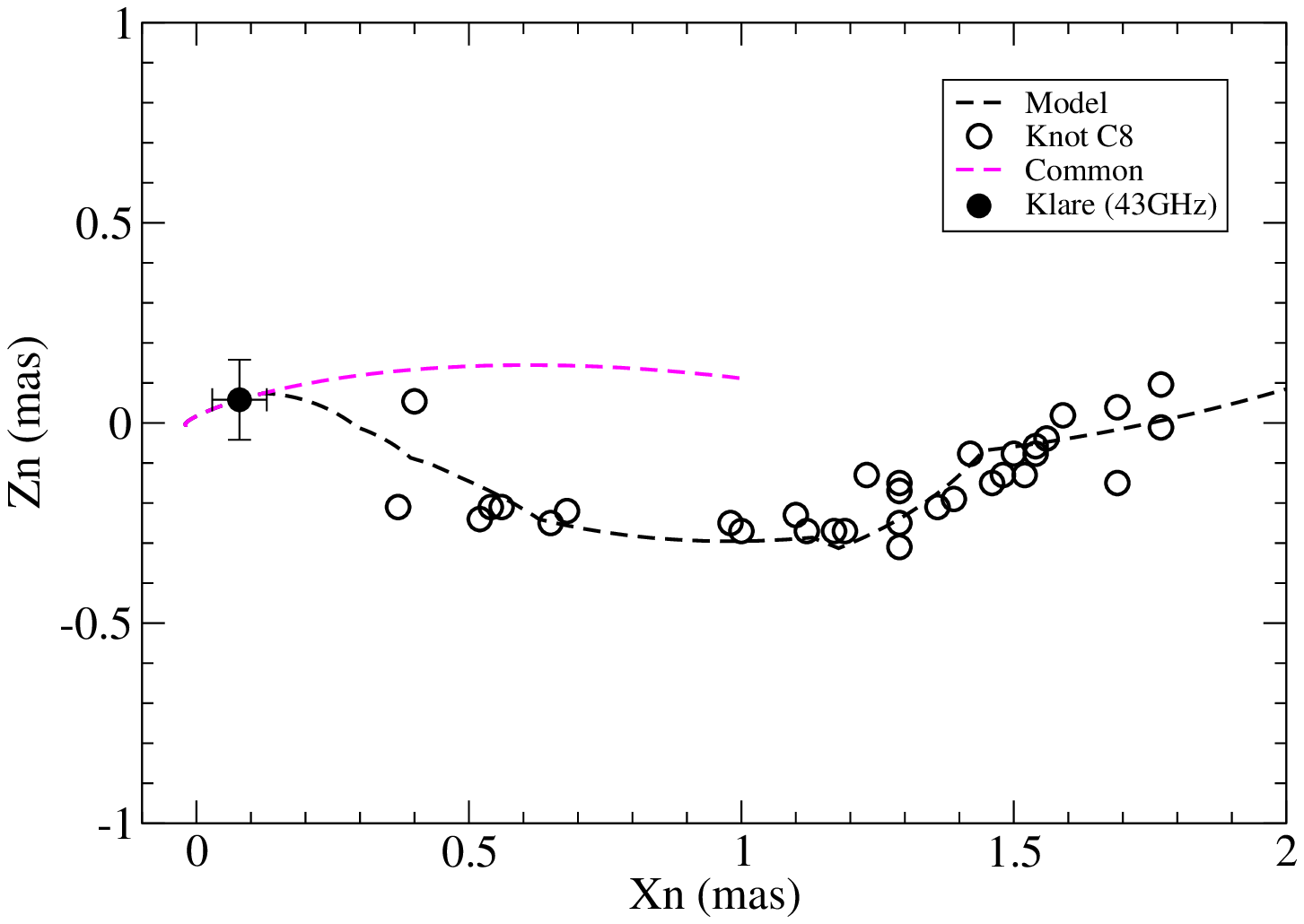}
    \includegraphics[width=5cm,angle=-90]{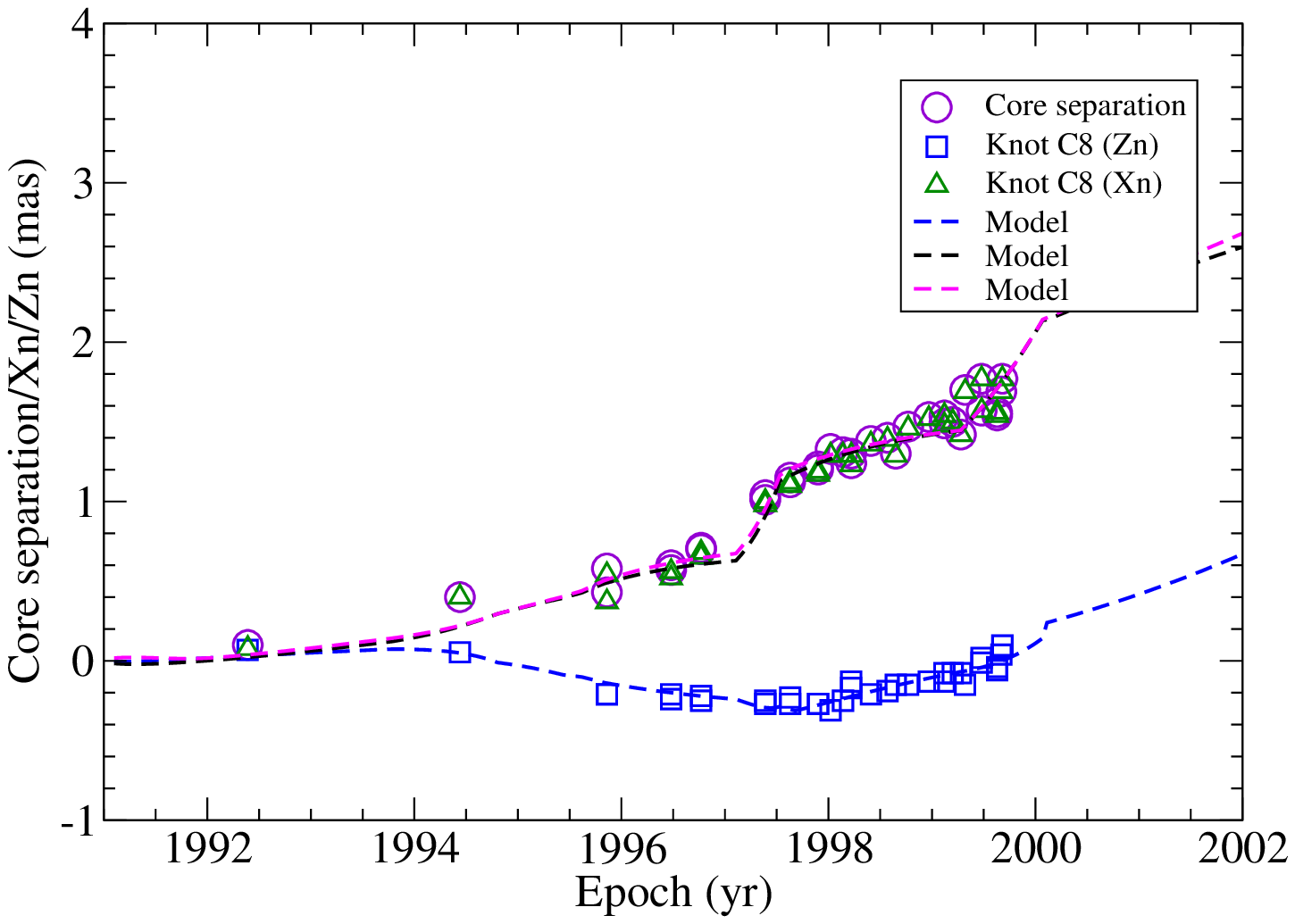}
    \caption{Knot C8: Model-fitting of its kinematics. The datapoint measured
     at 43GHz in 1992.39 (Klare \cite{Kl03}, Klare et al. \cite{Kl05}.) 
    indicates it moving along the precessing common helical trajectory 
    (red line; left panel) within $X_n{\leq}$0.12\,mas. Beyond $X_n$=0.12\,mas
     (or
    after 1993.78) it moved along its own individual track. It can be seen that
    its whole trajectory $Z_n(X_n)$ (right panel) was well model-fitted. Its
    core distance $r_n(t)$, coordinates $X_n(t)$ and $Z_n(t)$ 
    are also well fitted by the proposed model.}
    \end{figure*}
    \begin{figure*}
    \centering
    \includegraphics[width=5cm,angle=-90]{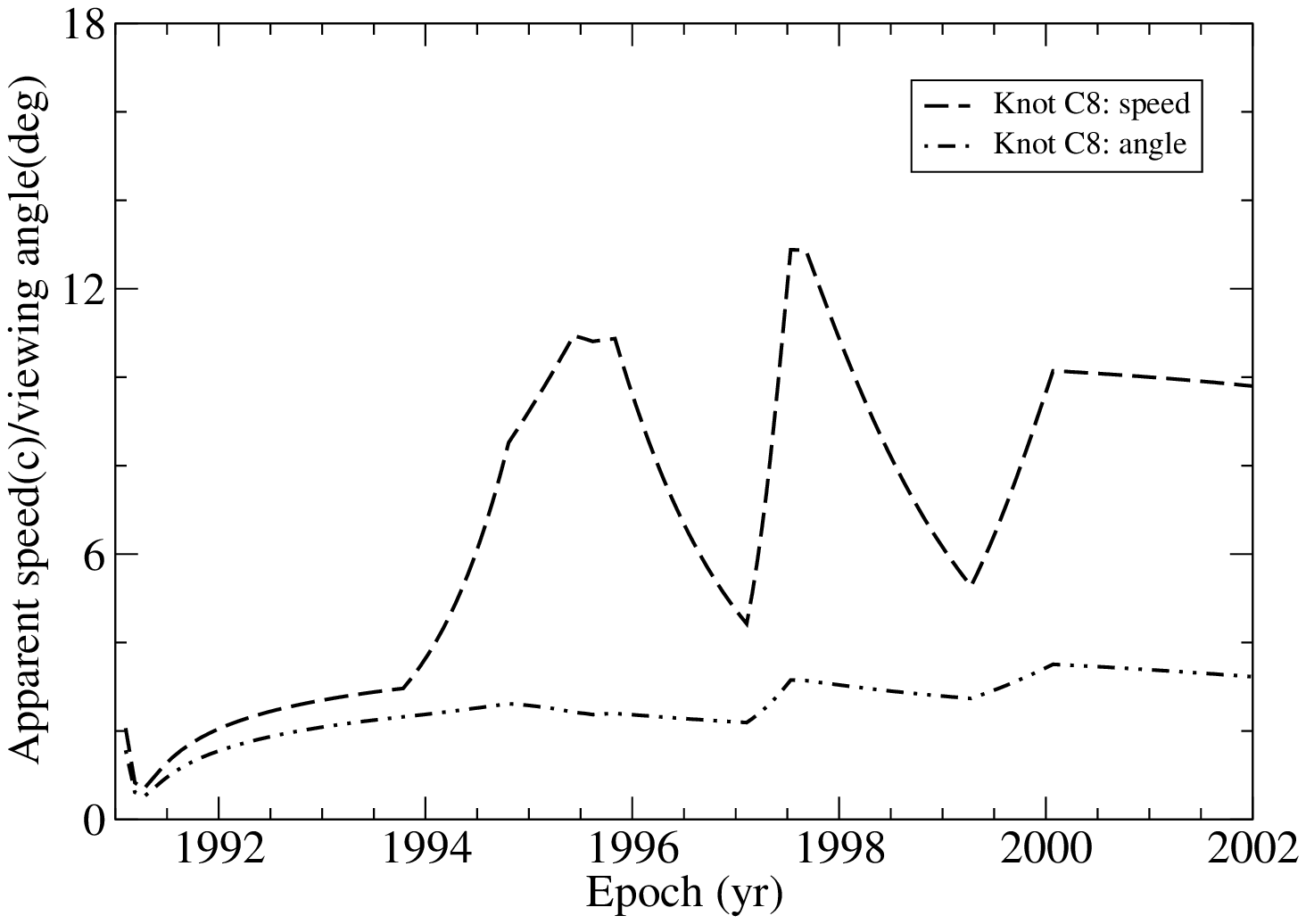}
    \includegraphics[width=5cm,angle=-90]{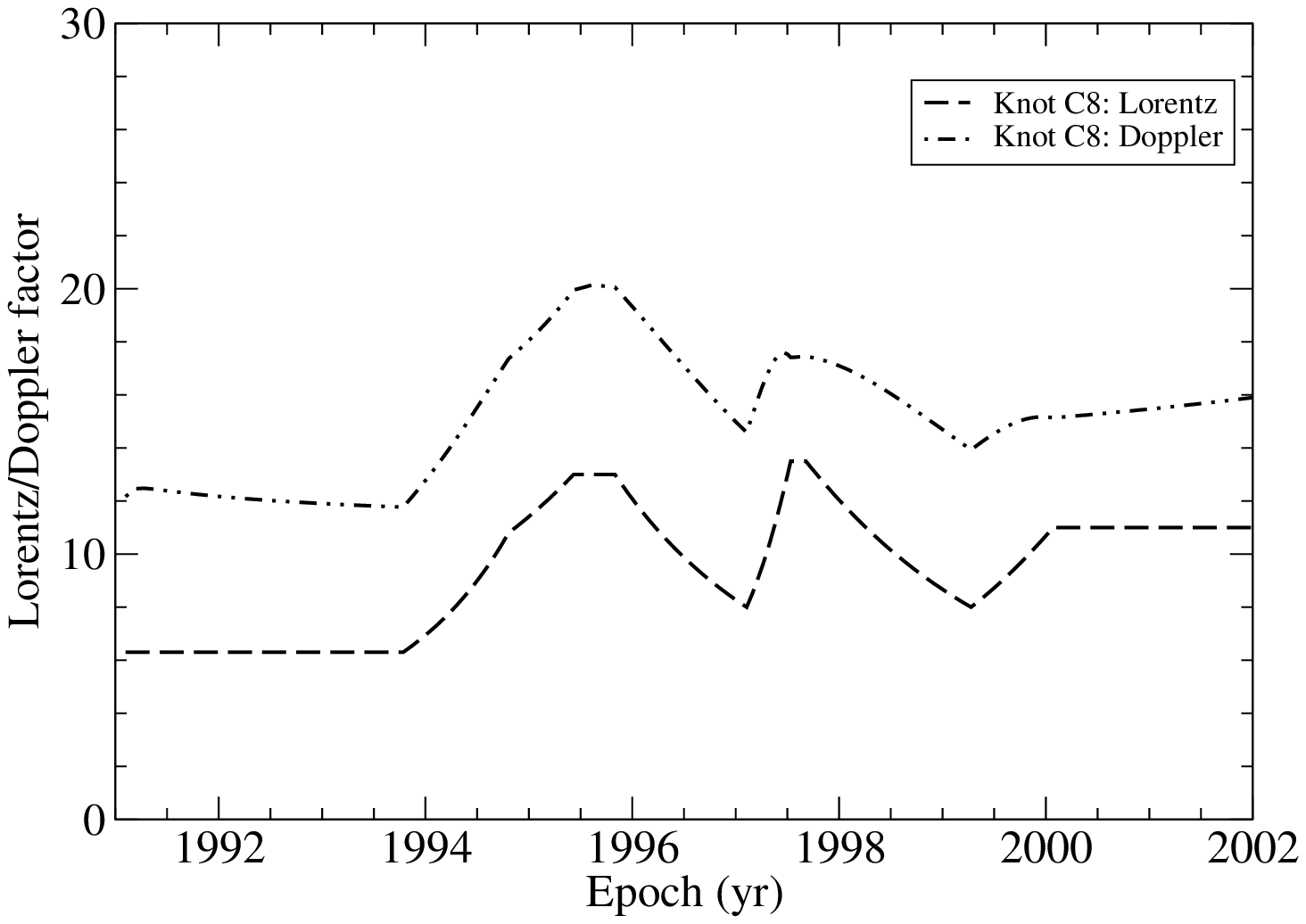}
    \caption{Knot C8: The model-derived apparent speed $\beta_{app}$(t) and
    viewing angle $\theta(t)$ are presented in the left panel, and 
    the model-derived bulk Lorentz factor and Doppler factor in the right panel.
    During 1994--1997 the apparent speed, bulk Lorentz factor and Doppler 
    factor all have a double-peak structure associated with  similar changes
    in its viewing angle. Thus knot C8 underwent a process of
     acceleration/deceleration/reacceleration/deceleration.} 
    \end{figure*}
    \begin{figure*}
    \centering
    \includegraphics[width=6cm,angle=-90]{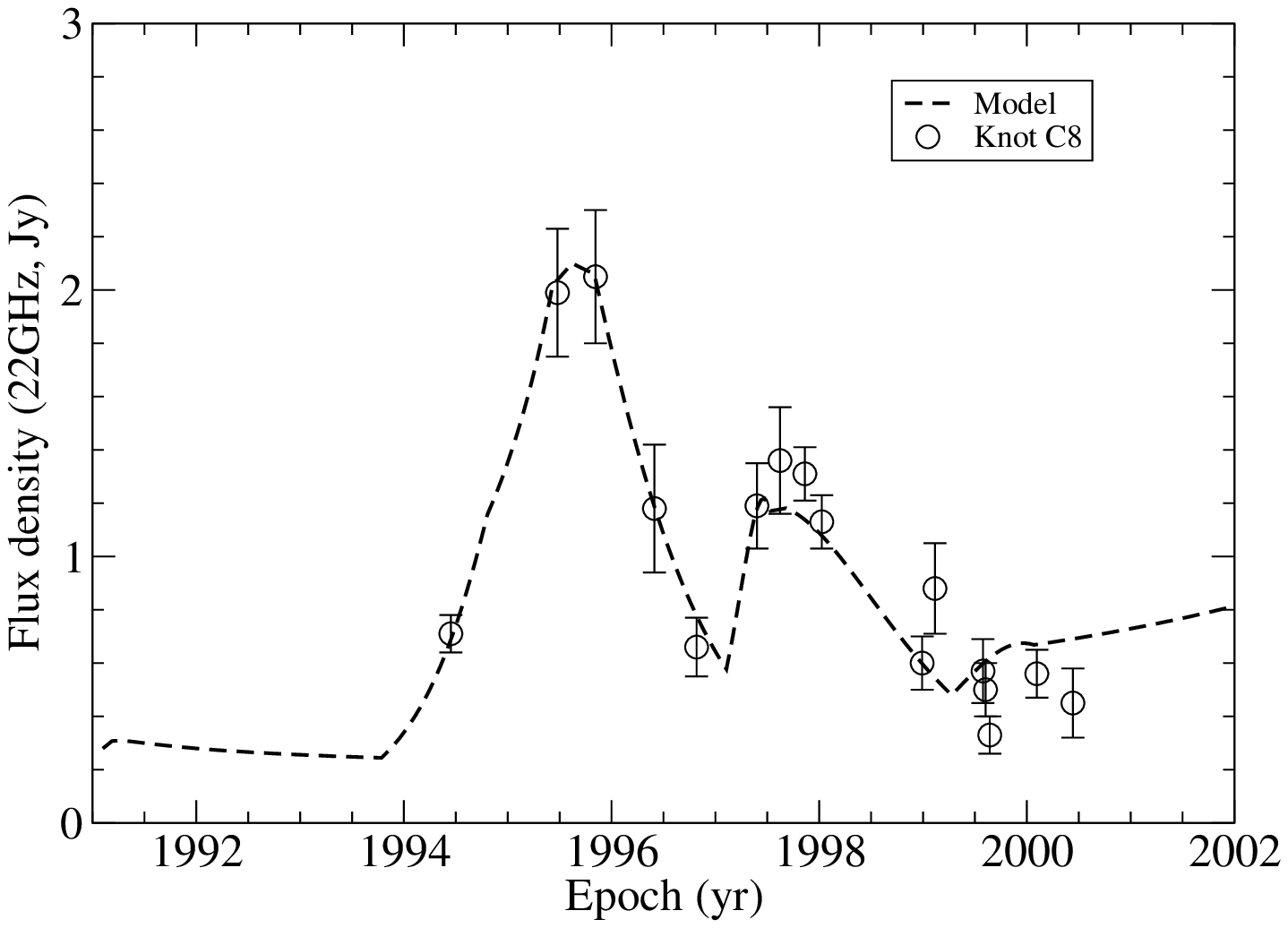}
    \caption{Knot C8: Doppler boosting effect and flux evolution.
     The model-derived Doppler factor with a double-peak structure
     leads to the 22\,GHz light curve well fitted by the  Doppler boosting
     profile ${S_{int}}{[\delta(t)]^{3+\alpha}}$; $S_{int}$=12.7$\mu$Jy and
     $\alpha$=1.0.}
    \end{figure*}
     \subsection{Model fits to the kinematics of knot C8}
   The model-fitting results of kinematics for knot C8 are shown in Figures
    15--17. The model-fitted parameters $\epsilon(t)$ and $\psi(t)$ indicate 
    that before 1993.78 knot C8 moved along the precessing common helical
    trajectory pattern, and after that epoch it moved along its own individual
    track. The transition occurred at core distance  $r_n$=0.142\,mas (or 
    $X_n$=0.122\,mas; corresponding traveled distance Z=6.0\,mas=39.9\,pc).
    Its whole trajectory $Z_n(X_n)$, core distance $r_n(t)$, coordinates 
    $X_n(t)$ and $Z_n(t)$ are well model-fitted by the precessing nozzle 
    scenario as shown in Figure 16.\\
    During 1994--1997 the model-derived bulk Lorentz factor $\Gamma(t)$, 
    Doppler factor  $\delta(t)$ and apparent speed $\beta_{app}(t)$ have a 
    double peak structure associated with a similar change in the
     viewing angle, demonstrating knot C8's double acceleration/deleceration 
    process (Figure 17).\\
    We emphasize that the VLBI-observation at 43\,GHz for C8 (Klare 
    \cite{Kl03}) is very instructive. In fact, if there were no
    measurement at 1992.39, we would not be able to recognize that knot C8 
      moved along its precessing common trajectory pattern within core 
    distance $\sim$0.15\,mas.  We expected that there may be more 
    superluminal components in 3C345 have similar kinematic behavior, as 
    described below for knots C20, C21, B5 and B7.
    \subsection{Knot C8: Doppler boosting effect and flux evolution}
    The model-derived double-peak structure for Doppler factor leads to a
    Doppler boosting profile ${S_{int}}{[\delta(t)]^{3+\alpha}}$ 
   with a double-peak  structure, which nicely explain the observed 22\,GHz 
    light cuvre as shown
    in Figure 18 ($S_{int}$=12.7$\mu$Jy and $\alpha$=1.0).
    \section{Interpretation of kinematics and flux evolution for knot C20}
    The kinematics and flux evolution of knot C20 were explained in the 
    framework of double jet scenario in Qian (\cite{Qi22b}), here we  
    apply the single jet scenario as proposed above to study its observed
     properties. According equation (21) we obtain its precession phase
     $\phi_0$=3.83+8$\pi$, corresponding to its ejection epoch $t_0$=2007.68.
    The model-fitting results are presented in Figures 19--22.
    \subsection{Knot C20: Model-fitting of the kinematics}
    The model-derived parameters $\epsilon(t)$ and $\psi(t)$ defining the 
    jet-axis, and its traveled distance Z(t) are shown in Figure 19, revealing
     that before 2008.23 (corresponding core distance $r_n$=0.076\,mas, 
     coordinate $X_n$=0.041\,mas, traveled distance Z=1.20\,mas=7.98\,pc)
    $\epsilon$=$2^{\circ}$ and $\psi$=$7.16^{\circ}$, knot C20 moved along 
    the precessing common helical trajectory pattern. After 2008.23 $\psi$
    started to decrease and it then moved along its own individual track.
    This inner precessing common track section is very short and was not
    observed.\\
    In Figure 20 its whole trajectory $Z_n(X_n)$ (black line, left panel; 
     the red-line
    indicating the precessing common track), and core distance $r_n(t)$, 
    coordinates $X_n(t)$ and $Z_n(t)$ (right panel) are well model-fitted .\\
     The model-derived bulk Lorentz factor and Doppler factor are presented in 
    Figure 21 (left panel), both showing a single-peak structure. Its
      apparent speed and viewing angle are  shown in the right panel. During
     2008--2010.4 its viewing angle varies following  a concave curve. \\
     \subsection{Knot C20: Doppler boosting effect and flux evolution}
     The model-derived Doppler factor as a continuous function of time (t)
     provides a valuable chance to study its flux evolution. As shown in Figure
     22, its 43GHz light curve measured  is well model-fitted by its
     Doppler boosting profile (${S_{int}}[\delta(t)]^{3+\alpha}$) (with 
     $S_{int}$=35.7$\mu$Jy, $\alpha$=0.50). Intrinsic variations superposes 
     on it  quite  prominently. 
     \begin{figure*}
     \centering
     \includegraphics[width=5cm,angle=-90]{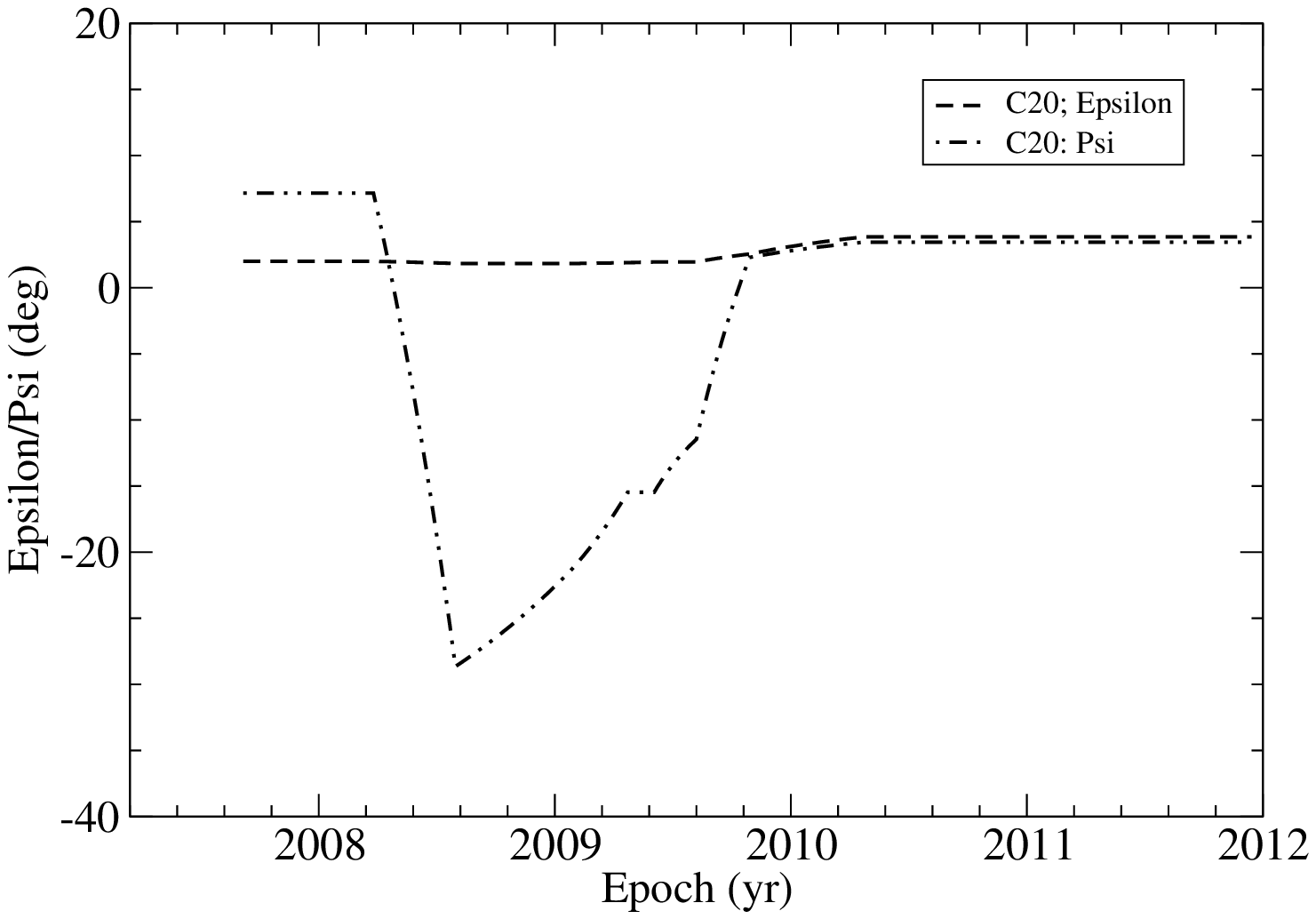}
     \includegraphics[width=5cm,angle=-90]{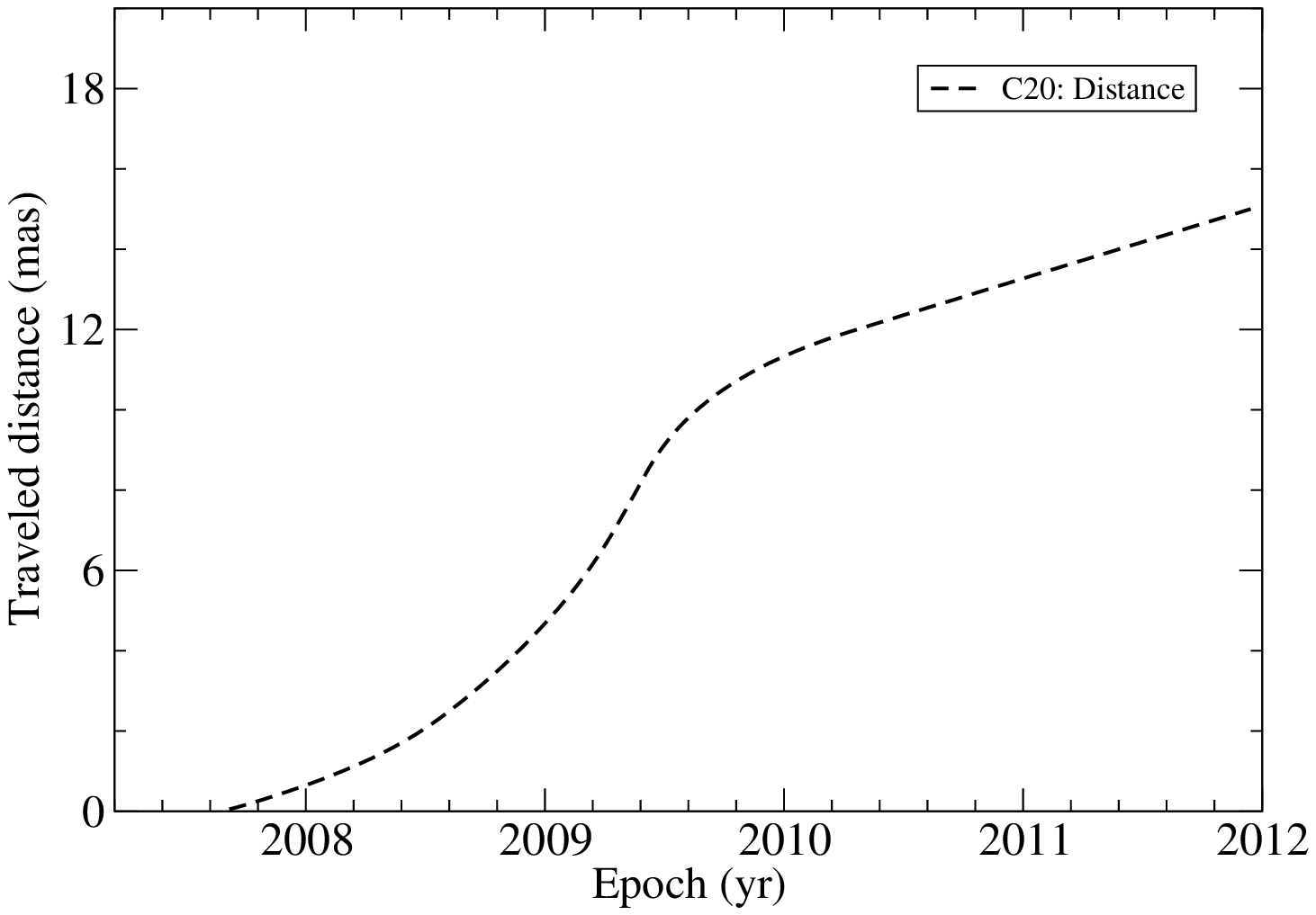}
     \caption{Knot C20: the model-derived parameters $\epsilon(t)$ and 
     $\psi(t)$ defining the jet-axis (left panel) and the traveled distance
      Z(t). Before 2008.23 $\epsilon$=$2^{\circ}$ and $\psi$=$7.16^{\circ}$
     (corresponding core distance $r_n$=0.076\,mas, $X_n$=0.041\,mas and
      traveled distance Z=1.20\,mas=7.98\,pc),
     knot C20 moved along the precesssing common helical trajectory pattern, 
     while after that epoch $\psi$ started to decrease and it then moved along
     its own individual track. Its trajectory-transit occurred near the core 
     and was not observed.}
     \end{figure*}
     \begin{figure*}
     \centering
     \includegraphics[width=5cm,angle=-90]{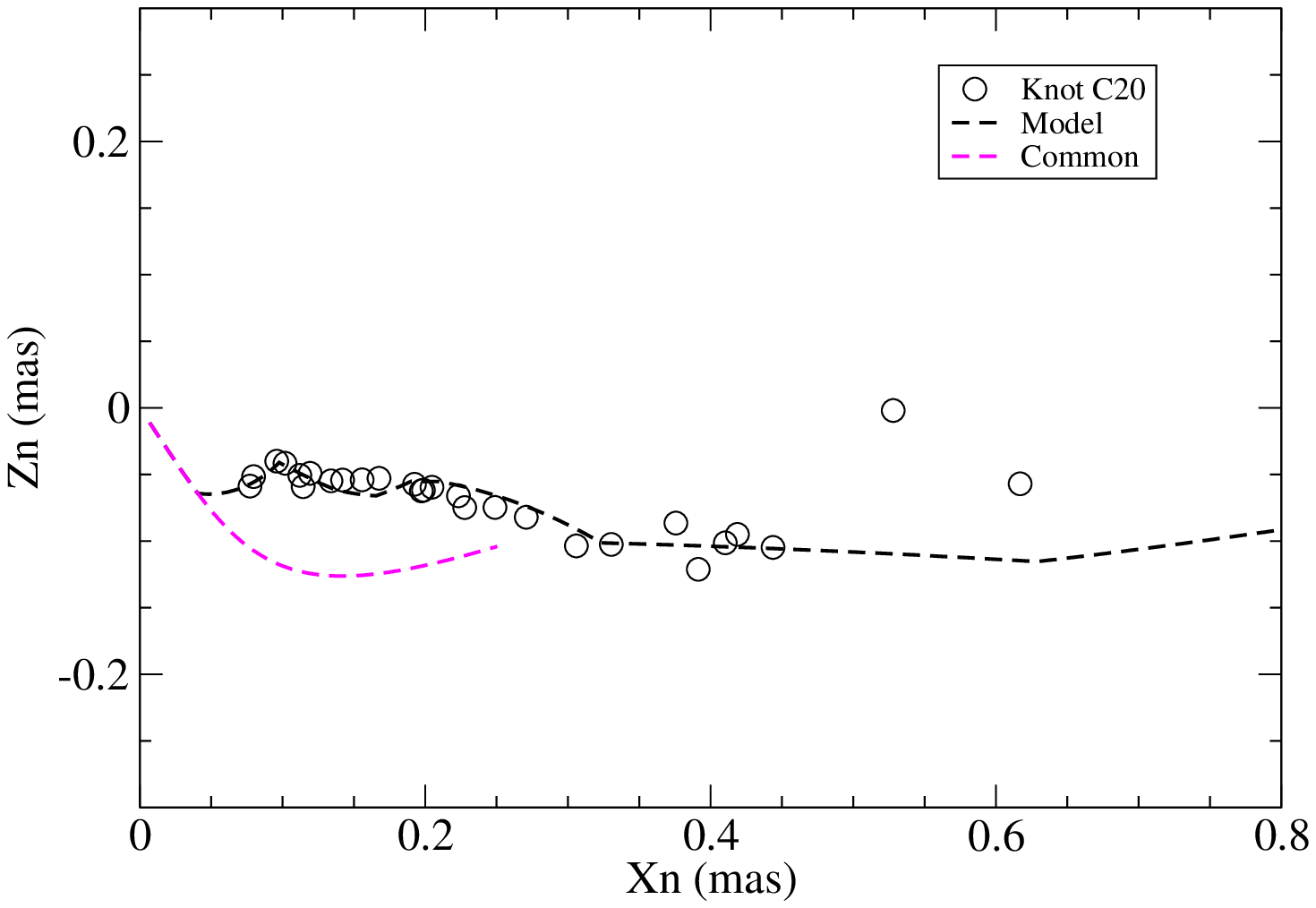}
     \includegraphics[width=5cm,angle=-90]{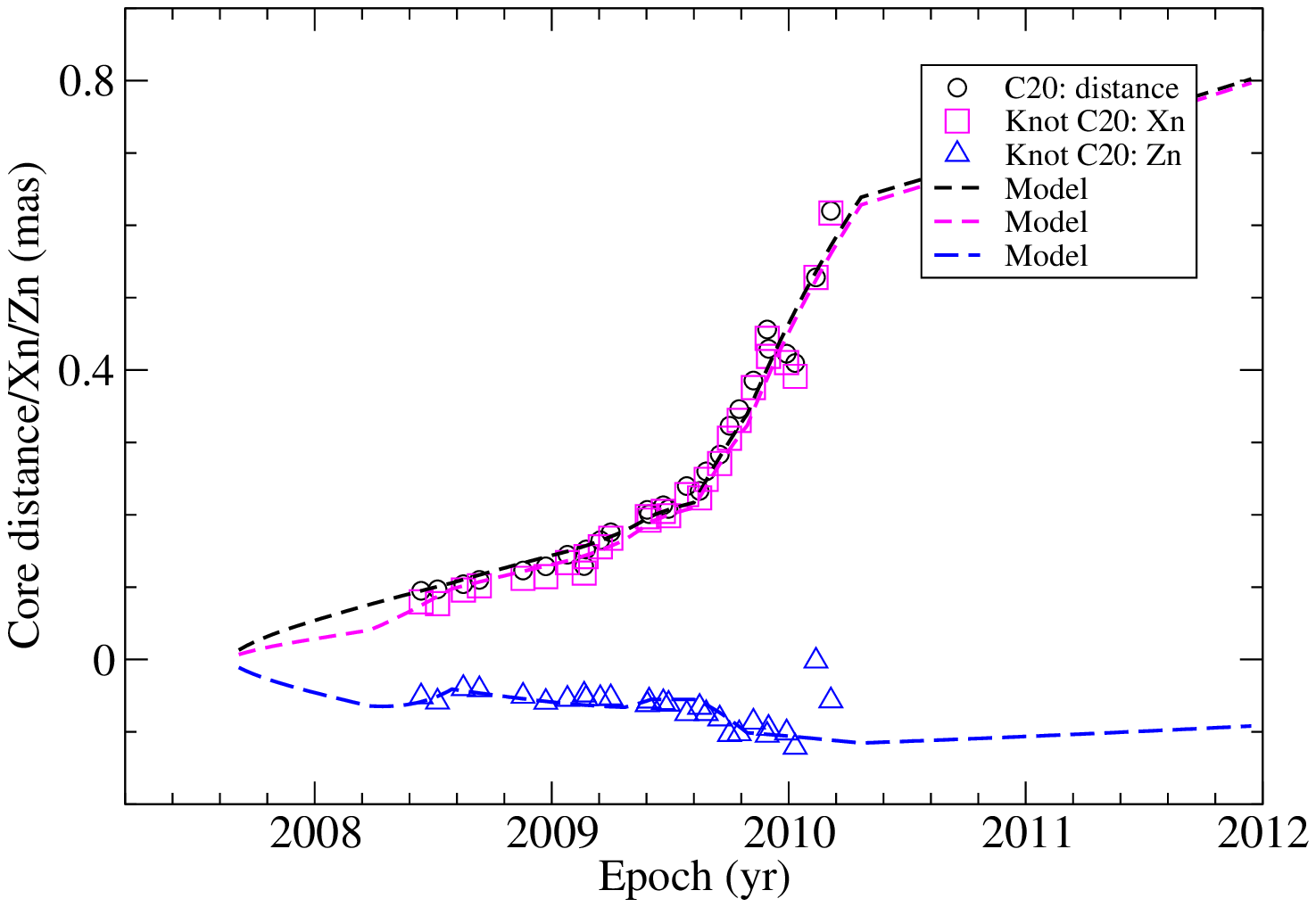}
     \caption{Knot C20. The whole trajectory $Z_n(X_n)$ is well model-fitted
      by the proposed model (left panel, black line; the red line indicating
       the precessing common track). The model-fits to the core distance
       $r_n(t)$, coordinates $X_n(t)$ and $Z_n(t)$ are shown in the 
     right panel.}
     \end{figure*}
     \begin{figure*}
     \centering
     \includegraphics[width=5cm,angle=-90]{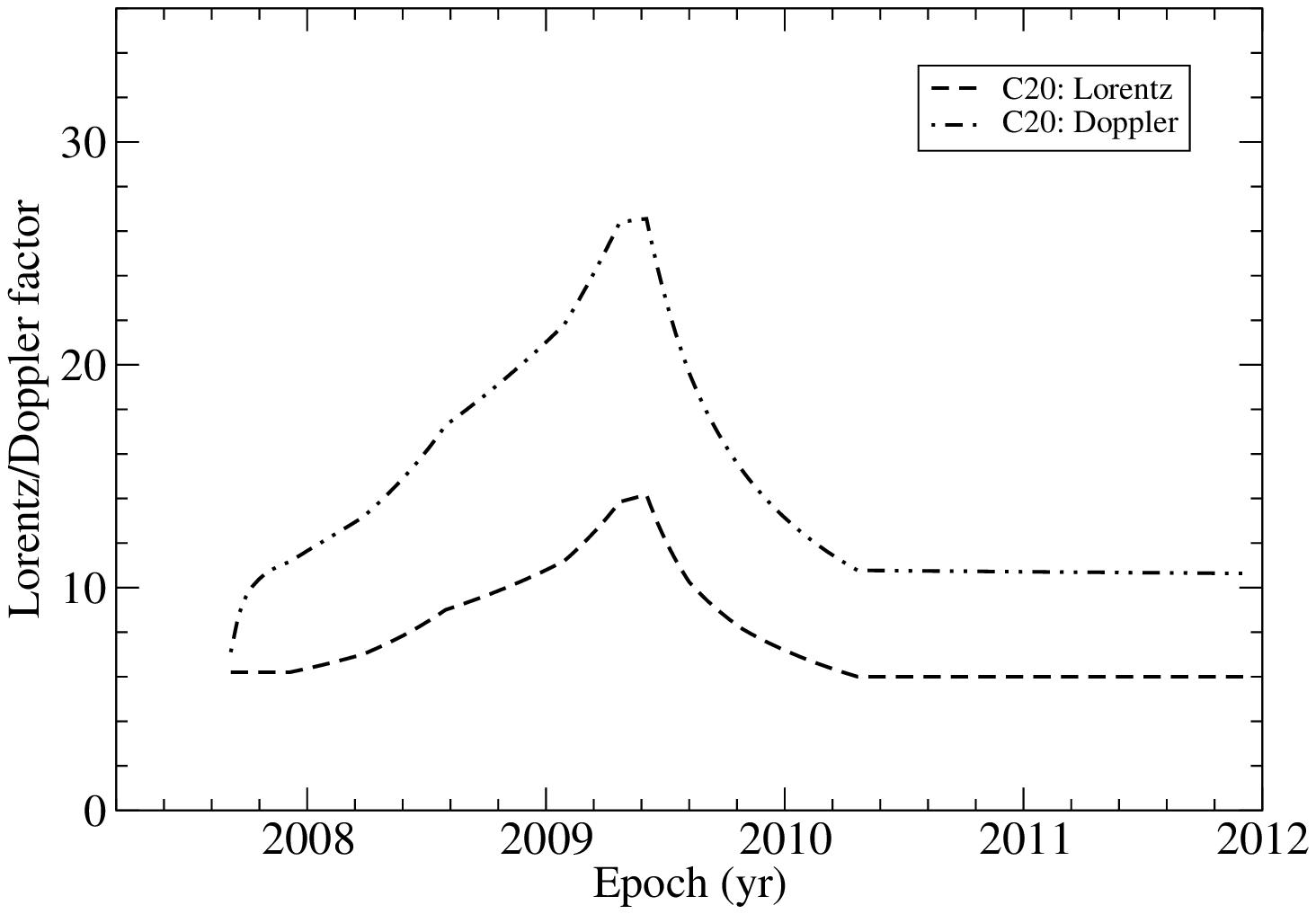}
     \includegraphics[width=5cm,angle=-90]{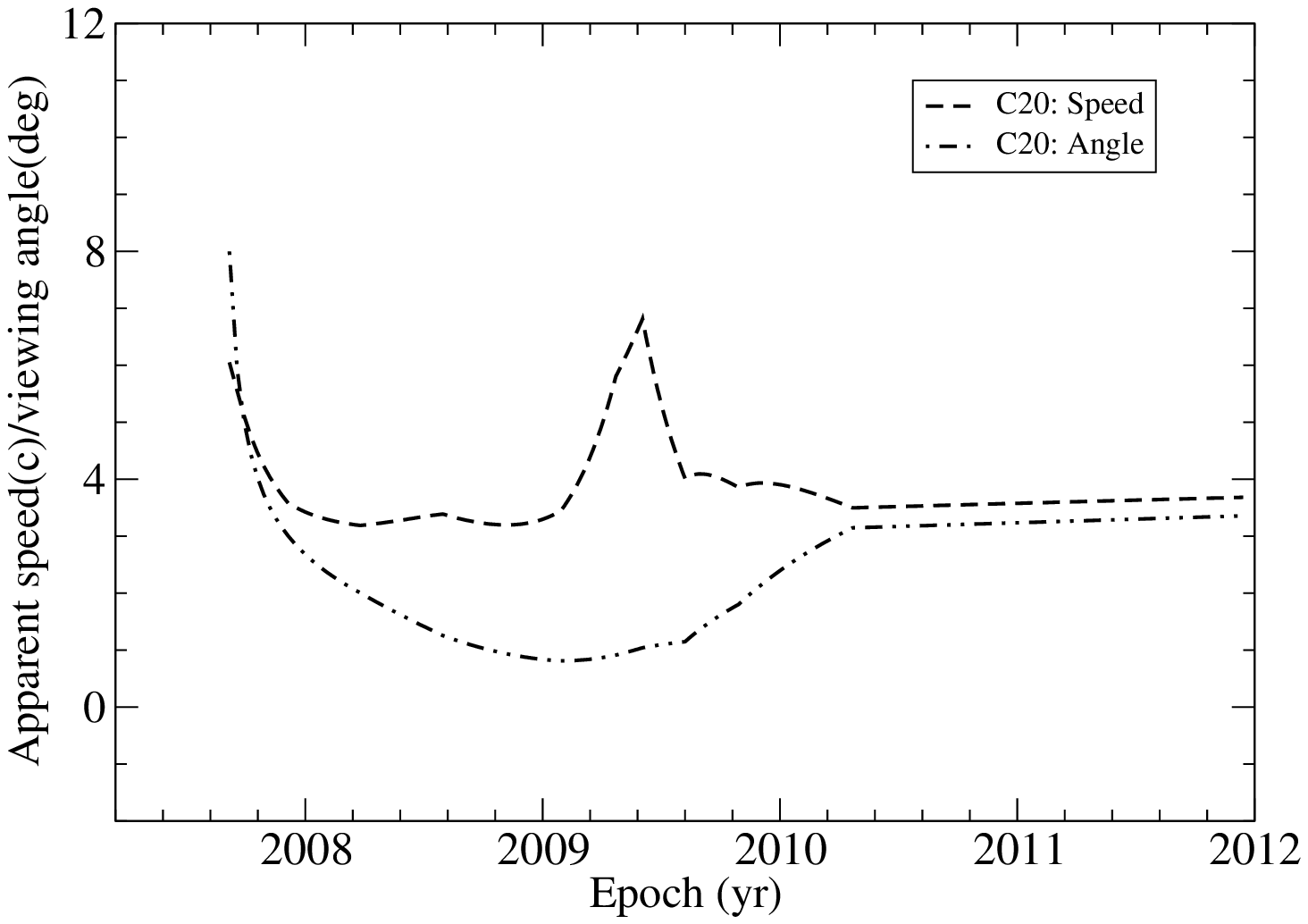}
     \caption{Knot C20. The model-derived bulk Lorentz factor $\Gamma(t)$
     and Doppler factor $\delta(t)$ show a broad peak structure 
      during $\sim$2008.4--2010.0 (left panel) associated with the increase
     in its core distance $r_n$ (Fig.20; right panel). 
     Correspondingly, the model-derived
       viewing angle $\theta(t)$ changes along a concave curve (right panel). 
     The model-derived apparent speed $\beta_{app}(t)$ only shows a 
     short-period peak during $\sim$2009.0--2009.6.}
     \end{figure*}
     \begin{figure*}
     \centering
     \includegraphics[width=6cm,angle=-90]{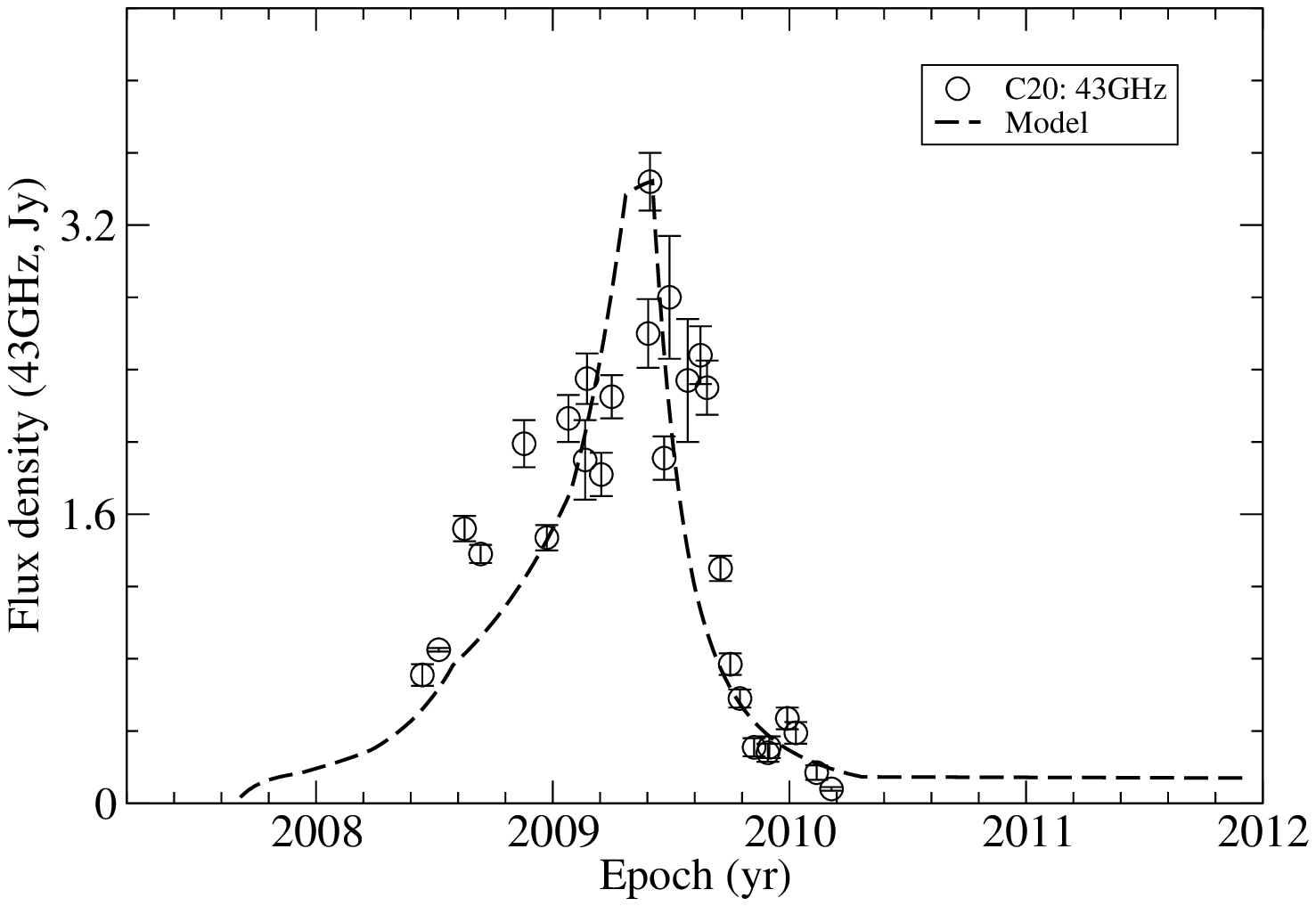}
     \caption{Knot C20. Its 43GHz light curve is well fitted by the 
     model-derived Doppler boosting profile ${S_{int}}[\delta(t)]^{3+\alpha}$
     with $S_{int}$=35.7$\mu$Jy and $\alpha$=0.50. Intrinsic variations are
     prominently superposed. }
     \end{figure*}
    \section{Interpretation of kinematics and flux evolution for knot C21}
    The model-fitting results of kinematics and flux evolution for knot C21
    are shown in Figures 23--26. Its precession phase $\phi_0$=3.52+8$\pi$
    corresponding to its ejection epoch $t_0$=2007.32 (Qian \cite{Qi22b}).
    \subsection{Knot C21: model-fitting of the kinematics}
    The model-derived parameters $\epsilon(t)$ and $\psi(t)$ defining the 
    jet-axis are shown in Figure 23 (left panel) and  indicate that before 
    2008.88 [corresponding core distance $r_n$=0.11\,mas, coordinate
    $X_n$=0.060\,mas, traveled distance Z=3.97\,mas=26.4\,pc (right panel)], 
    $\epsilon$=$2^{\circ}$ and $\psi$=$7.16^{\circ}$, knot C21 moved along 
    the precessing common helical trajectory pattern, while after 2008.88 
    $\psi$ started to change and it started to move along its own individual
    track. Its trajectory-transit occurred near the core and was not observed.\\
    Its whole trajectory $Z_n(X_n)$ was well fitted by the proposed model
    (Figure 24, left panel; the red line indicating the precessing common 
    trajectory section). The core distance $r_n$ and coordinates
    $X_n$ and $Z_n$ are well model-fitted (right panel).\\
    The model-derived bulk Lorentz factor $\Gamma(t)$ and Doppler fcator
    $\delta(t)$ are presented in Figure 25 (left panel), both having a
    double-peak structure during 2009.0--2010.4. The model-derived apparent
    speed $\beta_{app}(t)$ also has a double-peak structure, but the second
    peak is much more prominent.    
    \subsection{Knot C21. Doppler boosting effect and flux evolution}
    As shown in Figure 26 the doule-peak structure of its 43GHz light curve 
    is very well fitted by the Doppler boosting profile 
     ${S_{int}}[\delta(t)]^{3+\alpha}$ with $S_{int}$=25.6$\mu$Jy
     and $\alpha$=0.50. 
    \begin{figure*}
    \centering
    \includegraphics[width=5cm,angle=-90]{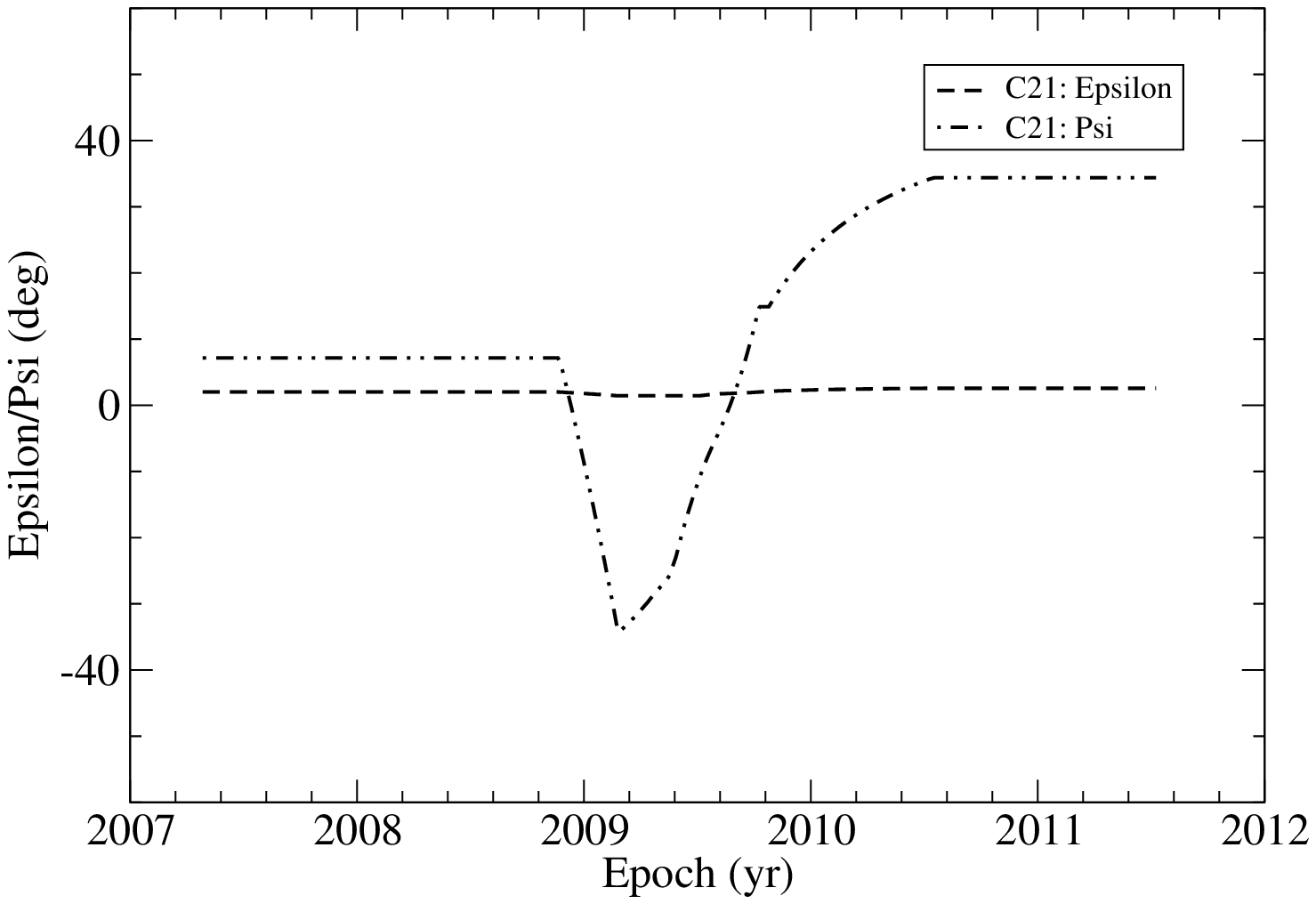}
    \includegraphics[width=5cm,angle=-90]{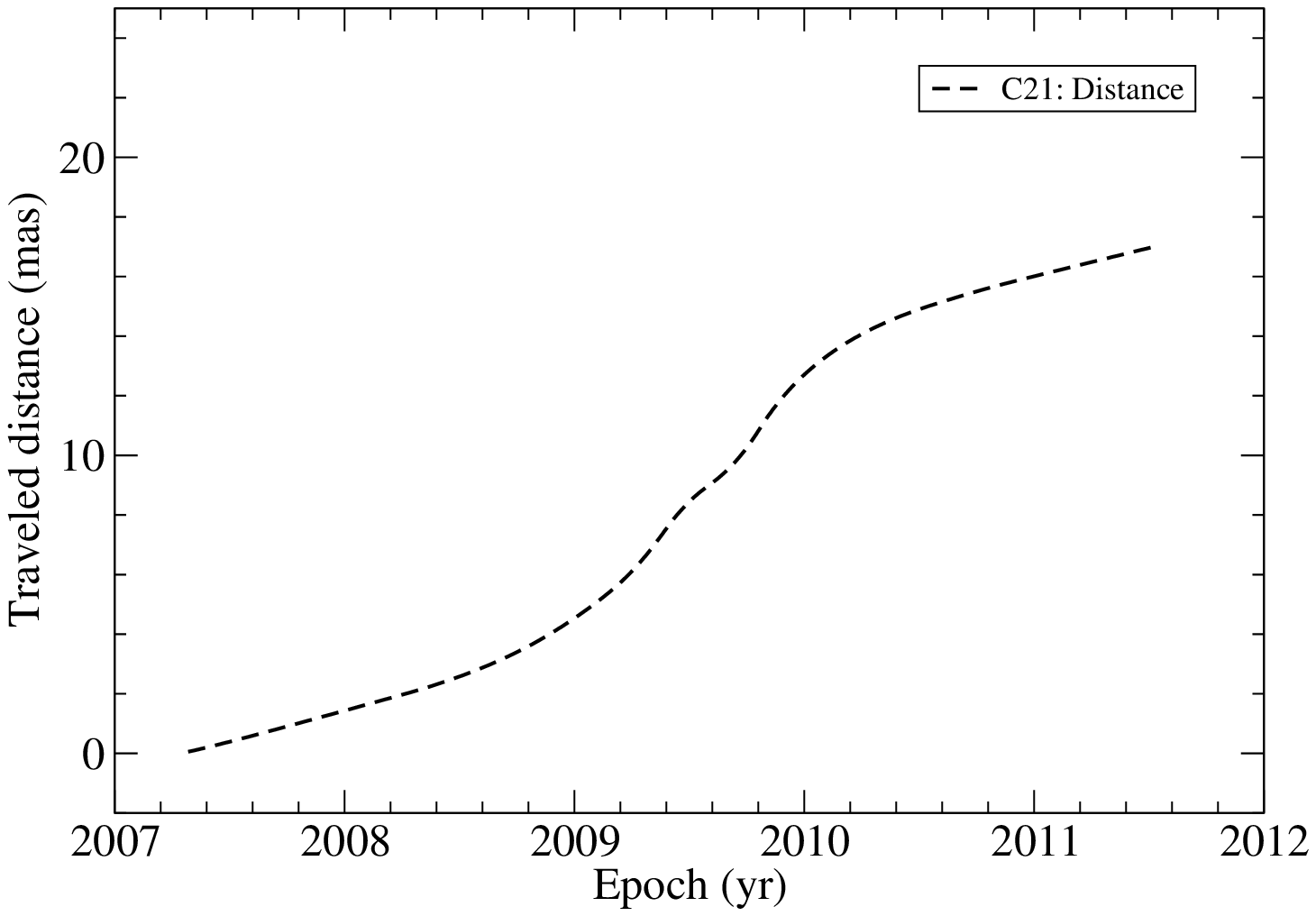}
    \caption{Knot C21. The model-derived curves of parameters $\epsilon$
    and $\psi$ (left panel) indicate that 
    $\epsilon$=$2^{\circ}$ and $\psi$=$7.16^{\circ}$ before 2008.88
    [corresponding 
    core distance $r_n$=0.11\,mas, coordinate $X_n$=0.060\,mas, traveled
     distance Z=3.97\,mas=26.4\,pc], knot C21 moved along the
    precessing common helical trajectory pattern, while after 2008.88 $\psi$
    decreases and it started to move along its own individual track.}
    \end{figure*}
    \begin{figure*}
    \centering
    \includegraphics[width=5cm,angle=-90]{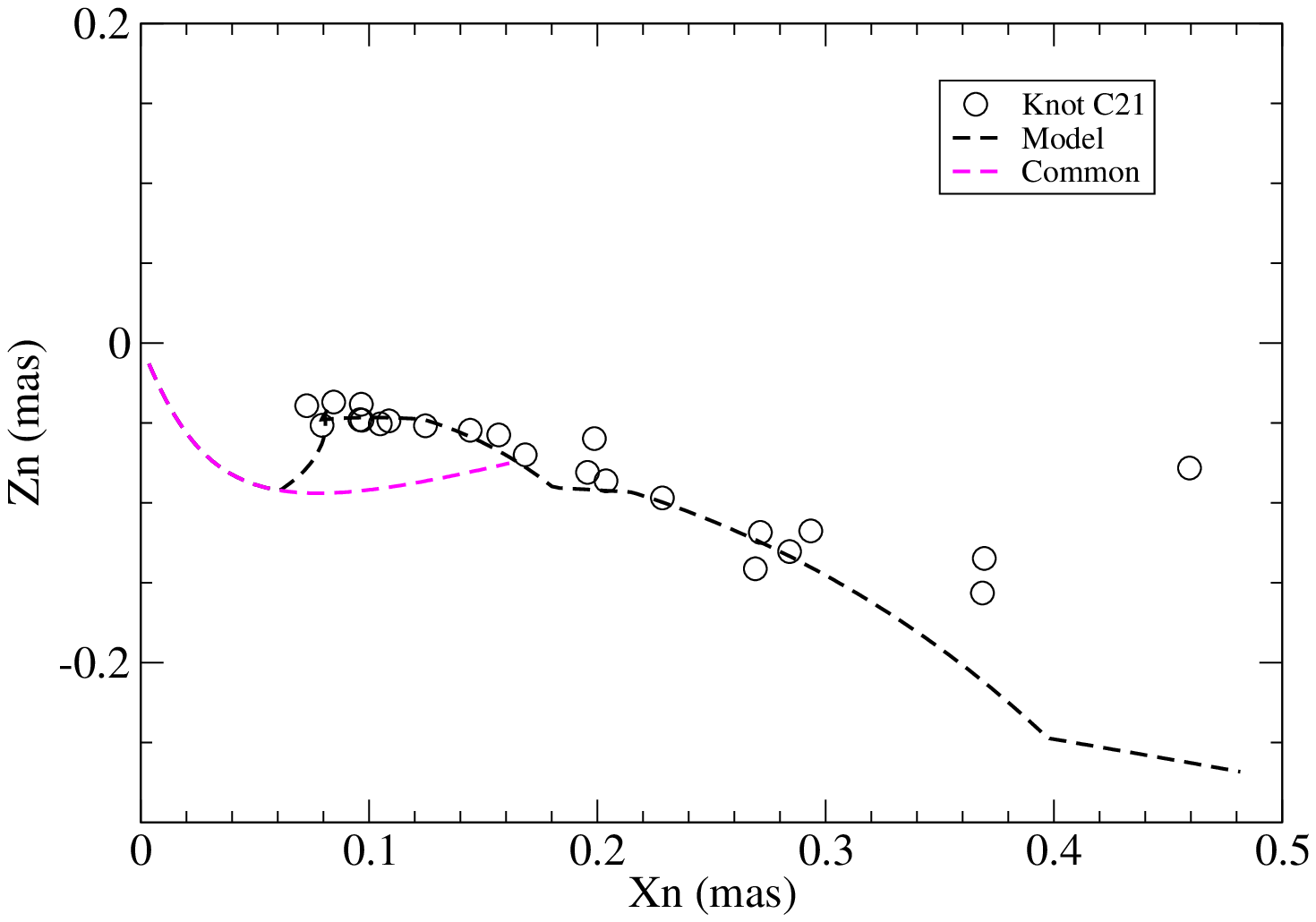}
    \includegraphics[width=5cm,angle=-90]{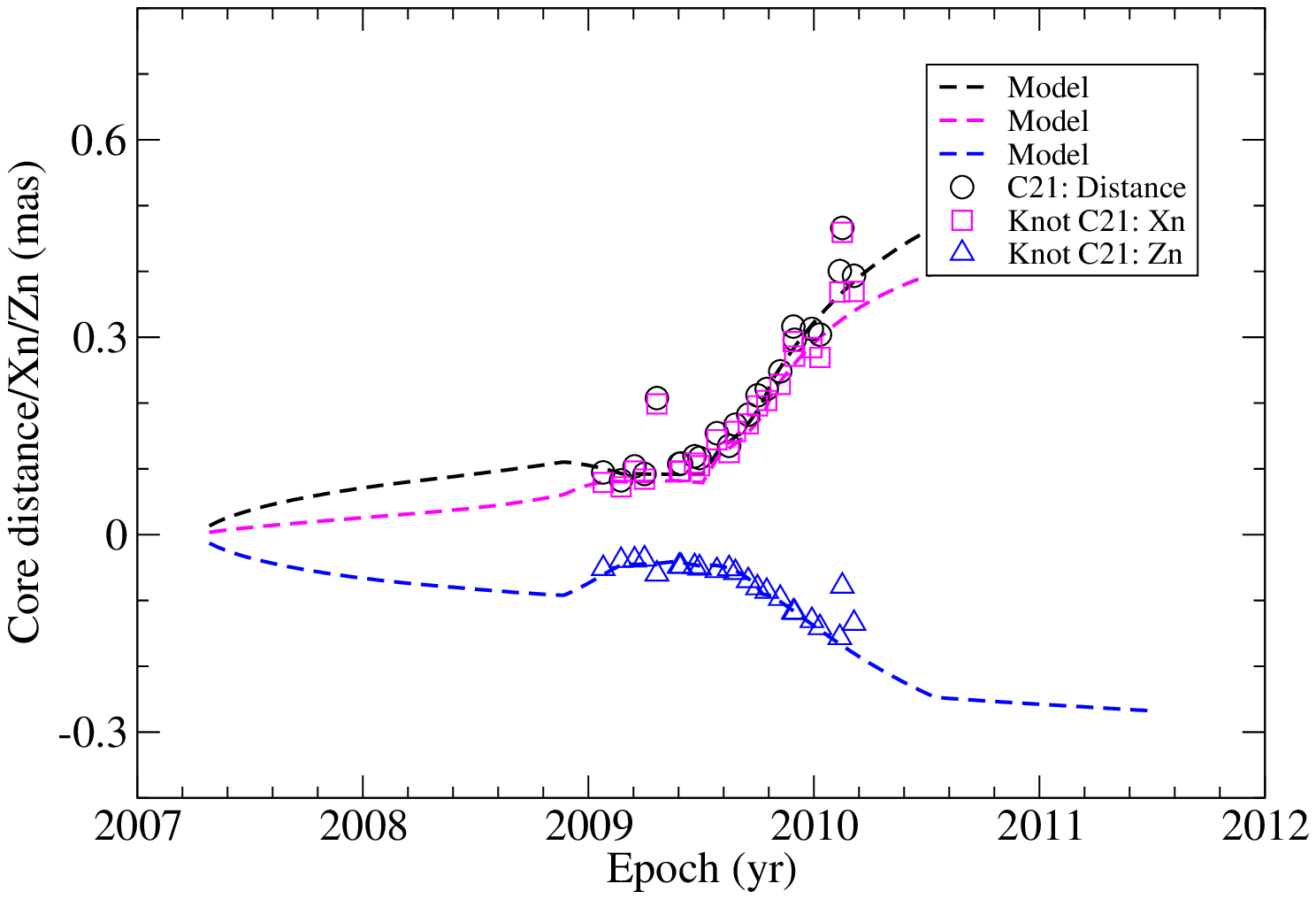}
    \caption{Knot C21. Its whole trajectory $Z_n(X_n)$ (left panel), 
     core distance $r_n$ and coordinates $X_n$ and $Z_n$ (right panel) 
     are well fitted by the proposed model. The red line indicates the 
     precessing common trajectory, showing the modeled trajectory-transit at
      $X_n$=0.06\,mas. }
    \end{figure*}
    \begin{figure*}
    \centering
    \includegraphics[width=5cm,angle=-90]{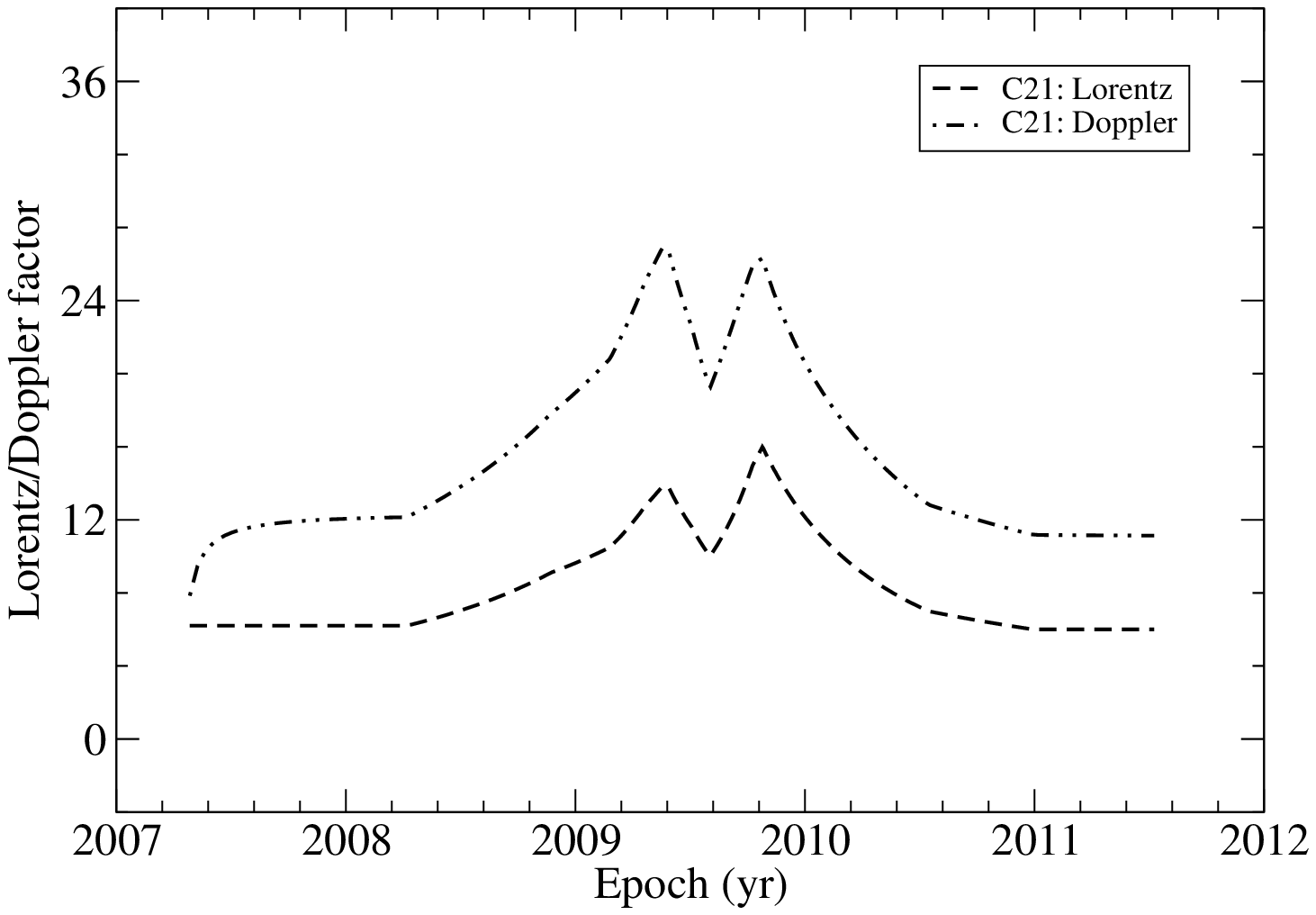}
    \includegraphics[width=5cm,angle=-90]{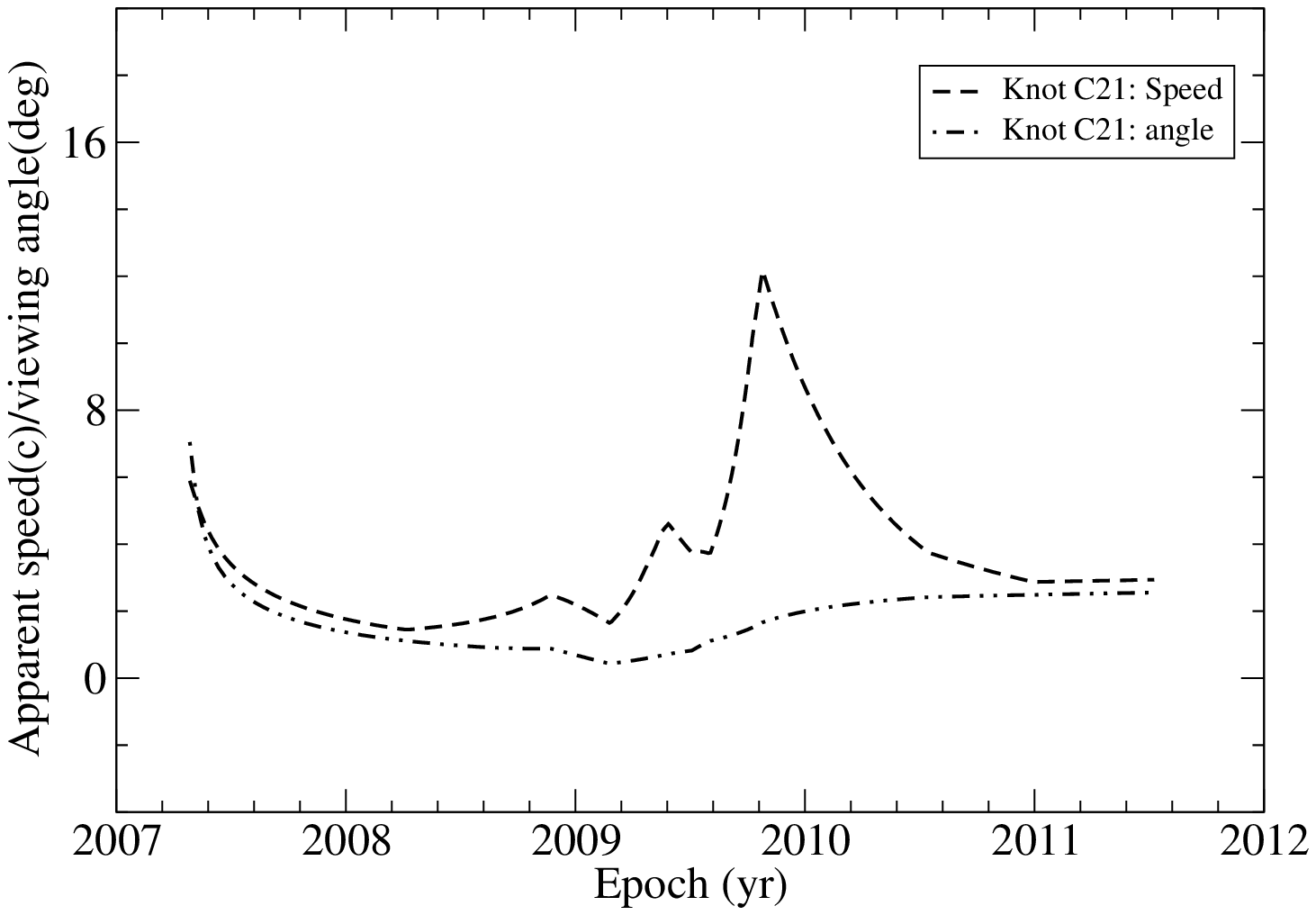}
    \caption{Knot C21. The model-derived bulk Lorentz factor $\Gamma(t)$
    and Doppler factor $\delta(t)$ having a double-peak structure during
     2009.0--2010.4 (left panel). The model-derived apparent speed 
    $\beta_{app}(t)$ also shows a double-peak structure, but the second one
    is much  more prominent. The viewing angle $\theta(t)$ varies along a
     concave curve.}
    \end{figure*}
    \begin{figure*}
    \centering
    \includegraphics[width=6.5cm,angle=-90]{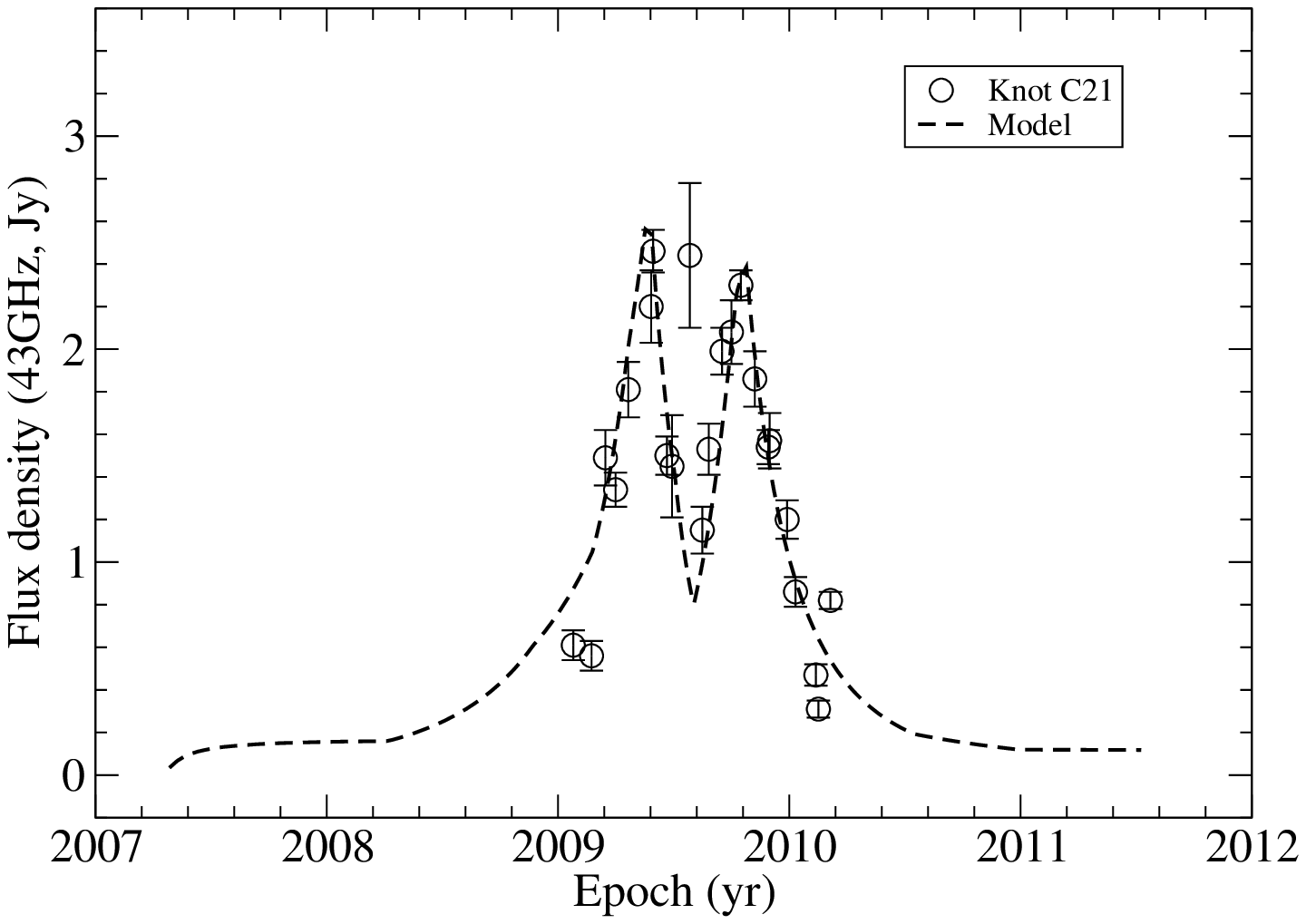}
    \caption{Knot C21. The 43\,GHz light curve is very well fitted by the
     model-derived Doppler boosting profile ${S_{int}}[\delta(t)]^{3+\alpha}$
      with $S_{int}$=25.6$\mu$Jy and $\alpha$=0.5.}
    \end{figure*}
    \section{Interpretation of kinematics and flux evolkution for knot B5}
    The model-fitting results of the knematics and flux evolution for knot B5
    are presented in Figures 27--30. Its precession phase $\phi_0$=6.24+8$\pi$
    corresponding to its ejection epoch $t_0$=2010.48 (Table 3; 
    Qian \cite{Qi22b}).
    \subsection{Model fits to the kinematics of knot B5}
    The model-derived parameters $\epsilon(t)$ and $\psi(t)$ defining the
    jet-axis are shown in Figure 27 (left panel), which indicates that its 
    trajectory-transit occurred at 2010.66. Before 2010.66 [corresponding
    core distance $r_n$=0.039\,mas, coordinate $X_n$=0.028\,mas, traveled
    distance Z=0.36\,mas=2.39\,pc (right panel)],
     $\epsilon$=$2^{\circ}$ and $\psi$=$7.16^{\circ}$, knot B5 moved along the
     precessing common helical trajectory, while after 2010.66 $\psi$ started
      to increase and it started to move along its own individual track.\\
     Its whole trajectory is well fitted by the proposed model as shown in 
     Figure 28 (left panel; the red-line indicating the precessing common
     trajectory with its trajectory-transit at $X_n$=0.039\,mas). The
     core distance $r_n$, coordinates $X_n$ and $Z_n$ are also very well
     model-fitted (right panel).\\
     The model-derived bulk Lorentz factor $\Gamma(t)$ and Doppler factor
     $\delta(t)$ are shown in Figure 29 (left panel), revealing  a double-peak
     structure. Correspondingly, the model-derived apparent speed 
    $\beta_{app}(t)$ (right panel) also has a similar double-peak structure.
     Its viewing angle varied along a concave curve (during the first peak)
     and then along a convex curve (during the second peak).
    \subsection{Knot B5: Doppler boosting effect and  flux evolution}
    Its 43\,GHz light curve is well fitted by the model-derived Doppler 
    boosting profile ${S_{int}}[\delta(t)]^{3+\alpha}$ with 
     $S_{int}$=37.0$\mu$Jy and $\alpha$=0.5, as shown in Figure 30. 
     Due to the increase in the viewing angle (along a convex
     curve) during 2012--2014.5 its second "flare" only shows low flux 
     variations. 
    \begin{figure*}
    \centering
    \includegraphics[width=5cm,angle=-90]{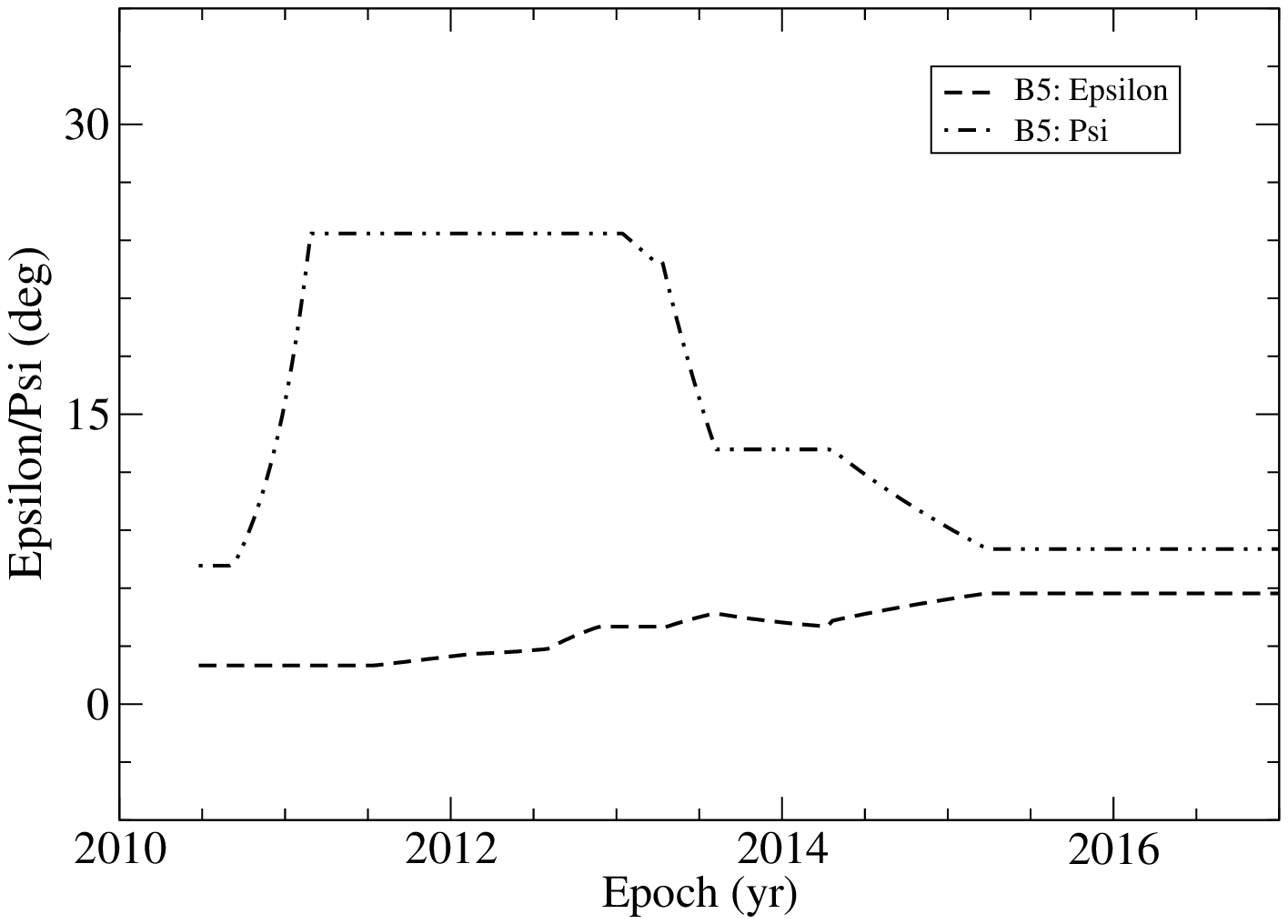}
    \includegraphics[width=5cm,angle=-90]{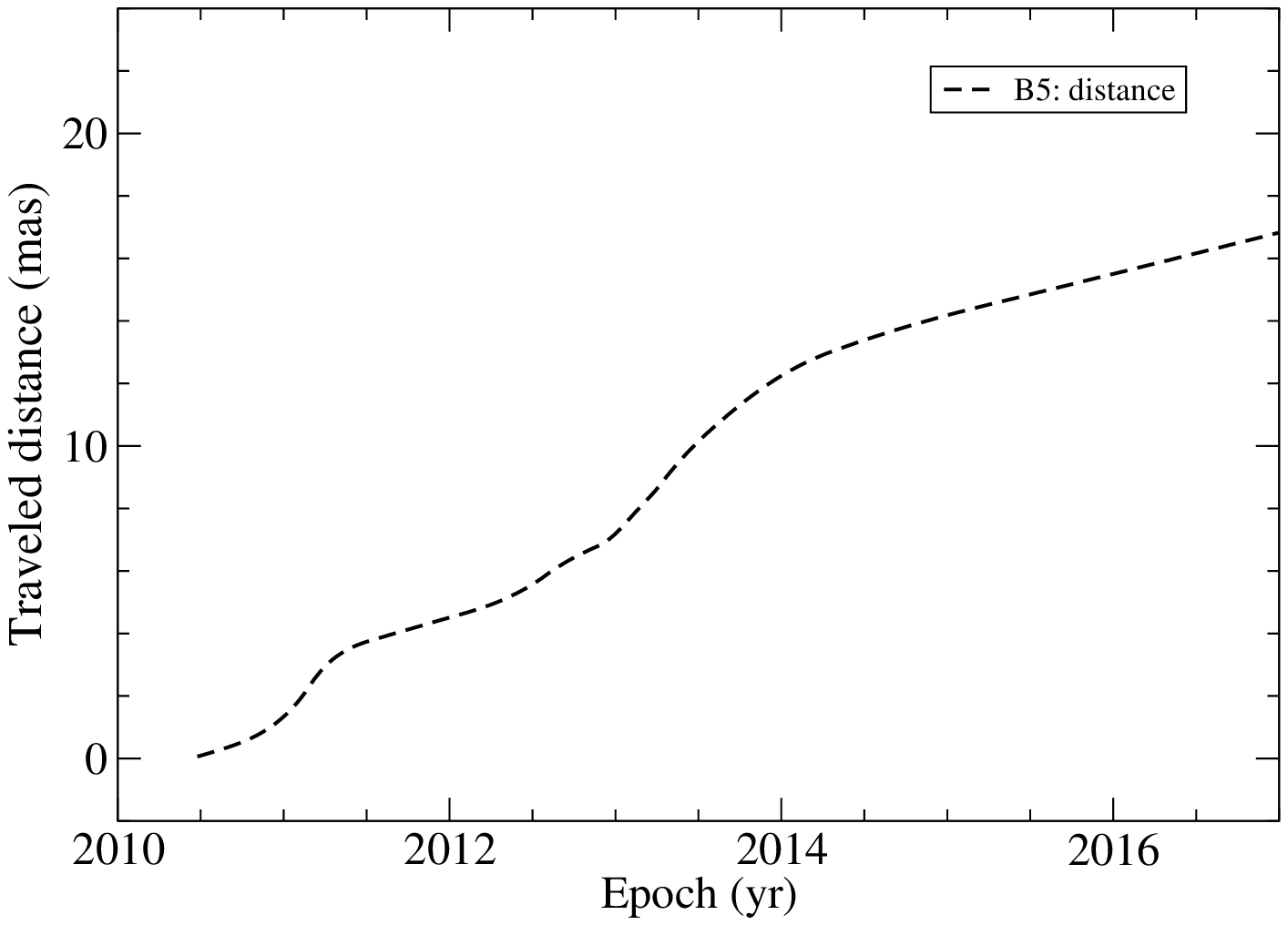}
    \caption{Knot B5. The model-derived parameters $\epsilon(t)$ and $\psi(t)$,
     indicating the trajectory-transit occurred at 2010.66, when the 
    corresponding core distance $r_n$=0.039\,mas, coordinate $X_n$=0.028\,mas,
    and the traveled distance Z=0.36\,mas=2.39\,pc (right panel). So,
    knot B5 moved along the precessing common helical trajectory pattern before
    2010.66, while after 2010.66 it started to  move along its own individual
    track.}
    \end{figure*}
    \begin{figure*}
    \centering
    \includegraphics[width=5cm,angle=-90]{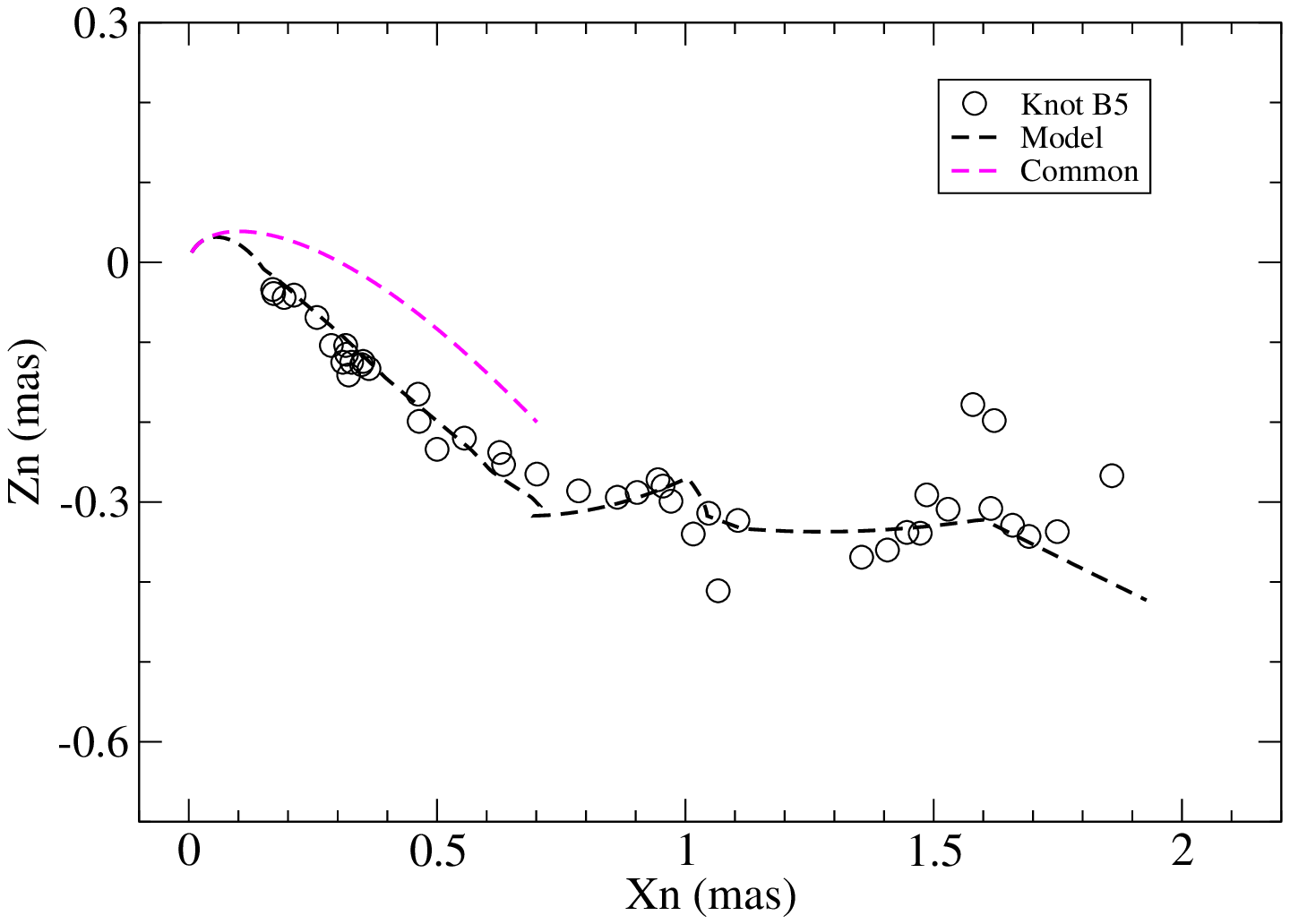}
    \includegraphics[width=5cm,angle=-90]{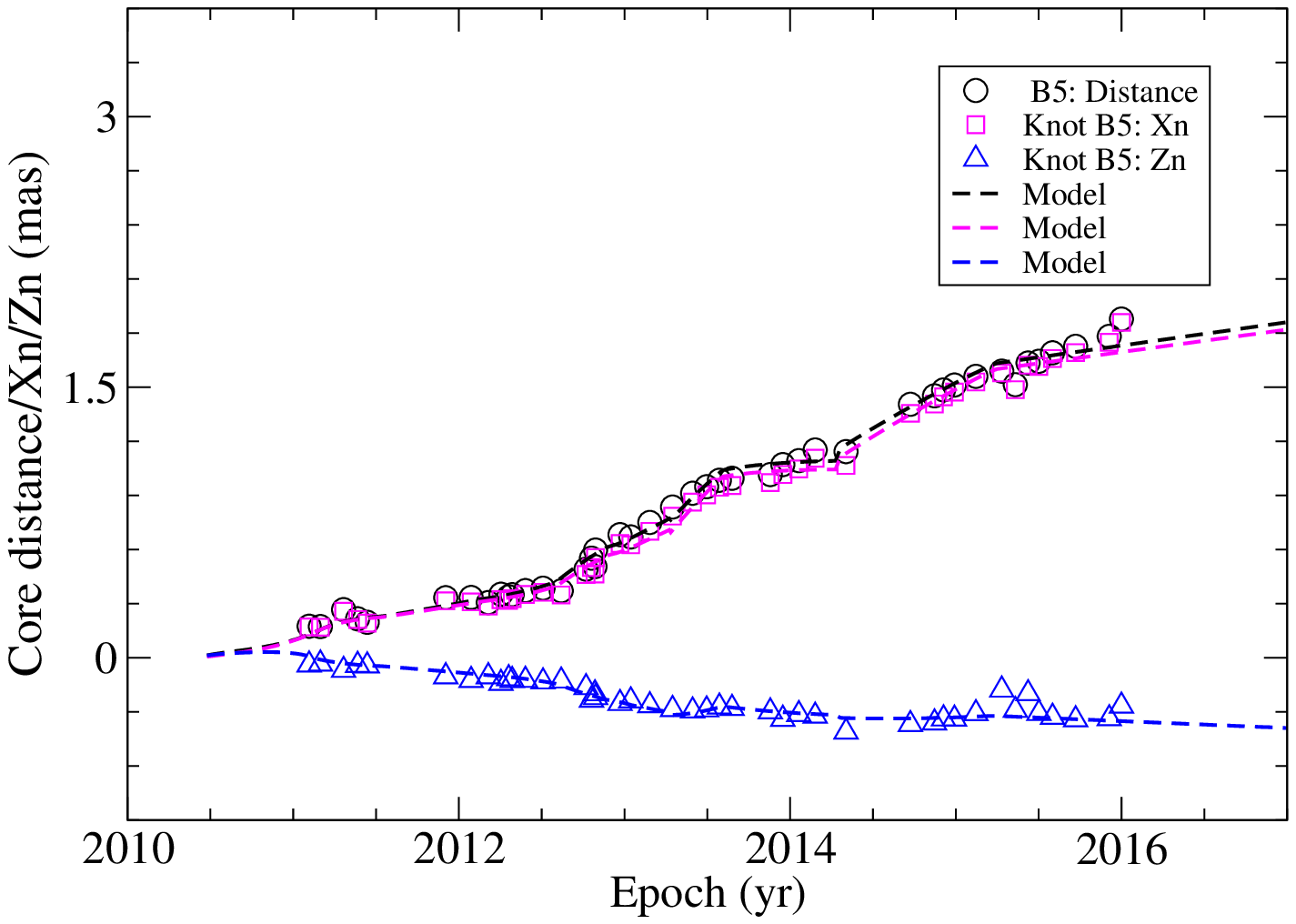}
    \caption{Knot B5. Its whole trajectory $Z_n(X_n)$ is well fitted by 
     the proposed model (black line, left panel; the red-line indicating
      the precessing 
     common trajectory and the trajectory transit at $X_n$=0.028\,mas).
     Its core distance $r_n$, coordinates $X_n$ and $Z_n$ are all well
     model-fitted (right pane).}
    \end{figure*}
    \begin{figure*}
    \centering
    \includegraphics[width=5cm,angle=-90]{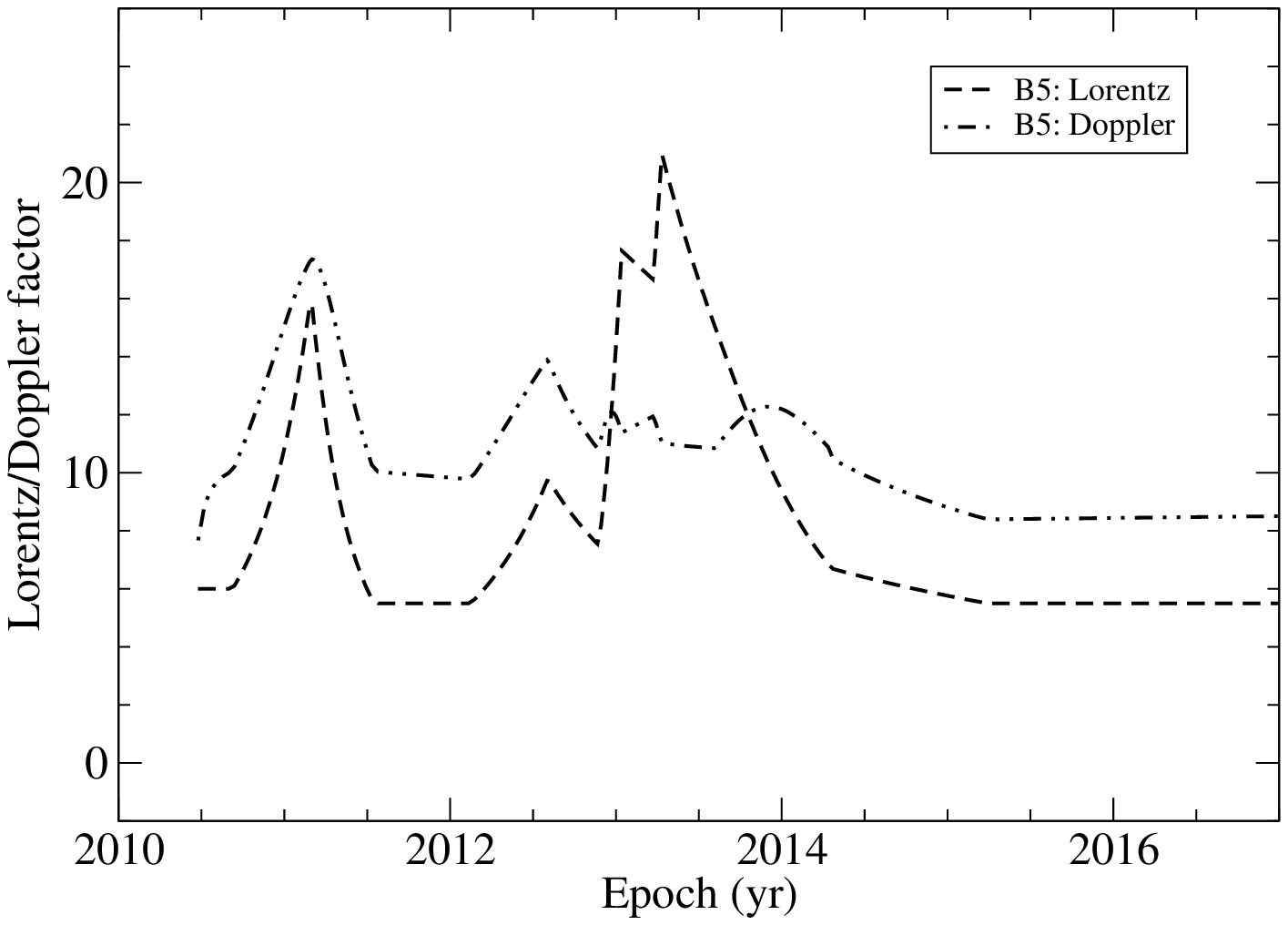}
    \includegraphics[width=5cm,angle=-90]{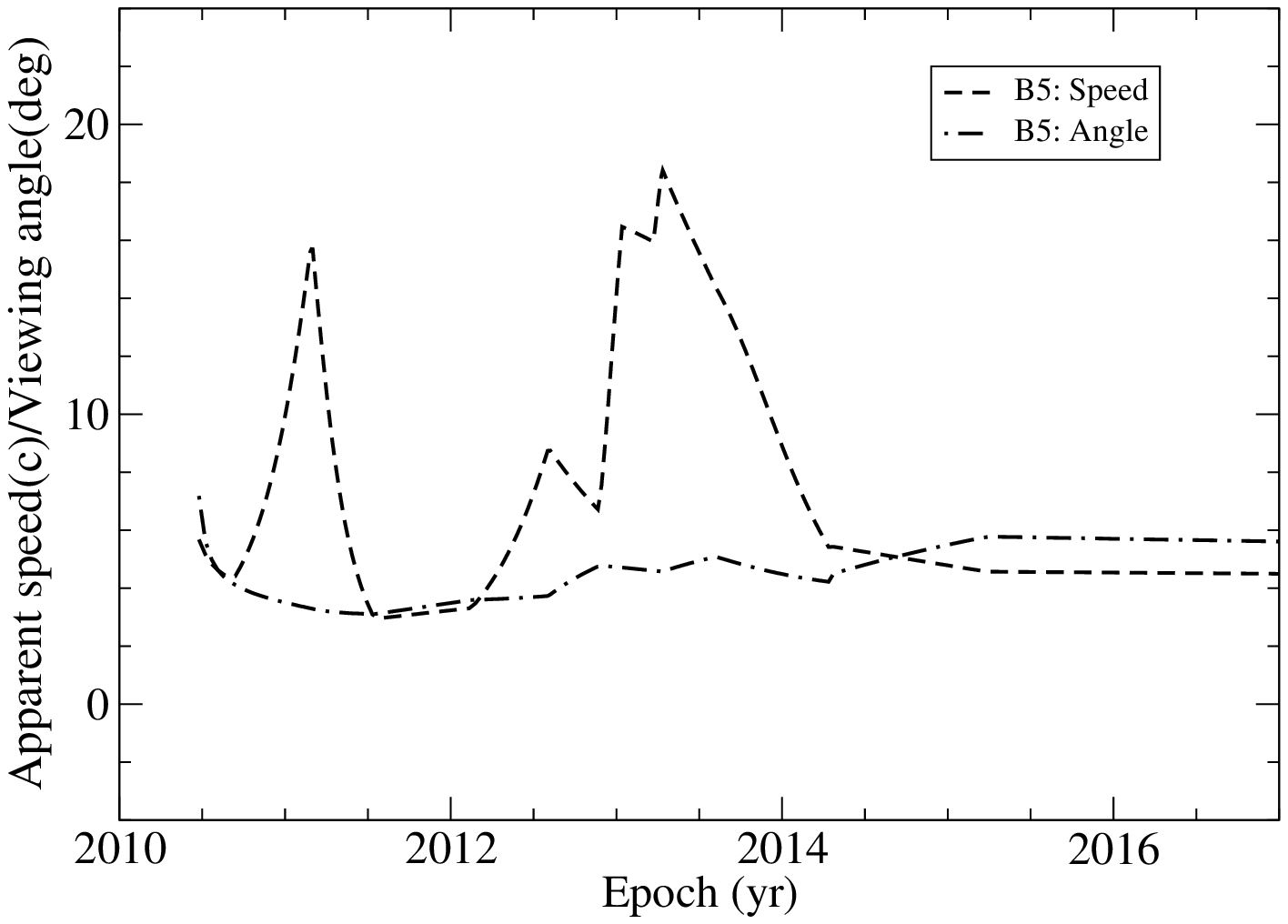}
    \caption{Knot B5. The model-derived bulk Lorentz factor $\Gamma(t)$ and 
     Doppler factor $\delta(t)$ are presented in left panel, both revealing 
     a double-peak structure, but the second peak in Doppler factor is much
     lower due to the increase in the viewing angle. Correspondingly, the 
     model-derived apparent speed has a similar double structure (right panel).
     The viewing angle varies along a convex curve, leading to the second
     peak of the apparent speed much broader than the first one.}
    \end{figure*}
    \begin{figure*}
    \centering
    \includegraphics[width=6.5cm,angle=-90]{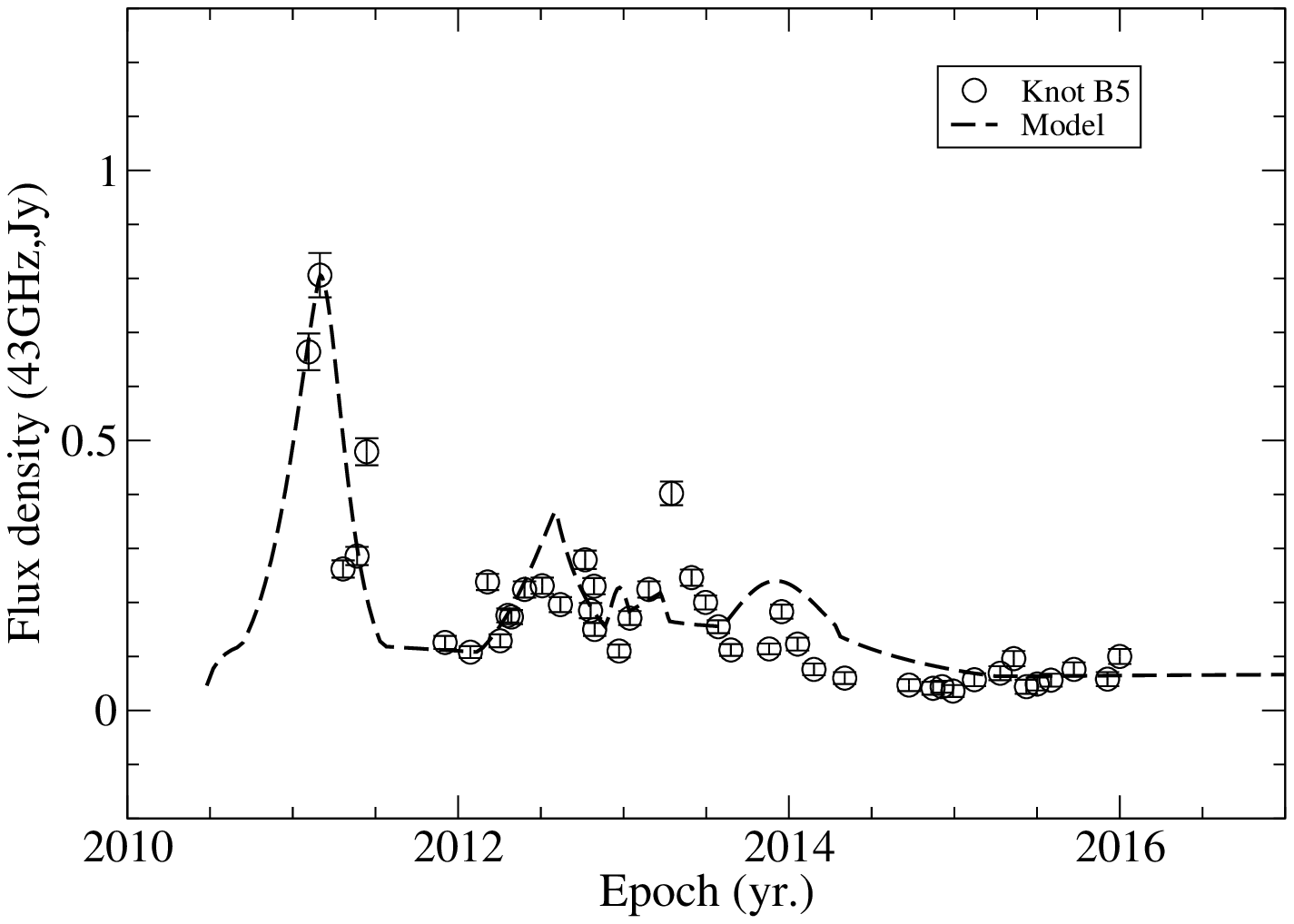}
    \caption{Knot B5. Doppler boosting effect and  Flux evolution.
     Corresponding to the double-peak structure, the model-derived 
    Doppler boosting profile ${S_{int}}[\delta(t)]^{3+\alpha}$ also has
    a similar double-peak structure and well fits the observed 43\,GHz
    light curve. The second "Doppler-boosted flare" is much weaker than the
    first one due to the increase in the viewing angle and the decrease in 
    Doppler factor during 2012.0--2014.5 (Fig.29). $S_{int}$=37.0$\mu$Jy and 
    $\alpha$=0.50.}
    \end{figure*}
    \begin{table*}
    \centering 
    \caption{Parameters of the precessing common trajectory for five 
    superluminal knots (C8, C20, C21, B5 and B7): precession phase
    $\phi_0$(rad), ejection epoch $t_0$, extension of the precessing common 
   trajectory from the core $r_{n,com}$ (in mas), the corresponding traveled 
   distance $Z_{com,mas}$ (in mas) and $Z_{com,pc}$ (in pc).}
   \begin{flushleft}
   \centering
   \begin{tabular}{llllll}
   \hline
   Knot & $\phi_0$ & $t_0$ & $r_{n,com}$ & $Z_{com,mas}$ & $Z_{com,pc}$\\
   \hline
   C8 & 2.13+4$\pi$ & 1991.10 & 0.14 &   6.0 & 39.9  \\
   C20 & 3.83+8$\pi$  & 2007.68 & 0.076    & 1.20     & 7.98      \\
   C21 & 3.52+8$\pi$   & 2007.32  & 0.11     & 3.97      & 26.4       \\
   B5 &  6.24+8$\pi$    &  2010.48  & 0.039    & 0.36     & 2.39        \\
   B7 &  0.92+10$\pi$  &   2011.60  & 0.14    & 3.00      & 20.0        \\
   \hline
   \end{tabular}
   \end{flushleft}
   \end{table*}
    \section{Interpretation of kinematics and flux evolution for knot B7}
    The model-fitting results of the kinematics and flux evolution for knot B7
     are shown in Figures 31--34. Its precession phase 
     $\phi_0$=0.92\,rad+10$\pi$,
     corresponding to its ejection epoch $t_0$=2011.60.
    \subsection{Model fits to the kinematics for knot B7.}
    The model-derived parameters $\epsilon (t)$ and $\psi(t)$ defining the
    jet axis are shown in Figure 31 (left panel), indicating that before 
    2012.99 $\epsilon$=$2^{\circ}$ and $\psi$=$7.16^{\circ}$, and knot B7 moved
    along the precessing common helical trajectory pattern [corresponding to 
    core distance $r_n{\leq}$0.14\,mas, coordinate ${X_n}\leq$0.11\,mas, 
    traveled distance Z${\leq}$3.0\,mas=20.0\,pc (right panel)].\\
      Its whole trajectory is well fitted by the proposed model as shown 
    in Figure 32 (left panel), where the red-line indicates the precessing 
    common trajectory and the trajectory-transit at $X_n$=0.11\,mas.
      Its core distance $r_n$, coordinates $X_n$ and $Z_n$ are very well
     model-fitted as shown in the right panel.\\
     The model-derived  bulk Lorentz factor $\Gamma(t)$ and Doppler factor 
     $\delta(t)$ as  continuous functions of time are shown in Figure 33
     (left panel), revealing a single peak structure duirng 2012.5--2014.0.
      Correspondingly, the model-derived apparent speed $\beta_{app}(t)$
      has a similar single peak structure is presented in the right panel.
      The model-derived viewing angle $\theta(t)$ is almost constant 
     ($\sim$$2.9^{\circ}$--$3.0^{\circ}$) and the peak structures in the
     Doppler factor and apparent speed are caused by the change in the
     Lorentz factor only, not due to the change in the viewing angle. 
      \subsection{Knot B7. Doppler boosting effect and flux evolution}
     As shown in Figure 34 its 43GHz light curve is very well fitted by 
    the Doppler boosting profile ${S_{int}}[\delta(t)]^{3+\alpha}$ 
    without any interference from its intrinsic
      flux variations. $S_{int}$=101.2$\mu$Jy and $\alpha$=0.50.
    \begin{figure*}
    \centering
    \includegraphics[width=5cm,angle=-90]{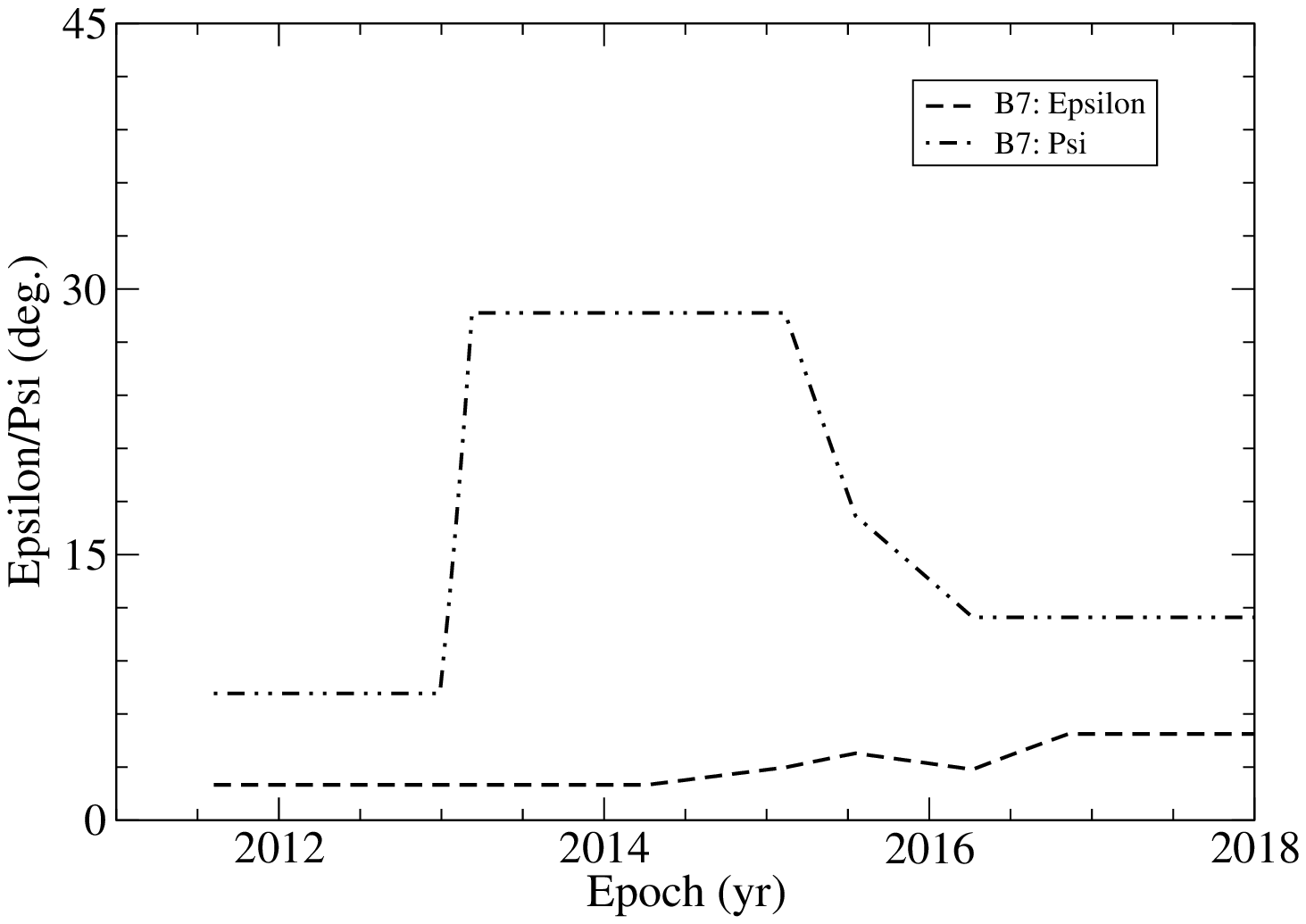}
    \includegraphics[width=5cm,angle=-90]{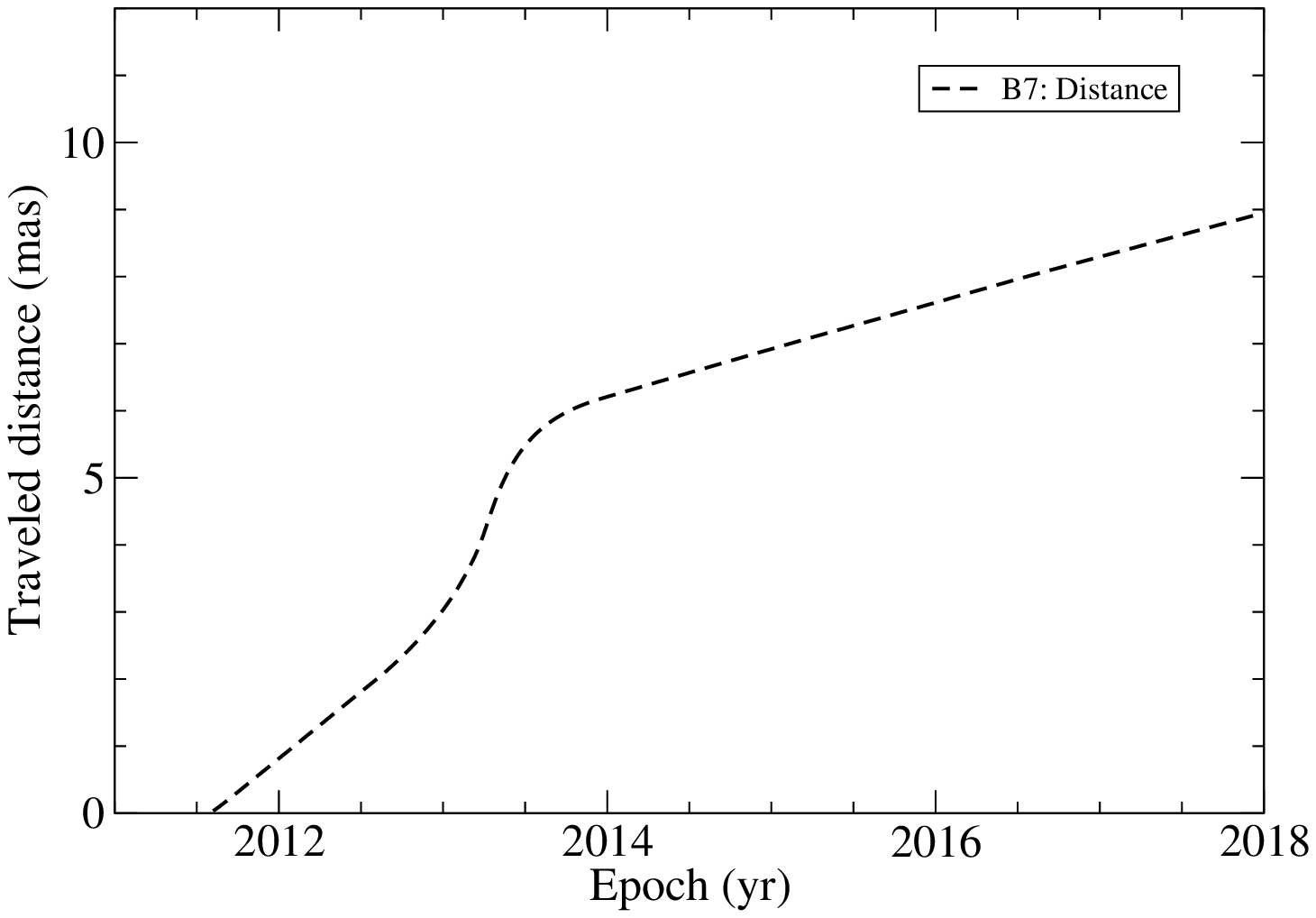}
    \caption{Knot B7. The model-derived curves for parameters $\epsilon(t)$
     and $\psi(t)$ defining the jet-axis (left panel), indicating the 
     trajectory transit occurred at 2012.99 [corresponding core distance
     $r_n$=0.14\,mas, coordinate $X_n$=0.11\,mas, traveled distance 
     Z=3.0\,mas=20.0\,pc (right panel)]. Thus before
     2012.99 knot B7 moved along the precessing common helical trajectory
     pattern, while after 2012.99 it moved along its own individual track.}
    \end{figure*}
    \begin{figure*}
    \centering
    \includegraphics[width=5cm,angle=-90]{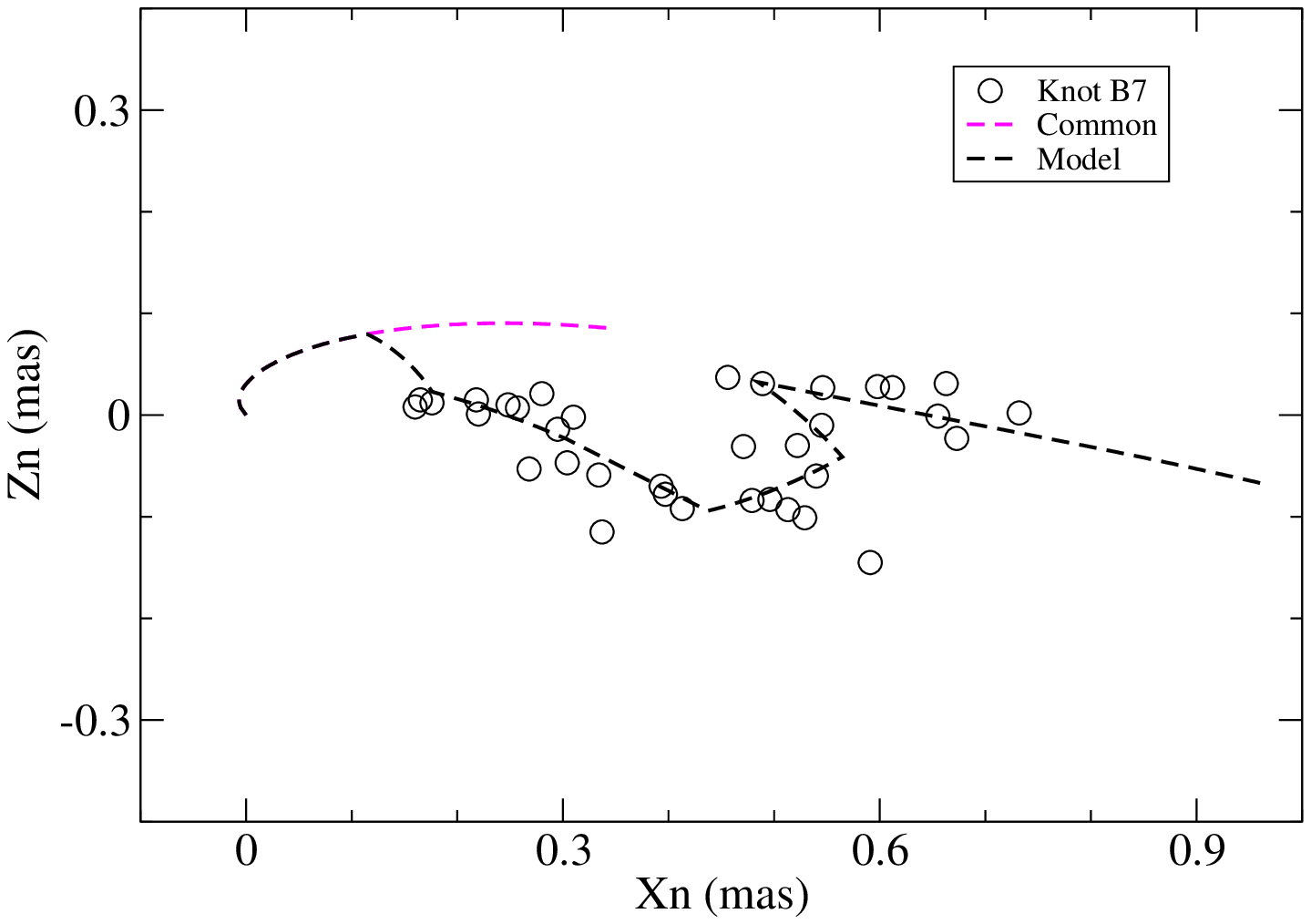}
    \includegraphics[width=5cm,angle=-90]{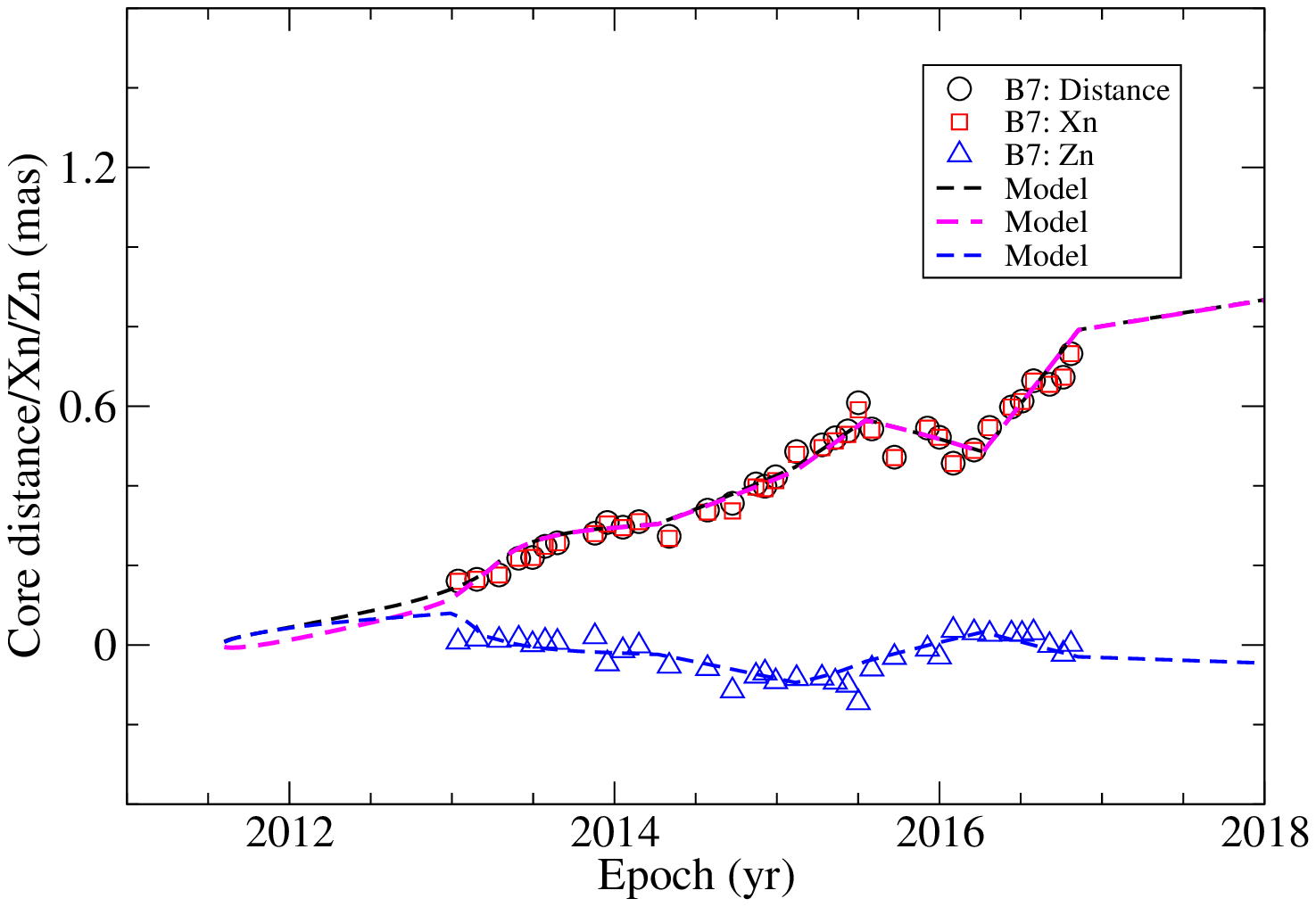}
    \caption{Knot B7. Its whole trajectory $Z_n(X_n)$ is well model-fitted as 
    shown in the left panel. The red-line indicates the precessing common 
    trajectory, demonstrating the trajectory-transit at $X_n$=0.11\,mas.
     Its core distance $r_n$, coordinates $X_n$ and $Z_n$ are very well 
    model-fitted (right panel). }
    \end{figure*}
    \begin{figure*}
    \centering
    \includegraphics[width=5cm,angle=-90]{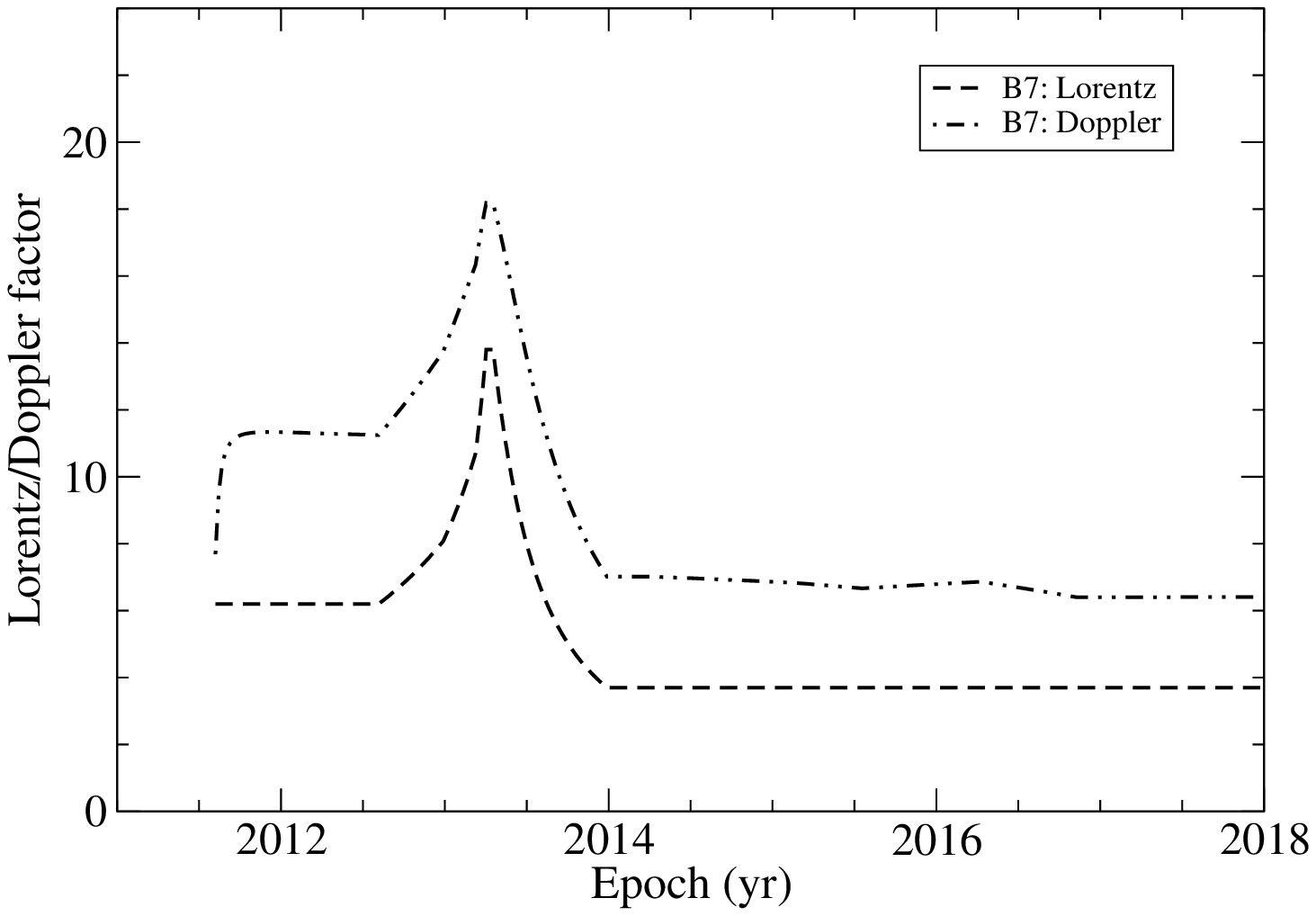}
    \includegraphics[width=5cm,angle=-90]{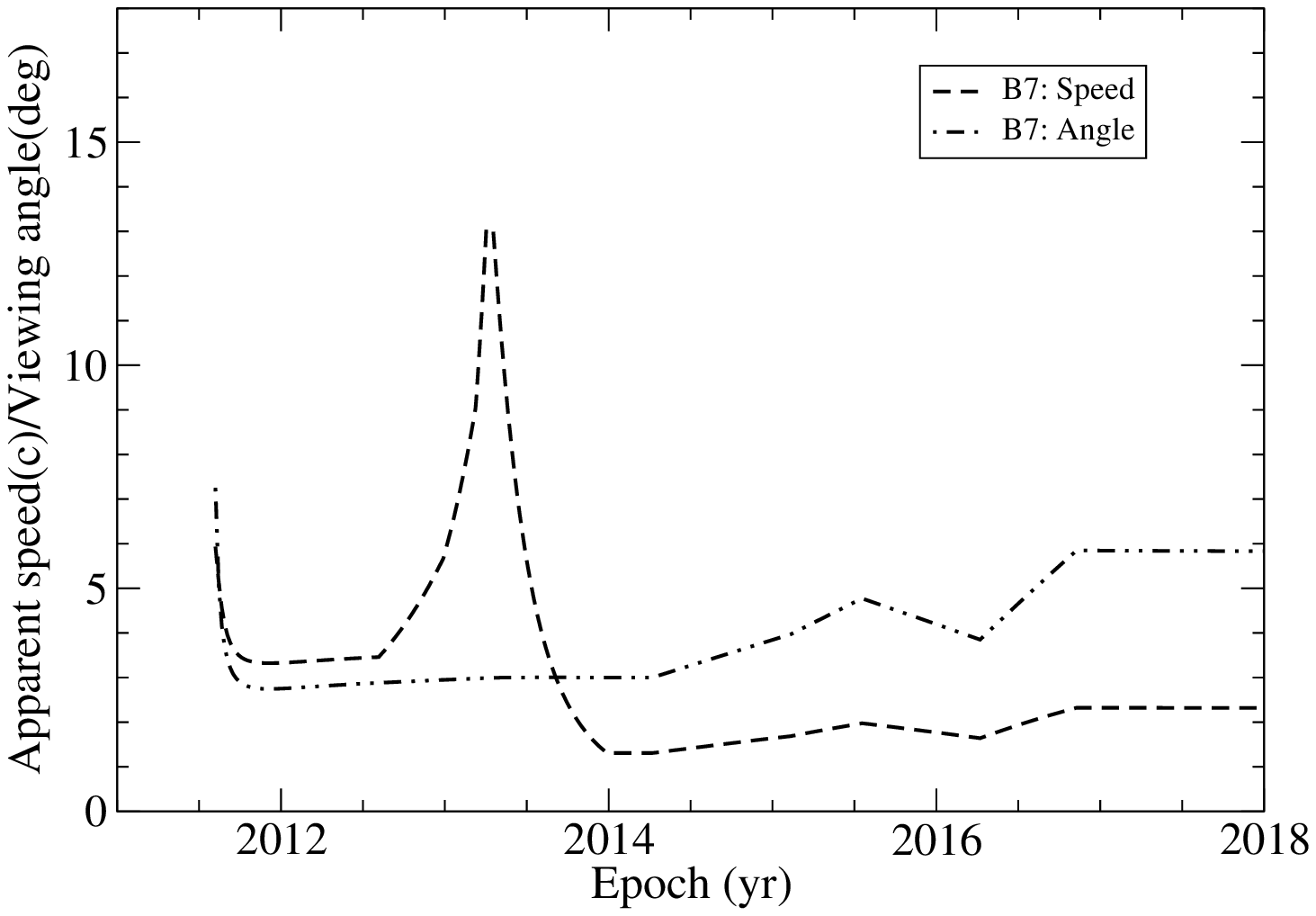}
    \caption{Knot B7. The model-derived bulk Lorentz factor $\Gamma(t)$ and 
    Doppler factor $\delta(t)$ are presented in the left panel, showing a 
    single-peak structure during 2012.5--2014.0. The model-derived apparent
    speed $\beta_{app}(t)$  also has a single-peak structure, while the
    model-derived viewing angle $\theta(t)$ changes only a little 
    ($\sim$$2.9^{\circ}$--$3.0^{\circ}$) during 2012.5--2014.0. Thus the
    single peak structure in the apparent speed and Doppler factor are 
     completely due to the acceleration and deceleration of knot B7 
     (increase and decrease in its bulk Lorentz factor).}
    \end{figure*}
    \begin{figure*}
    \centering
    \includegraphics[width=6.5cm,angle=-90]{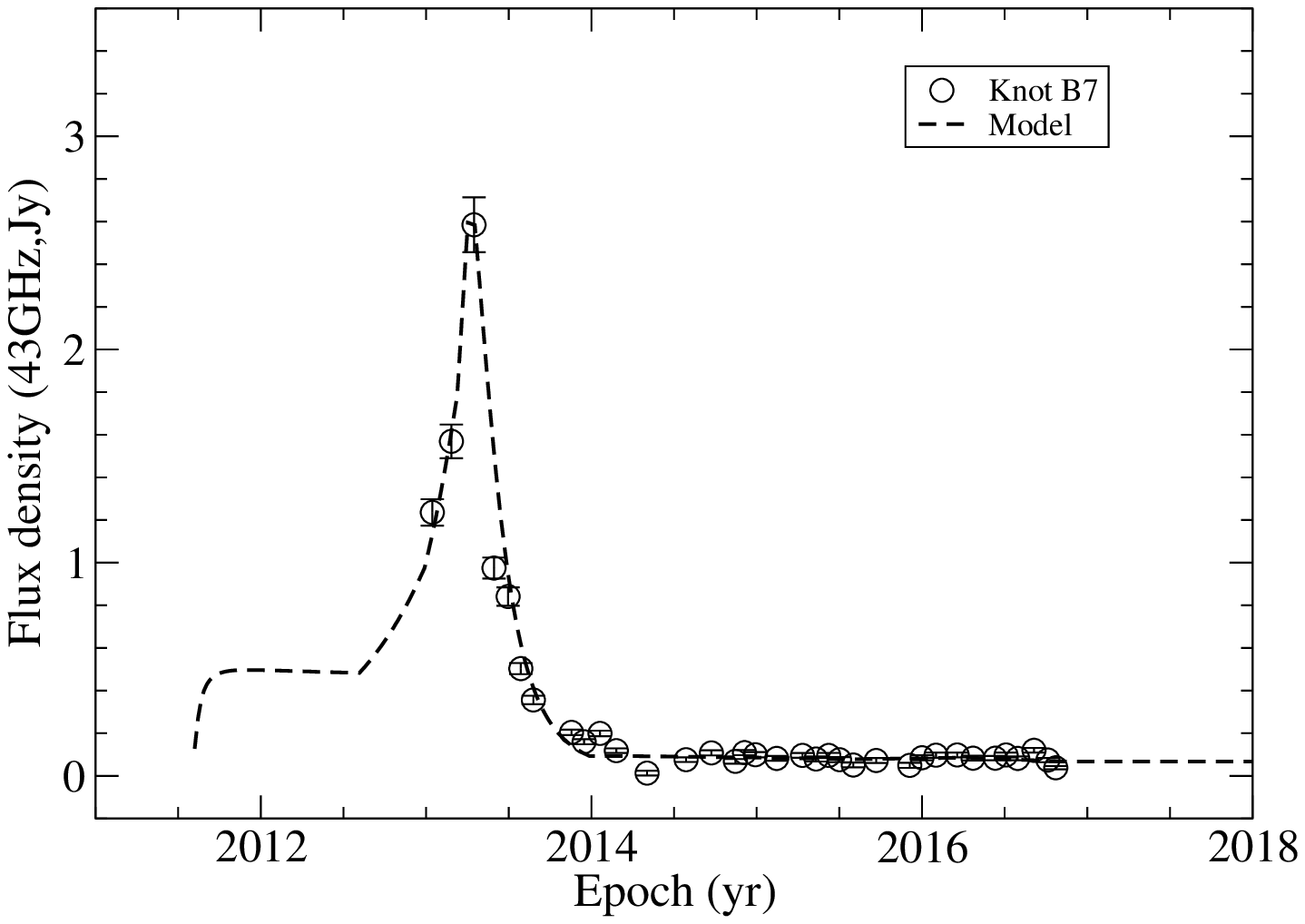}
    \caption{Knot B7. The 43\,GHz light curve is very well model-fitted by
    the Doppler boosting profile ${S_{int}}[\delta(t)]^{3+\alpha}$ with
     $S_{int}$=101.2$\mu$Jy and $\alpha$=0.50.}
    \end{figure*}
     \section{Discussion and Conclusions}
    In the previous work (Qian \cite{Qi22a}) we suggested that the superluminal
   components measured in 3C345 might be divided into two groups: group-A 
   (consisting of 13 knots: C4--C14 and C22-C23) and gorup-B (consisting of 
   14 knots: C15, C15a, C16--C21, B5--B8 and B11--B12). We interpreted the 
   kinemtics of these knots by assuming that the knots of group-A and group-B 
   are ejected from thier respective jets (jet-A and jet-B), which have 
   the same precession period of 7.3\,yr with the same sense. We also assumed 
   that jet-A and jet-B have their individual  precessing common helical 
   trajectory patterns along which the knots of group-A and group-B moved 
   respectively in their inner trajectory regions, while in their outer 
   trajectory regions they move along their own individual tracks.\\
    It was shown that the kinematic behavior of all the knots could be well
    explained in this framework of  precessing nozzle scenario with a 
    double-jet structure. And in the work (Qian \cite{Qi22b})
    the flux evolutions of  superluminal knots (C4, C5, C9, C10 and C22
    of group-A) were well explained in terms of their Doppler boosting effects,
    which were model-derived from the model-fitting of their kinematics. 
    Similarly, in the work (Qian \cite{Qi23}) the flux density variations of 
    the superluminal knots (C19, C20, C21, B5 and B7 of group-B) were also
    well explained in terms of their Doppler boosting effects.\\
     Thus, based on the assumption of jet-nozzle precession for blazar 3C345,
    the precessing nozzle scenario with a double-jet structure 
    as a hypothesized workingframe is useful to understand the
    kinematics and emission poroperties of the superluminal knots in 3C345.\\
    In this paper we have considered a precessing nozzle scenario with a
     single jet. We assumed that the superluminal knots of group-B
     also moved along the same precessing common helical trajectory pattern
    as the knots of group-A. In this case the trajectory-transits of 
    these knots (C20, C21, B5 and B7) occurred within core distances
     $r_n{<}$0.05--0.15\,mas, so only their outer trajectories were observed, 
    and their flux evolution associated with their superluminal motion could
     be also well explained.
    We expect that the kinematics and flux evolution of the remained knots
    (C15, C15a, C16--C19, B6, B8, B11 and B12) could be interpreted as well.\\
    We would like to emphasize that both the scenarios are based on the the
     possible evidence of the jet (or jets) in 3C345 precssing with a certain
     period (Klare et al. \cite{Kl05}, Qian et al.\cite{Qi09}). Both the
     precessing nozzle scenarios with a single jet or double-jets have been 
     applied  to interpret the kinematics and flux evolution of the 
     superluminal components in 3C345.  Higher resolution VLBI-observations 
     deep into the core (within core distance $r_n{<}$0.05--0.1\,mas)
      with mm-VLBI and Space-VLBI are required to test them.\\
     Finally, we would emphasize that the explanation of the light curves
     of superluminal components associated with their superluminal
     motion in terms of Doppler beaming effects indicates: the 
     light-travel-time 
     effect is important not only for interpreting the apparently superluminal 
     motion of knots observed in blazars, but also for understanding their
     intrinsic variability, because in these cases the radiation of knots 
     is emitted from the "emitting elements" distributed in a region
     with a size of c$\Gamma{\delta}{\Delta{t_{var}}}$ 
     along the line of sight ($\Delta{t_{var}}$--variability time-scale;
     also cf. Sec.3.2). And continuous injection mechanisms should be
     taken into account (e.g., Qian \cite{Qi96b}). 
    \acknowledgements{Qian sincerely thanks Drs. S.G.~Jorstad and Z,~Weaver
   (Boston University, USA) for kindly providing the 43GHz data on the 
   superluminal components measured during the period 2007--2018, which 
   are from the VLBA-BU Monitoring Program, funded by NASA through the 
    Fermi Guest Investigator Program.}

     \end{document}